%% file: ME.tex
\documentclass[a4paper,11pt]{article}
\pdfoutput=1 
\usepackage{jcappub}
\usepackage{amsmath}
\usepackage{graphicx}
\usepackage{latexsym}
\usepackage{xspace}
\usepackage{color}
\usepackage{hyperref} 
\usepackage{bm}
\usepackage{relsize}
\usepackage{tabularx}
\usepackage{multirow}
\usepackage{amssymb}
\usepackage[table]{xcolor}
\usepackage{braket}
\usepackage[utf8]{inputenc}
\usepackage{slashed}
\usepackage{empheq}
\usepackage{cancel}
\usepackage{soul}

\usepackage[margin=1in,includefoot]{geometry}
\usepackage{xfrac} 
\usepackage[compat=1.1.0]{tikz-feynman}

\allowdisplaybreaks

\makeatletter
\gdef\@fpheader{}
\g@addto@macro\bfseries{\boldmath}
\makeatother

\makeatletter
\DeclareRobustCommand*\uell{{\mathpalette\@uell\relax}}
\newcommand*\@uell[2]{
	\setbox0=\hbox{$#1\ell$}
	\setbox1=\hbox{\rotatebox{10}{$#1\ell$}}
	\dimen0=\wd0 \advance\dimen0 by -\wd1 \divide\dimen0 by 2
	\mathord{\lower 0.1ex \hbox{\kern\dimen0\unhbox1\kern\dimen0}}
}

\newcommand{\ds}{\displaystyle}

\input{newcommands}

\title{Benchmarking the cosmological master equations}

\author[a,b,c]{Thomas Colas,}
\affiliation[a]{Universit\'e Paris-Saclay, CNRS, Institut d'Astrophysique Spatiale, 91405, Orsay, France}
\emailAdd{thomas.colas@universite-paris-saclay.fr}

\author[a,b]{Julien Grain,}
\affiliation[b]{Laboratoire Astroparticule et Cosmologie, CNRS Universit\'e Paris Cit\'e, 10 rue
	Alice Domon et L\'eonie Duquet, 75013 Paris, France}
\emailAdd{julien.grain@universite-paris-saclay.fr}

\author[c,b]{Vincent Vennin}
\affiliation[c]{Laboratoire de Physique de l'\'Ecole Normale Sup\'erieure, ENS, Universit\'e PSL, CNRS, Sorbonne Universit\'e, Universit\'e Paris Cit\'e, F-75005 Paris, France}
\emailAdd{vincent.vennin@ens.fr}

\date{today}

\begin{document}
	\sloppy
	
	\abstract{Master equations are commonly employed in cosmology to model the effect of additional degrees of freedom, treated as an ``environment'', onto a given ``system''. However, they rely on assumptions that are not necessarily satisfied in cosmology, where the environment may be out of equilibrium and the background is dynamical. In this work, we apply the master-equation program to a model that is exactly solvable, and which consists of two linearly coupled scalar fields evolving on a cosmological background. The light field plays the role of the system and the heavy field is the environment.  By comparing the exact solution to the output of the master equation, we can critically assess its performance. We find that the master equation exhibits a set of ``spurious'' terms that explicitly depend on the initial conditions, and which arise as a consequence of working on a dynamical background. Although they cancel out in the perturbative limit of the theory (\ie at leading orders in the interaction strength), they spoil resummation. However, when those terms are removed, the master equation performs impressively well to reproduce the power spectra and the amount of the decoherence of the light field, even in the strongly decohered regime. We conclude that master equations are able to perform late-time resummation, even though the system is far from the Markovian limit, provided spurious contributions are suppressed.}
	
	
	
	\maketitle
	
	\flushbottom
	
	\section{Introduction}	
	\label{sec:intro}
 
According to the standard model of cosmology, all structures in our universe emerge from the gravitational amplification of vacuum quantum fluctuations at early times. This idea is supported by the data, \eg~the measurements of the cosmic microwave background anisotropies~\cite{Planck:2018jri}, which reveal that primordial fluctuations are almost scale invariant, quasi Gaussian and adiabatic. Those observations are consistent with a phase of primordial inflation, driven by a single scalar field along a smooth potential.

However, most physical setups that have been proposed to embed inflation contain a large number of additional degrees of freedom~\cite{Martin:2013tda}. Even if they provide negligible contributions to the dynamics of the universe expansion, they may affect the emergence of cosmic structures in various ways. For instance, they could lead to entropic fluctuations, or to deviations from Gaussian statistics, that future cosmological surveys might be able to detect~\cite{Amendola:2016saw, Maartens:2015mra}. They may also contribute to processes occurring after inflation (such as the production of curvature perturbations~\cite{Lyth:2001nq}, dark matter~\cite{Enqvist:2017kzh}, or dark energy~\cite{Ringeval:2010hf, Kiefer:2010pb}) but that crucially depend on the way those extra fields are excited during inflation. At the more fundamental level, additional degrees of freedom may also alter the quantum state in which primordial density fluctuations are placed, in particular through the mechanism of decoherence~\cite{Brandenberger:1992sr, Barvinsky:1998cq, Lombardo:2005iz, Kiefer:2006je, Martineau:2006ki, Burgess:2006jn, Prokopec:2006fc, Nelson:2016kjm, Martin:2018zbe, Martin:2018lin}. Decoherence~\cite{Zurek:1981xq, Zurek:1982ii, Joos:1984uk} is usually associated with the erasure of genuine quantum signatures so this may affect our ability to prove or disprove that cosmic structures are of quantum-mechanical origin~\cite{Hollowood:2017bil, Martin:2021znx}.

For those reasons, it has become of increasing importance to design reliable tools to model the presence of additional degrees of freedom in the early universe~\cite{Koks:1996ga, Anastopoulos:2013zya, Fukuma:2013uxa, Cheung:2007st, Chen:2009zp,Senatore:2010wk, Assassi:2013gxa, Arkani-Hamed:2015bza, Shandera:2017qkg, Akhtar:2019qdn, Burgess:2020tbq, Banerjee:2020ljo, Pinol:2021aun, Oppenheim:2022xjr, Pimentel:2022fsc, Jazayeri:2022kjy, Brahma:2022yxu}. One such approach is the so-called master equation program (see for instance \Refs{Breuer:2002pc, Calzetta:2008iqa}), where an effective equation of motion is obtained for the reduced density matrix of a ``system'' of interest, once the degrees of freedom contained in the ``environment'' have been traced out. One of its appealing advantages is its ability to resum late-time secular effects \cite{Burgess:2014eoa, Boyanovsky:2015tba, Boyanovsky:2015jen, Burgess:2015ajz, Kaplanek:2021fnl, Chaykov:2022zro, Chaykov:2022pwd}, hence to go beyond standard perturbation theory and implement non-perturbative resummations in cosmology. 

However, master equations were primarily developed in the context of quantum optics, so they rely on assumptions (\eg that the environment comprises a large reservoir in thermal equilibrium) that are not necessarily satisfied in cosmology. There, since the background is dynamical, the Hamiltonian is time-dependent \cite{Martin:2007bw} and the environment is generally out-of-equilibrium \cite{Hsiang:2021vgx}. This is why, in this work, we want to understand under which conditions the master-equation program can be employed in cosmology, and what physical insight one shall expect to get out of it. 

We address this issue by considering a toy model that is exactly solvable, such that the output of master equations can be compared to the exact result and examined in a critical way. This allows us to benchmark master equations. In practice, we consider two linearly coupled scalar field evolving on a homogeneous and isotropic universe. The model has been solved exactly in \Refs{Colas:2021llj, Banerjee:2021lqu}, where it has been shown that each Fourier sector is placed in a four-mode squeezed state, which is a Gaussian state. By tracing over the heaviest field, one obtains the reduced state of the lightest field, which follows a non-unitary evolution, and which can be compared with the predictions of different approaches, such as master equations or standard perturbative techniques. In this model, the environment does not reach thermal equilibrium, and as we will show the Markovian limit~\cite{Lindblad:1975ef} is not attained either. This is why it is a priori challenging for conventional master-equation approaches to properly describe its dynamics. 

The rest of this article is organised as follows. In \Sec{sec:CME}, we introduce the master-equation formalism, and clarify the levels at which the different approximations enter the calculation. In \Sec{sec:model}, we introduce the cosmological model mentioned above, and show how it can be solved exactly. We then apply the master-equation program to this setting, and find that it exhibits a set of terms that we dub ``spurious''. These terms do not exist in the perturbative limit of the theory, and they prevent resummation due to their dependence on the initial conditions. In \Sec{sec:sol} we then analyse the ability of the master equation to reproduce the power spectra of the model, as well as to predict the amount of quantum decoherence, when spurious terms are removed ``by hand''. We find that master equations are impressively efficient in that case, even in the strongly decohered regime, and that they perform much better than standard perturbative methods (such as \eg the in-in formalism). This also leads us to draw a few conclusions as to whether a heavy scalar field can efficiently decohere cosmological fluctuations. In \Sec{sec:discuss}, we summarise our main findings and further discuss the status of the spurious terms. The paper ends by a few technical appendices, to which we defer the derivation of some of the results given in the main text.

\section{The master-equation bestiary}\label{sec:CME}
	
The master-equation program proposes to describe the quantum state of a system when it weakly couples to an environment. In practice, one considers a Hamiltonian of the form
\bea\label{eq:Hgen}
\widehat{\calH} = \widehat{\calH}_{\mathrm{S}} + \widehat{\calH}_{\mathrm{E}} + g\widehat{\calH}_{\mathrm{int}}\, ,
\eea
where $\widehat{\calH}_{\mathrm{S}}$ and $ \widehat{\calH}_{\mathrm{E}}$ respectively denote the Hamiltonians for the system and the environment in the absence of interactions, and $g\widehat{\calH}_{\mathrm{int}}$ is the interaction term, controlled by the coupling constant $g$. The system alone is described by the reduced density matrix, which is obtained from the full density matrix by tracing over the environmental degrees of freedom,
\bea
\label{eq:rho:red}
\widehat{\rho}_{\mathrm{red}} = \Tr_{\mathrm{E}}(\widehat{\rho}  )\, .
\eea
An evolution equation for $\widehat{\rho}_{\mathrm{red}}$ can be derived with different levels of approximation, corresponding to as many different master equations.
In this section, we review the most common master equations, see \Refa{Breuer:2002pc} for further details (readers already familiar with the master-equation basic tools may want to skip this section and jump to \Sec{sec:model}).
		
	\subsection{An exact master equation: the Nakajima-Zwanzig equation}
	
Our first step is to derive an exact, formal master equation, before applying an approximation scheme. 		
Hereafter we work in the interaction picture, where quantum states evolve with the interaction Hamiltonian $g\widehat{\calH}_{\mathrm{int}}$ and operators evolve with the free Hamiltonian, \ie the Hamiltonian in the absence of interactions $\widehat{\calH}_0\equiv \widehat{\calH}_{\mathrm{S}} + \widehat{\calH}_{\mathrm{E}}$. Operators in the interaction picture are denoted with an overall tilde, in order to make the distinction with the Schr\"odinger and Heisenberg pictures where they carry an overall hat. The link between the Schr\"odinger and the interaction picture is given by 
\bea
\label{eq:rho:Hint}
\widetilde{\rho}(\eta) = \widehat{\mathcal{U}}^{\dag}_0(\eta) \widehat{\rho}(\eta)\widehat{\mathcal{U}}_0(\eta)  
\quad \text{and} \quad
\widetilde{\mathcal{H}}_{ \mathrm{int}} (\eta) = \widehat{\mathcal{U}}^{\dag}_0(\eta) \widehat{\mathcal{H}}_{\mathrm{int}}(\eta) \widehat{\mathcal{U}}_0(\eta)\, ,  
\eea
where $\eta$ denotes time and where we have introduced the free evolution operator 
\bea
 \widehat{\mathcal{U}}_0(\eta)= \mathcal{T} \exp\left[{-i \int_{\eta_0}^\eta \widehat{\calH}_0({\eta}')\dd{\eta}'}\right]= \mathcal{T} \exp\left[{-i \int_{\eta_0}^\eta \widehat{\calH}_{\mathrm{S}}({\eta}')\dd{\eta}'}\right] \otimes \mathcal{T} \exp\left[{-i \int_{\eta_0}^\eta \widehat{\calH}_{\mathrm{E}}({\eta}')\dd{\eta}'}\right] ,
 \eea
with $\mathcal{T} $ indicating time ordering (time arguments increase from right to left). 
In this work we employ natural units where $\hbar=c=1$. As mentioned above, in the interaction picture the total density matrix evolves with the interaction Hamiltonian,
		\bea \label{eq:LVN}
		\frac{\dd \widetilde{\rho}}{\dd \eta}=
		-ig \left[ \widetilde{\mathcal{H}}_{\mathrm{int}}(\eta),\widetilde{\rho}(\eta) \right] \equiv g \mathcal{L}(\eta)\widetilde{\rho}(\eta) \, ,
		\eea
which defines the Liouville--Von-Neumann super-operator\footnote{In this work, following \Refa{Breuer:2002pc}, ``super-operator'' denotes an operation which maps positive operators to positive operators.} $\mathcal{L}$. 		

Let us now introduce the projection super-operator $\mathcal{P}$, defined as  
		\begin{align}
		 \mathcal{P}  \widetilde{\rho}  = \mathrm{Tr}_{\mathrm{E}} \left(\widetilde{\rho} \right) \otimes \widetilde{\rho}_{\mathrm{E}}\, ,
		\end{align}
where $\widetilde{\rho}_{\mathrm{E}}$ is a fixed reference state in the environment. In practice, it is taken as the state of the environment in the absence of interactions with the system, which is indeed constant in the interaction picture. One can check that $\mathcal{P}$ is a projector, \ie $\mathcal{P}^2 = \mathcal{P} $, and that $ \mathcal{P}  \widetilde{\rho}$ contains the relevant information to reconstruct the reduced state~\eqref{eq:rho:red} of the system. Upon applying the super-projector $\mathcal{P}$ and its complementary projector $\mathcal{Q}=\mathrm{Id}-\mathcal{P}$ to \Eq{eq:LVN}, one obtains
		\begin{align}
			\frac{\partial }{\partial \eta} \mathcal{P}\widetilde{\rho}(\eta) &= g \mathcal{P} \mathcal{L}(\eta) \widetilde{\rho} (\eta), \\ 
			\label{eq:eomenv}		\frac{\partial }{\partial \eta} \mathcal{Q}\widetilde{\rho}(\eta) &= g \mathcal{Q} \mathcal{L}(\eta) \widetilde{\rho} (\eta).
		\end{align}
Here we have used that since the reference state $\widetilde{\rho}_{\mathrm{E}}$ is independent of time, $\mathcal{P}$ and $\mathcal{Q}$ commute with $\partial/\partial\eta$. Inserting the identity $\mathrm{Id} = \mathcal{P} + \mathcal{Q}$ between the Liouville operator and the density matrix, one obtains
		\begin{align}
			\label{eq:Prhoeom}	\frac{\partial }{\partial \eta} \mathcal{P}\widetilde{\rho}(\eta) &= g \mathcal{P} \mathcal{L}(\eta)\mathcal{P} \widetilde{\rho} (\eta) + g \mathcal{P} \mathcal{L}(\eta)\mathcal{Q} \widetilde{\rho} (\eta), \\ 
			\label{eq:Qrhoeom}
			\frac{\partial }{\partial \eta} \mathcal{Q}\widetilde{\rho}(\eta) &= g \mathcal{Q} \mathcal{L}(\eta)\mathcal{P} \widetilde{\rho} (\eta) +g \mathcal{Q} \mathcal{L}(\eta)\mathcal{Q} \widetilde{\rho} (\eta).
		\end{align}
		A formal solution of \Eq{eq:Qrhoeom} is given by
		\begin{align}\label{eq:Qrho}
			\mathcal{Q}\widetilde{\rho}(\eta) = \mathcal{G}_{\mathcal{Q}}(\eta, \eta_0) \mathcal{Q}\widetilde{\rho}(\eta_0) +  g \int^{\eta}_{\eta_0} \dd \eta' \mathcal{G}_{\mathcal{Q}}(\eta, \eta') \mathcal{Q} \mathcal{L}(\eta')\mathcal{P} \widetilde{\rho} (\eta'),
		\end{align}
		where $\eta_0$ is some initial time and $\mathcal{G}_{\mathcal{Q}}(\eta, \eta')$ is the propagator defined as 
		\begin{align}\label{eq:Qpropag}
			\mathcal{G}_{\mathcal{Q}}(\eta, \eta') \equiv \mathcal{T} \exp \left[g\int_{\eta'}^{\eta} \dd \eta''  \mathcal{Q} \mathcal{L}(\eta'')\right] .
		\end{align}	
Plugging \Eq{eq:Qrho} into \Eq{eq:Prhoeom}, one then obtains a closed equation for the time evolution of the projected density matrix $\mathcal{P}\widetilde{\rho}$, namely
		\begin{align}
			\frac{\partial }{\partial \eta} \mathcal{P}\widetilde{\rho}(\eta) &= g \mathcal{P} \mathcal{L}(\eta) \mathcal{G}_{\mathcal{Q}}(\eta, \eta_0) \mathcal{Q}\widetilde{\rho}(\eta_0) + g \mathcal{P} \mathcal{L}(\eta)\mathcal{P} \widetilde{\rho} (\eta) + g^2 \int^{\eta}_{\eta_0} \dd \eta' \mathcal{P} \mathcal{L}(\eta) \mathcal{G}_{\mathcal{Q}}(\eta, \eta') \mathcal{Q} \mathcal{L}(\eta')\mathcal{P} \widetilde{\rho} (\eta') .
		\end{align}
This is the Nakajima-Zwanzig equation. Although formal, it provides an exact master equation for the reduced state of the system. It can be further simplified by assuming that the initial state does not contain correlations between the system and the environment, \ie $\widetilde{\rho}(\eta_0)= \mathrm{Tr}_{\mathrm{E}}(\widetilde{\rho})\otimes \mathrm{Tr}_{\mathrm{S}}(\widetilde{\rho})=\mathrm{Tr}_{\mathrm{E}}(\widetilde{\rho})\otimes \widetilde{\rho}_{\mathrm{E}}$, hence $\mathcal{Q}\widetilde{\rho}(\eta_0) = 0$. Moreover, without loss of generality one can assume that the expectation value of the interaction Hamiltonian vanishes in the reference state, \ie $\Tr_{\mathrm{E}} (\widetilde{\calH}_{\mathrm{int}} \widetilde{\rho}_{\mathrm{E}}) = 0$ [if this is not satisfied, one simply redefines $\widetilde{\calH}_{\mathrm{S}}$ by adding  $g \Tr_{\mathrm{E}} (\widetilde{\calH}_{\mathrm{int}} \widetilde{\rho}_{\mathrm{E}})\otimes \mathrm{Id}_{\mathrm{E}}$ to it]. This leads to $\mathcal{P} \mathcal{L}(\eta)\mathcal{P} = 0$,\footnote{This can be shown by computing 
\bea
\mathcal{P} \mathcal{L}\mathcal{P} \widetilde{\rho}= -i \mathcal{P} \left[\widetilde{\calH}_{\mathrm{int}} ,\mathcal{P} \widetilde{\rho}\right]
 = -i \mathcal{P} \left[\widetilde{\calH}_{\mathrm{int}} ,\Tr_{\mathrm{E}}(\widetilde{\rho})\otimes \widetilde{\rho}_{\mathrm{E}}\right]
 =-i \left[\Tr_{\mathrm{E}}\left(\widetilde{\calH}_{\mathrm{int}}  \widetilde{\rho}_{\mathrm{E}}\right), \Tr_{\mathrm{E}} \left(\widetilde{\rho}\right)\right]\otimes \widetilde{\rho}_{\mathrm{E}}=0\, .
\eea
\label{footnote:PLP}
} so the Nakajima-Zwanzig equation reduces to
		\begin{align}\label{eq:NZ}
			\frac{\partial}{\partial  \eta} \mathcal{P}\widetilde{\rho}(\eta) = g^2\int_{\eta_0}^{\eta} \dd \eta' \mathcal{K}(\eta, \eta') \mathcal{P}\widetilde{\rho}(\eta'),
		\end{align}  
where we have introduced the memory kernel $\mathcal{K}(\eta, \eta') $ defined as
		\begin{align}
		\label{eq:calK:def}
			\mathcal{K}(\eta, \eta') =   \mathcal{P} \mathcal{L}(\eta) \mathcal{G}_{\mathcal{Q}}(\eta, \eta') \mathcal{Q} \mathcal{L}(\eta')\mathcal{P}.
		\end{align}
In this form, the master equation is as difficult to solve as the Liouville equation~\eqref{eq:LVN} of the full setup. However, it allows efficient approximation schemes to be designed, as we shall now see. The first approximation relies on the assumption of weak coupling between the system and the environment and is discussed in \Sec{sec:Born}, the second approximation concerns properties of the environment itself and is developed in \Sec{sec:Lindblad}.
		\subsection{Born approximation: the time-convolutionless cumulant expansion}		
		\label{sec:Born}
An effective description of the system alone is in general possible only when it weakly couples to its environment. This naturally provides a small parameter, namely the interaction strength, in which to perform an expansion. This is the so-called Born approximation, which also addresses one of  the difficulties inherent to the Nakajima-Zwanzig equation~\eqref{eq:NZ}, namely the fact that it is non-local in time, \ie the time derivative of $\mathcal{P}\widetilde{\rho}(\eta)$ depends on its past history $\mathcal{P}\widetilde{\rho}(\eta')$ for $\eta' < \eta$. 
		The Time-ConvolutionLess projection operator method (TCL in the following) consists in expanding the dynamics of the system in powers of the coupling constant $g$, rendering the equation local in time (while preserving its non-Markovian nature\footnote{In this work, following \Refa{Breuer:2002pc}, the dynamical map $\widetilde{\rho}(\eta)\to\widetilde{\rho}(\eta')=\mathcal{M}_{\eta\to\eta'}\widetilde{\rho}(\eta)$ is said to be Markovian if its generators form a semi-group, \ie $\mathcal{M}_{\eta\to\eta'}=\mathcal{M}_{\eta''\to\eta'}\mathcal{M}_{\eta\to\eta''}$. Note that a Markovian master equation is necessarily local in time, but the reverse is not necessarily true.\label{footnote:Markovian}}). One thus obtains an equation of the form
\begin{align}
\label{eq:cumulant}
			\frac{\partial}{\partial \eta}  \mathcal{P}\widetilde{\rho}(\eta) = \sum_{n=2}^{\infty} g^n\mathcal{K}_n(\eta) \mathcal{P}\widetilde{\rho}(\eta)\, ,
\end{align}
where the $\mathcal{K}_n$ operators are called the TCL${}_n$ operators and can be computed iteratively. This can be done by expanding \Eq{eq:Qpropag} in $g$, and by using \Eq{eq:cumulant} to express $\mathcal{P}\widetilde{\rho}(\eta')$ in terms of $\mathcal{P}\widetilde{\rho}(\eta)$ in the right-hand side of \Eq{eq:NZ}, at the required order. For instance, at leading order in $g$, $ \mathcal{G}_{\mathcal{Q}}(\eta, \eta') =\mathrm{Id}$, see \Eq{eq:Qpropag}, so \Eq{eq:calK:def} leads to $\mathcal{K}(\eta,\eta')=\mathcal{P} \mathcal{L}(\eta) \mathcal{Q} \mathcal{L}(\eta')\mathcal{P} = \mathcal{P} \mathcal{L}(\eta) \mathcal{L}(\eta')\mathcal{P}$, where we have used that $ \mathcal{Q}= 1-\mathcal{P}$ and that $\mathcal{P}\mathcal{L}\mathcal{P}=0$, see footnote~\ref{footnote:PLP}. At that order, \Eq{eq:NZ} also indicates that $\mathcal{P}\widetilde{\rho}$ is constant hence
		\begin{align}
		\label{eq:K2}
			\mathcal{K}_2 (\eta) &= \int_{\eta_0}^{\eta} \dd \eta' \mathcal{P} \mathcal{L}(\eta) \mathcal{L} (\eta') \mathcal{P}\, ,
		\end{align}	
and truncating \Eq{eq:cumulant} at order $n=2$ leads to the TCL${}_2$ master equation
\begin{align}
\label{eq:TCL2exp}
			\frac{\dd \widetilde{\rho}_{\text{red}}}{\dd \eta} = - g^2 \int_{\eta_0}^{\eta} \dd \eta' \Tr_{\mathrm{E}} \left[\widetilde{\mathcal{H}}_{\text{int}}(\eta),\left[ \widetilde{\mathcal{H}}_{\text{int}}(\eta'),  \widetilde{\rho}_{\text{red}}(\eta) \otimes \widetilde{\rho}_{\mathrm{E}}\right]\right].
\end{align}
			
This expansion can be carried on. At order $n=3$, one needs to expand the memory kernel $\mathcal{K}(\eta,\eta')$ at order $g$ and keep $\mathcal{P}\widetilde{\rho}(\eta')\simeq \mathcal{P}\widetilde{\rho}(\eta) $ in the right-hand side of \Eq{eq:NZ}, given that $\mathcal{P}\widetilde{\rho}(\eta')- \mathcal{P}\widetilde{\rho}(\eta) =\order{g^2}$ as shown above. One obtains $\mathcal{K}_3 (\eta) = \int_{\eta_0}^\eta \dd\eta' \int_{\eta'}^\eta\dd\eta'' \mathcal{P}\mathcal{L}(\eta)\mathcal{Q}\mathcal{L}(\eta'')\mathcal{Q}\mathcal{L}(\eta')\mathcal{P}$, so
\bea
\mathcal{K}_3 (\eta) = \int_{\eta_0}^\eta \dd\eta' \int_{\eta'}^\eta\dd\eta'' \mathcal{P}\mathcal{L}(\eta)\mathcal{L}(\eta'')\mathcal{L}(\eta')\mathcal{P}
\eea 
where we have used again that $ \mathcal{Q}= 1-\mathcal{P}$ and that $\mathcal{P}\mathcal{L}\mathcal{P}=0$. Note that, if the odd moments of the interaction Hamiltonian vanish in the environment (as will be the case for the model studied in the rest of this work), \ie $\Tr_{\mathrm{E}}[\mathcal{H}_{\mathrm{int}}(\eta_1) \cdots \mathcal{H}_{\mathrm{int}}(\eta_{2p+1}) \widetilde{\rho}_{\mathrm{E}} ]=0$, a similar calculation as the one performed in footnote~\ref{footnote:PLP} for $p=1$ then shows that $\mathcal{P} \mathcal{L}(\eta_1)\cdots\mathcal{L}(\eta_{p+1})\mathcal{P}=0$. This implies that $\mathcal{K}_3$ vanishes, as well as all odd TCL${}_n$ generators.

In that case, the leading correction comes from TCL${}_4$, which receives two contributions. The first one comes from the term of order $g^2$ in the memory kernel $\mathcal{K}(\eta,\eta')$ while keeping $\mathcal{P}\widetilde{\rho}(\eta')\simeq \mathcal{P}\widetilde{\rho}(\eta) $ in the right-hand side of \Eq{eq:NZ}. The second contribution comes from keeping the memory kernel at leading order but expand $\mathcal{P}\widetilde{\rho}(\eta') $ at order $g^2$. The latter can be formally obtained from the TCL${}_2$ equation, the solution of which reads $\mathcal{P}\widetilde{\rho}(\eta')= \mathcal{P}\widetilde{\rho}(\eta_0)+g^2 \int_{\eta_0}^{\eta' }\mathcal{K}_2(\eta'')  \mathcal{P}\widetilde{\rho}(\eta'')\dd\eta''$. Together with \Eq{eq:K2}, this leads to 
		\begin{align}\label{eq:TCL4gen}
			\mathcal{K}_4 (\eta) &= \int_{\eta_0}^{\eta} \dd \eta_1 \int_{\eta_0}^{\eta_1} \dd \eta_2 \int_{\eta_0}^{\eta_2} \dd \eta_3 \Big[ \mathcal{P} \mathcal{L}(\eta) \mathcal{L} (\eta_1)  \mathcal{L}(\eta_2) \mathcal{L} (\eta_3) \mathcal{P} - \mathcal{P} \mathcal{L}(\eta) \mathcal{L} (\eta_1) \mathcal{P}  \mathcal{L}(\eta_2) \mathcal{L} (\eta_3) \mathcal{P} \nonumber \\ 
			-& \mathcal{P} \mathcal{L}(\eta) \mathcal{L} (\eta_2) \mathcal{P}  \mathcal{L}(\eta_1) \mathcal{L} (\eta_3) \mathcal{P} - \mathcal{P} \mathcal{L}(\eta) \mathcal{L} (\eta_3) \mathcal{P}  \mathcal{L}(\eta_1) \mathcal{L} (\eta_2) \mathcal{P} \Big] .
		\end{align}
This expansion can be carried on to the required level of accuracy, which allows one to work out \Eq{eq:cumulant} when truncated at the corresponding order TCL${}_n$. Note that, even if the TCL${}_2$ order may be sufficient for practical purposes, the derivation of the fourth-order generator is useful to control the validity of the cumulant expansion, by evaluating the error estimate $g^2 ||\mathcal{K}_4 ||/||\mathcal{K}_2 ||$ and checking that it is indeed small.
		\subsection{Markovian approximation: the Lindblad equation}
		\label{sec:Lindblad}
The TCL${}_2$ master equation~\eqref{eq:TCL2exp} is in general not Markovian in the sense given in footnote~\ref{footnote:Markovian}, since it involves a convolution over the past history through the integral over $\eta'$.
However, a further approximation can be performed that renders the dynamics Markovian. This leads to the so-called Gorini--Kossakowski--Sudarshan--Lindblad equation, in short Lindblad equation in what follows. It can be obtained by first decomposing the interaction Hamiltonian as
		\begin{align}
			\widehat{\mathcal{H}}_{\mathrm{int}}(\eta) = \sum_i \widehat{\bs{O}}^{(\mathrm{S})}_i(\eta) \otimes \widehat{\bs{O}}^{(\mathrm{E})}_i(\eta)\, ,
		\end{align}
where $\widehat{\bs{O}}^{(\mathrm{S})}_i$ and $\widehat{\bs{O}}^{(\mathrm{E})}_i$ form a basis of operators acting on the system and the environment respectively. Plugging this decomposition into \Eq{eq:TCL2exp}, the TCL${}_2$ master equation reads
		\begin{align}\label{eq:TCL2expdev}
			\frac{\dd \widetilde{\rho}_{\mathrm{red}}}{\dd \eta} =& - \sum_{i,j} g^2  \int_{\eta_0}^{\eta} \dd \eta' \Bigg\{ \Rea\left[\bs{\mathcal{K}}_{ij}^{>}(\eta,\eta')\right] \left[\widetilde{\bs{O}}^{(\mathrm{S})}_i(\eta),\left[\widetilde{\bs{O}}^{(\mathrm{S})\dagger}_j(\eta'), \widetilde{\rho}_{\mathrm{red}} (\eta) \right]\right] \nonumber \\ 
			&+ i \Ima\left[\bs{\mathcal{K}}_{ij}^{>}(\eta,\eta')\right] \left[\widetilde{\bs{O}}^{(\mathrm{S})}_i(\eta),\left\{\widetilde{\bs{O}}^{(\mathrm{S})\dagger}_j(\eta'), \widetilde{\rho}_{\mathrm{red}} (\eta) \right\}\right] \Bigg\},
		\end{align}
		where $\left\{A,B\right\} \equiv AB+BA$ denotes the anticommutator and the memory kernel $\bs{\mathcal{K}}_{ij}^{>}(\eta,\eta')$ is defined as
		\begin{align} \label{eq:memory}
			\bs{\mathcal{K}}_{ij}^{>}(\eta,\eta') = \Tr_{\mathrm{E}} \left[\widehat{\bs{O}}^{(\mathrm{E})}_i(\eta) \widehat{\bs{O}}^{(\mathrm{E})\dagger}_j(\eta') \widehat{\rho}_{\mathrm{E}}\right].
		\end{align}
This expression is given in the Heisenberg picture.
It involves the two-point correlation functions of the $\widehat{\bs{O}}^{(\mathrm{E})}_i$ operators in the environment, and thus depends on the environment properties. 

Typical environments contain a large number of degrees of freedom, hence they behave as reservoirs in which these correlation functions quickly decay with $\vert\eta-\eta'\vert$. More precisely, if the relaxation time of the environment is small compared to the typical time scales over which the system evolves, one may coarse-grain the evolution of the system on scales larger than the environment relaxation time. The memory kernel is then sharply peaked, such that the integral over $\eta'$ only receives contributions close to its upper bound $\eta$. In this limit, the past history ($\eta'<\eta$) is not involved in the dynamics anymore, which therefore becomes Markovian. 

Formally, if $\bs{\mathcal{K}}_{ij}^{>}(\eta,\eta') \propto \delta(\eta-\eta')$, in the Schr\"odinger picture \Eq{eq:TCL2expdev} takes the form
		\begin{align}\label{eq:GKSL}
		\frac{\dd \widehat{\rho}_{\mathrm{red}}}{\dd \eta}  &=-i\left[\widehat{\mathcal{H}}_{\mathrm{S}}(\eta),\widehat{\rho}_{\mathrm{red}}(\eta)\right]+\ds\sum_{i,j}\bs{\mathcal{D}}_{ij}\left[\widehat{\bs{O}}^{(\mathrm{S})}_i\widehat{\rho}_{\mathrm{red}}(\eta) \widehat{\bs{O}}^{\dag (\mathrm{S})}_j-\frac{1}{2}\left\{\widehat{\bs{O}}^{\dag (\mathrm{S})}_j\widehat{\bs{O}}^{(\mathrm{S})}_i,\widehat{\rho}_{\mathrm{red}}(\eta)\right\}\right] ,
		\end{align}
where the dissipator matrix $\bs{\mathcal{D}}_{ij}$ is a positive semi-definite matrix. This entails that it can be diagonalised by a unitary transformation (due to the hermiticity implied by the positive semi-definiteness), and in this basis \Eq{eq:GKSL} becomes\footnote{Another approximation known as the rotating-wave approximation is sometimes performed to obtain the Lindblad equation. Since the evolution of the system is coarse-grained over time scales larger than those describing the dynamics of the environment, this approximation consists in removing the quickly oscillating terms appearing in the master equation, for consistency. The implementation of this approach is however challenging in cosmology, where the dynamical background prevents the existence of a natural frequency basis~\cite{Kaplanek:2022xrr}.}
		\begin{align}		
			\frac{\dd \widehat{\rho}_{\mathrm{red}}}{\dd \eta}  &=-i\left[\widehat{\mathcal{H}}_{\mathrm{S}}(\eta),\widehat{\rho}_{\mathrm{red}}(\eta)\right]+\ds\sum_{k}\gamma_k\left[\widehat{\bs{L}}_k\widehat{\rho}_{\mathrm{red}}(\eta) \widehat{\bs{L}}^{\dag}_k-\frac{1}{2}\left\{\widehat{\bs{L}}^{\dag}_k \widehat{\bs{L}}_k,\widehat{\rho}_{\mathrm{red}}(\eta)\right\}\right]
		\end{align} 
where $\widehat{\bs{L}}_k$ are the so-called jump operators and $\gamma_k$ are the positive eigenvalues of the dissipator matrix. This is called a Lindblad equation and is the most generic form of a Markovian dynamical equation that preserves trace, Hermiticity and positivity of the density matrix~\cite{Lindblad:1975ef}. This is why Lindblad equations play a key role when studying environmental effects. However, they rely on strong hypotheses regarding the decay rate of the memory kernel in the environment, which may or may not be always satisfied. Indeed, in the cosmological context, fields evolve on a dynamical background, which implies that the environment does not necessarily reach a stationary state in which fluctuations swiftly decay. One of the goals of this article is to check the reliability of the master-equation approach for cosmological systems.
       \subsection{Link with perturbative methods}\label{sec:inin}
Later on in this work, we will investigate the extent to which TCL master equations go beyond perturbative effects and enable some non-perturbative resummation. At this stage however, it is important to stress that, when solved perturbatively, they reduce to standard perturbative results. This is because, when deriving the TCL${}_n$ equation, no contribution of order lower than $g^n$ has been dropped.  

More explicitly, the Liouville--Von-Neumann equation~\eqref{eq:LVN} can be formally solved as
    	\begin{align}\label{eq:formal}
    		\widetilde{\rho}(\eta)  &= \vert\cancel{0}\rangle\langle\cancel{0}\vert -i g \int_{-\infty}^{\eta}\dd \eta' \left[\widetilde{\mathcal{H}}_{\mathrm{int}}(\eta'),\widetilde{\rho}(\eta')\right] ,
    	\end{align}
where $\left|\cancel{0}\right>$ denotes the initial state of the combined system-environment setup. By recursively evaluating $\widetilde{\rho}$ in the right-hand side with \Eq{eq:formal} itself, one obtains
        \begin{align}
        \label{eq:rho:iterative:SPT}
            \widetilde{\rho}(\eta) = \sum_{n=0}^\infty (-ig)^n \int_{-\infty}^{\eta}\dd \eta_1\int_{-\infty}^{\eta_1}\dd \eta_2 \cdots \int_{-\infty}^{\eta_{n-1}}\dd \eta_n
             \left[\widetilde{\mathcal{H}}_{\mathrm{int}}(\eta_1),\left[ \widetilde{\mathcal{H}}_{\mathrm{int}}(\eta_2),\cdots \left[\widetilde{\mathcal{H}}_{\mathrm{int}}(\eta_n),\vert\cancel{0}\rangle\langle\cancel{0}\vert \right] \cdots \right] \right] , 
        \end{align}
which displays all contributions to the quantum state order-by-order in $g$. In turn, this allows one to compute corrections to the observables at all orders, as in the in-in formalism.\footnote{This can also be shown in the in-in formalism, where the expectation value of an operator $\widehat{O}$ at time $\eta$ reads
		\begin{align}
		\label{eq:in-in:meanO:gen}
		    \langle \widehat{O} \rangle(\eta) = \left<\cancel{0}\right| \overline{\mathcal{T}}\left[\ee^{i g \int_{- \infty }^{\eta} \dd \eta' \widetilde{\mathcal{H}}_{ \mathrm{int}}(\eta') } \right]\widetilde{O}(\eta)\mathcal{T}\left[\ee^{-i g \int_{- \infty }^{\eta} \dd \eta'' \widetilde{\mathcal{H}}_{ \mathrm{int}}(\eta'') } \right] \left|\cancel{0}\right>\, ,
		\end{align}
where $\overline{\mathcal{T}}$ denotes anti time-ordering. By Taylor expanding the exponential functions, one obtains
	\begin{align}\label{eq:pertinin}
		  \langle\widehat{O}\rangle(\eta)	 &=\sum_{n=0}^{\infty}( ig)^n  \int_{-\infty}^{\eta}\dd \eta_1\int_{-\infty}^{\eta_1}\dd \eta_2 \cdots \int_{-\infty}^{\eta_{n-1}}\dd \eta_n
		   \left<\cancel{0}\right|\left[\widetilde{\mathcal{H}}_{\mathrm{int}}(\eta_n),\left[\widetilde{\mathcal{H}}_{\mathrm{int}}(\eta_{n-1}), \cdots \left[\widetilde{\mathcal{H}}_{\mathrm{int}}(\eta_1),\widetilde{O}(\eta) \right]\cdots \right] \right] \left|\cancel{0}\right> .
		\end{align}
		Using that $\langle\widehat{O}\rangle(\eta)=\mathrm{Tr}[\widetilde{O}(\eta)\widetilde{\rho}(\eta) ]$, together with $\mathrm{Tr}[\widetilde{O}(\eta)[\widetilde{\mathcal{H}}_{\mathrm{int}}(\eta_i),\vert\cancel{0}\rangle\langle\cancel{0}\vert]]=-\langle\cancel{0}\vert[\widetilde{\mathcal{H}}_{\mathrm{int}}(\eta_i),\widetilde{O}(\eta)]\vert\cancel{0}\rangle$, this is indeed consistent with \Eq{eq:rho:iterative:SPT}.}

Let us see how this compares with a perturbative solution of TCL${}_n$. For TCL${}_2$, since the right-hand side of \Eq{eq:TCL2exp} is proportional to $g^2$, one has $\widetilde{\rho}_{\mathrm{red}}(\eta)\otimes\widetilde{\rho}_{\mathrm{E}}=\widetilde{\rho}_{\mathrm{red}}(\eta_0)\otimes\widetilde{\rho}_{\mathrm{E}}+\order{g^2}=\vert\cancel{0}\rangle \langle\cancel{0}\vert+\order{g^2}$, and \Eq{eq:TCL2exp} leads to 
\bea
\widetilde{\rho}_{\mathrm{red}}(\eta)=\vert\cancel{0}\rangle\langle\cancel{0}\vert-g^2\int_{\eta_0}^{\eta} \dd \eta' \Tr_{\mathrm{E}} \left[\widetilde{\mathcal{H}}_{\mathrm{int}}(\eta),\left[ \widetilde{\mathcal{H}}_{\text{int}}(\eta'),  \vert\cancel{0}\rangle\langle\cancel{0}\vert\right]\right]+\order{g^4}\, .
\eea
Assuming that  $\Tr_{\mathrm{E}} (\widetilde{\calH}_{\mathrm{int}} \widetilde{\rho}_{\mathrm{E}}) = 0$ as done above \Eq{eq:NZ}, this reduces to \Eq{eq:rho:iterative:SPT} when traced over the environmental degrees of freedom and truncated at order $g^2$. This shows that solving TCL${}_2$ at order $g^2$ is equivalent to Standard Perturbation Theory (SPT hereafter) at that same order. Likewise, one can show that solving TCL${}_n$ perturbatively at order $g^n$ is equivalent to SPT${}_n$. Therefore, TCL${}_n$ contains \emph{all} terms of order $g^n$, and \emph{some} terms of order $g^{m>n}$.\footnote{Let us stress that since the TCL expansion is organised differently from the one of SPT, it does not admit a straightforward diagrammatic representation. In this sense it is more comparable to the Dynamical Renormalisation Group (DRG) resummation \cite{Boyanovsky:1998aa, Burgess:2009bs, Green:2020txs} where diagrams are partially resummed.\label{footnote:DRG}}

This is why TCL is at least as good as SPT, and one of our goals is to determine how much better it is when employed in a cosmological context. 
In other words, when master equations are used as \textit{bona fide} dynamical maps (\ie when they are taken \textit{per se} and solved without further perturbative expansion), we want to investigate their ability to resum late-time secular effects in situations of cosmological interest~\cite{Boyanovsky:2015tba, Burgess:2015ajz, Brahma:2021mng}.

	\section{Curved-space Caldeira-Leggett model}\label{sec:model} 
	Let us now apply the master-equation program to two massive test fields $\varphi$ and $\chi$ in a Friedmann-Lema\^itre-Robertson-Walker geometry, described by the metric
	\begin{eqnarray}\label{eq:metric}
	\mathrm{d}s^2 = a^2(\eta) \left(- \mathrm{d}\eta^2 + \dd\vec{x}^2\right), 
	\end{eqnarray}	
where $a$ is the scale factor and $\eta$ is conformal time. For convenience we restrict the analysis to a de-Sitter background for which $a (\eta) \equiv -1/(H  \eta)$, where $H$ is the constant Hubble parameter and $\eta$ varies between $-\infty$ and $0$. We consider the case where the fields are minimally coupled to gravity and where their self-interaction is quadratic, so the action is of the form
	\begin{eqnarray}
	\label{eq:action:CCL}
	\kern-0.5em S = - \kern-0.5em \int   \kern-0.2em  \mathrm{d}^4x \sqrt{-\det g} \left[\left(\frac{1}{2} g^{\mu \nu}\partial_\mu \varphi \partial_\nu \varphi +\frac{1}{2}m^2 \varphi^2\right) + \left(\frac{1}{2} g^{\mu \nu}\partial_\mu \chi \partial_\nu \chi +\frac{1}{2}M^2 \chi^2\right) + \lambda^2 \varphi \chi \right] .
	\end{eqnarray}
In this expression, $m$ and $M$ are the masses of the two fields and we assume that they satisfy $m<3H/2<M$. So $\varphi$ and $\chi$ can be respectively considered as light and heavy, in the cosmological sense. Having in mind possible applications to cosmological perturbations, where the adiabatic degree of freedom is light, in what follows they will respectively play the role of the system and of the environment. The parameter $\lambda$, which also has dimension of a mass, controls their interaction. If those fields were to describe cosmological perturbations, higher-order interaction terms would be parametrically suppressed, and this setting would correspond to the leading order in cosmological perturbation theory. This model, refereed to as the curved-space Caldeira-Leggett model~\cite{Caldeira:1981rx, Caldeira:1982uj, Caldeira:1982iu, Banerjee:2021lqu}, is therefore of physical interest, and as we shall now see it has the advantage to be exactly solvable. 

The quantum state of the fields $\varphi$ and $\chi$ was studied in details in \Refs{Colas:2021llj, Choudhury:2022btc}, where it was shown that each Fourier sector is placed in a four-mode squeezed state. On super-Hubble scales, the dynamical background leads to the creation of pairs of particles with opposite wave-momenta in each field, and the interaction then entangles these particles, leading to correlations between the two fields. Four-mode squeezed states are Gaussian states, and since the action~\eqref{eq:action:CCL} is quadratic Gaussianity is indeed preserved throughout the evolution. Such states are fully described by their covariance matrix (\ie their quantum two-point expectation values). This is why our goal is now to compute the covariance matrix of the system.
	
\subsection{Exact description}\label{subsec:exact}
	
The action~\eqref{eq:action:CCL} being quadratric, different Fourier modes decouple on a homogeneous background, which makes it useful to introduce 
	\begin{align}\label{eq:Fourier}
		v_{\varphi}(\eta,\bs{k}) \equiv  a(\eta)\int_{\mathbb{R}^{3}} \frac{\mathrm{d}^3 \boldmathsymbol{x}}{(2\pi)^{3/2}} \varphi(\bs{x}) e^{-i\boldsymbol{k}.\bs{x}}
		\quad\text{and}\quad 
		v_{\chi}(\eta,\bs{k}) \equiv  a(\eta)\int_{\mathbb{R}^{3}} \frac{\mathrm{d}^3 \boldmathsymbol{x}}{(2\pi)^{3/2}} \chi(\bs{x}) e^{-i\boldsymbol{k}.\bs{x}}.
	\end{align}	
An additional prefactor $a$ is introduced in these expressions for later convenience. The conjugate momenta can be obtained from \Eq{eq:action:CCL} and read
	\begin{align}
	\label{eq:momenta:def}
	    p_\varphi = v'_\varphi - \frac{a'}{a} v_\varphi 
	    \quad\text{and}\quad
	    p_\chi = v'_\chi - \frac{a'}{a} v_\chi\, ,
	\end{align}
where hereafter a prime denotes derivation with respect to the conformal time $\eta$. A Legendre transform gives the Hamiltonian
	\begin{eqnarray}\label{eq:Hamiltu}
	H = \int_{\mathbb{R}^{3+}} \mathrm{d}^3\boldmathsymbol{k} \bs{z}^{\dag}\boldsymbol{H}(\eta)\bs{z}\, ,
	\end{eqnarray}
where the phase-space variables have been arranged into the vector $\bs{z} \equiv (v_\varphi, p_\varphi, v_\chi, p_\chi)^{\mathrm{T}}$, and where $\boldsymbol{H}$ is a four-by-four matrix given by
	\bea\label{eq:Hmat}
	\boldsymbol{H}(\eta)=\left(\begin{array}{cc}
		\boldsymbol{H}^{(\varphi)} & \boldsymbol{V} \\
		\boldsymbol{V} & \boldsymbol{H}^{(\chi)}
	\end{array}\right),
	\eea
with
	\begin{align}
	\label{eq:Hvarphimat}			\boldsymbol{H}^{(\varphi)}(\eta) = \begin{pmatrix}
					k^2 + m^2 a^2 &\frac{a'}{a} \\ 
					\frac{a'}{a} & 1
				\end{pmatrix} ,
				\quad
				\boldsymbol{H}^{(\chi)}(\eta) = \begin{pmatrix}
					k^2 + M^2 a^2 & \frac{a'}{a} \\ 
					 \frac{a'}{a} & 1
				\end{pmatrix} ,
				\quad
			\bs{V}(\eta) \equiv \begin{pmatrix}
					\lambda^2 a^2 & 0 \\ 
					0 & 0
				\end{pmatrix}.
			\end{align}
Note that, since $\varphi$ and $\chi$ are real fields, one has 	$\bs{z}^{*}(\eta,\bs{k}) = \bs{z}(\eta,-\bs{k})$. This explains why, in order to avoid double counting, the integral in \Eq{eq:Hamiltu} is performed over $\mathbb{R}^{3+}\equiv\mathbb{R}^2\times\mathbb{R}^+$.
			
Following the canonical quantisation prescription, field variables are promoted to quantum operators. In order to work with hermitian operators, we split the fields into real and imaginary components, that is 
	\begin{align}
		\widehat{\boldmathsymbol{z}} &= \frac{1}{\sqrt{2}}\left(\widehat{\boldmathsymbol{z}}^{\mathrm{R}} + i \widehat{\boldmathsymbol{z}}^{\mathrm{I}}\right) , 
	\end{align}
such that $\widehat{\boldmathsymbol{z}}^s$ is Hermitian for $s = \mathrm{R}, \mathrm{I}$. These variables are canonical since $[\widehat{v}^{s}_{i}(\bs{k}),\widehat{p}^{s\prime}_{j}(\bs{q}) ] = i \delta^3(\boldsymbol{k}-\bs{q})\delta_{i,j}\delta_{s,s\prime}$ where $i,j=\varphi,\chi$. In this basis, the Hamiltonian takes the same form as in \Eq{eq:Hamiltu}, \ie 
\bea
\label{eq:Hamiltonian:k:s}
\widehat{H} =\frac{1}{2}\sum_{s=\mathrm{R},\mathrm{I}} \int_{\mathbb{R}^{3+}} \mathrm{d}^3\boldmathsymbol{k} \left(\widehat{\bs{z}}^s\right)^\mathrm{T}\boldsymbol{H}(\eta)\widehat{\bs{z}}^s\, .
\eea
Being separable, there is no mode coupling nor interactions between the $\mathrm{R}$ and $\mathrm{I}$ sectors and the state is factorisable in this decomposition. Hence, from now on, we focus on a given wavenumber $\bs{k}$ and a given $s$-sector, and to make notations lighter we leave the $\bs{k}$ and $s$ dependence implicit.  
	
A further factorisation can be performed under the field-space rotation
	\bea
	\boldmathsymbol{\widehat{z}}= \underbrace{
		\begin{pmatrix}
			\cos\theta & 0 & -\sin\theta  & 0\\
			0 &  \cos\theta & 0 & -\sin\theta\\
			\sin\theta & 0 & \cos\theta & 0\\
			0 & \sin\theta & 0 &\cos\theta
	\end{pmatrix}}_{\boldmathsymbol{P}}
	\boldmathsymbol{\widehat{z}}_{\uell-\mathrm{h}}  \, ,
	\quad\text{where}\quad
	\theta =  \frac{1}{2}\arctan\left(\frac{2 \lambda^2}{m^2-M^2}\right) ,
	\label{eq:thetamix}
	\eea
where $\uell$ and $\mathrm{h}$ stand for ``light'' and ``heavy'' respectively. In this basis the two fields decouple, and their masses are given by
	\begin{eqnarray}
	\label{eq:meff1}	m_{\uell}^2 
	&=&\frac{1}{2} \left[m^2+M^2-\left(M^2-m^2\right)\sqrt{1+\left(\frac{2\lambda^2}{M^2-m^2}\right)^2}\right],\\
	\label{eq:meff2}	m_{\mathrm{h}}^2 
	&=&\frac{1}{2} \left[m^2+M^2+\left(M^2-m^2\right)\sqrt{1+\left(\frac{2\lambda^2}{M^2-m^2}\right)^2}\right] .
	\end{eqnarray}
These expressions imply that
$
	m_{\uell}^2 <m^2<M^2<m_{\mathrm{h}}^2
$
so after the field-space rotation it remains true that $m_{\uell}^2<9H^2/4<m_{\mathrm{h}}^2$, hence the notation.

In this basis, the problem reduces to the dynamics of two uncoupled free fields evolving in a de-Sitter background. In the Heisenberg picture, this can be cast in terms of the mode-function decomposition
	\begin{align}\label{eq:modefctdecomp}
	    \widehat{v}_{i}(\eta) = v_{i}(\eta) \widehat{a}_{i} +  v^{*}_{i}(\eta) \widehat{a}^{\dag}_{i}
	\end{align}
	for $i = \uell, \mathrm{h}$ and where $\widehat{a}_{i}$ and $\widehat{a}^{\dag}_{i}$ are the creation and annihilation operators of the uncoupled fields. Heisenberg's equation yield  the classical equation of motion for the mode functions, \ie
	\begin{eqnarray}
	\label{eq:dyn1} v''_{\uell} + \left( k^2 -  \frac{\nu^2_{\uell}-\frac{1}{4}}{\eta^2}\right)v_{\uell} = 0
\quad\text{and}\quad
	\label{eq:dyn2} v''_{\mathrm{h}} + \left( k^2 -  \frac{\nu^2_{\mathrm{h}}-\frac{1}{4}}{\eta^2}\right)v_{\mathrm{h}} = 0\, .
	\end{eqnarray}
In these expressions, $\nu_{\uell}= \frac{3}{2}\sqrt{1 - \left(\frac{2m_{\uell}}{3H}\right)^2}$ and $\nu_{\mathrm{h}} = \frac{3}{2}\sqrt{1 - \left(\frac{2m_{\mathrm{h}}}{3H}\right)^2} \equiv i \mu_{\mathrm{h}}$. By normalising the mode functions to the Bunch-Davies vacuum~\cite{Bunch:1978yq} in the asymptotic, sub-Hubble past, one obtains\footnote{Note that, since all mass parameters (including $\lambda$) are negligible compared to $k/a$ in the asymptotic past, the Bunch-Davies vacuum can be set both in the $\varphi-\chi$ and in the $\uell-\mathrm{h}$ basis~\cite{Grain:2019vnq}. The vacuum state being invariant under rotations (see \Refa{Colas:2021llj}), those two prescriptions are identical.\label{footnote:BunchDavies}}
	\begin{align}
    \label{eq:modefctl}	v_{\uell}(\eta) = \frac{1}{2}\sqrt{\frac{\pi z}{k}} \ee^{i\frac{\pi}{2}\left(\nu_{\uell}+\frac{1}{2}\right)} H_{\nu_{\uell}}^{(1)}(z) 
    \quad\text{and}\quad
   v_{\mathrm{h}}(\eta) =\frac{1}{2}\sqrt{\frac{\pi z}{k}} \ee^{-\frac{\pi}{2}\mu_{\mathrm{h}}+i\frac{\pi}{4}} H_{i\mu_{\mathrm{h}}}^{(1)}(z) \, .
	\end{align}
In these expressions, $z \equiv - k \eta$ and $H^{(1)}_{\nu}$ is the Hankel function of the first kind and of order $\nu$. The mode functions of the momenta operators can be obtained by using \Eq{eq:momenta:def}, which still applies in the $\uell-\mathrm{h}$ basis, and one finds
	\begin{align}
    \label{eq:modefctpl}	p_{\uell}(\eta) &=-\frac{1}{2}\sqrt{\frac{k \pi}{z}} \ee^{i\frac{\pi}{2}\left(\nu_{\uell}+\frac{1}{2}\right)} \left[ \left(\nu_{\uell}+\frac{3}{2}\right) H_{\nu_{\uell}}^{(1)}(z)-zH_{\nu_{\uell}+1}^{(1)}(z) \right], \\
    \label{eq:modefctph}	p_{\mathrm{h}}(\eta) &=-\frac{1}{2}\sqrt{\frac{k \pi}{z}}\ee^{-\frac{\pi}{2}\mu_{\mathrm{h}}+i\frac{\pi}{4}}\left[ \left(i\mu_{\mathrm{h}}+\frac{3}{2}\right) H_{i\mu_{\mathrm{h}}}^{(1)}(z)-zH_{i\mu_{\mathrm{h}}+1}^{(1)}(z) \right]. 
	\end{align}
As mentioned above, the state being Gaussian, it is fully characterised by the covariance matrix
	\begin{align}
    \label{eq:covdef}		\boldmathsymbol{\Sigma}(\eta) = \frac{1}{2} \Tr \left[ \left\{ \boldmathsymbol{\widehat{z}} (\eta) , \boldmathsymbol{\widehat{z}}^{\mathrm{T}} (\eta) \right\} \widehat{\rho}_0\right],
	\end{align}
where $\widehat{\rho}_0$ is the Schr\"odinger state at initial time, $\widehat{\rho}_0 = \widehat{\rho}(\eta_0)$. In the uncoupled basis, this leads to a block-diagonal covariance matrix of the form
	\bea
	\label{eq:Sigma:l-h}
	\boldmathsymbol{\Sigma}_{\uell-\mathrm{h}}(\eta) =
	\begin{pmatrix}
		\boldmathsymbol{\Sigma}_{\uell}(\eta) & \boldmathsymbol{0}\\
		\boldmathsymbol{0} & \boldmathsymbol{\Sigma}_{\mathrm{h}}(\eta)
	\end{pmatrix}
	\quad\text{where}\quad
	\boldmathsymbol{\Sigma}_{i} (\eta)=
	\begin{pmatrix}
		\left\vert v_{i}(\eta) \right\vert^2 & \Rea\left[v_{i}(\eta)   p^{*}_{i}(\eta) \right]\\
		\Rea\left[v_{i}(\eta)   p^{*}_{i}(\eta) \right] & \left\vert  p_{i}(\eta) \right\vert^2
	\end{pmatrix}
	\eea
    for $i = \uell, \mathrm{h}$. In the $\varphi-\chi$ basis, the covariance matrix can be readily obtained by performing the rotation
	\bea
	\boldmathsymbol{\Sigma}(\eta)=\boldmathsymbol{P}\cdot \boldmathsymbol{\Sigma}_{\uell-\mathrm{h}}(\eta) \cdot \boldmathsymbol{P}^{\mathrm{T}}
 \equiv 
	\begin{pmatrix}
		\boldmathsymbol{\Sigma}_{\varphi\varphi}(\eta) & \boldmathsymbol{\Sigma}_{\varphi\chi}(\eta)\\
		\boldmathsymbol{\Sigma}_{\varphi\chi}(\eta) & \boldmathsymbol{\Sigma}_{\chi\chi}(\eta)
	\end{pmatrix} \label{eq:Covdef}\, , 	\eea
with
	\begin{align}
\label{eq:exactCovphiphi}	    \boldmathsymbol{\Sigma}_{\varphi\varphi}(\eta) &=\cos^2(\theta)\boldmathsymbol{\Sigma}_{\uell}(\eta) + \sin^2(\theta)\boldmathsymbol{\Sigma}_{\mathrm{h}}(\eta) ,\\
	    \boldmathsymbol{\Sigma}_{\varphi\chi}(\eta) &=\cos(\theta)\sin({\theta})\left[\boldmathsymbol{\Sigma}_{\uell}(\eta) -\boldmathsymbol{\Sigma}_{\mathrm{h}}(\eta)\right] ,\\
	    \boldmathsymbol{\Sigma}_{\chi\chi}(\eta) &=\cos^2(\theta)\boldmathsymbol{\Sigma}_{\mathrm{h}}(\eta) + \sin^2(\theta)\boldmathsymbol{\Sigma}_{\uell} (\eta) .
	\end{align}

Finally, the reduced state of the system $\varphi$ is obtained by tracing out the $\chi$ field, see \Eq{eq:rho:red}. It is still a Gaussian state, with covariance matrix given by $\boldmathsymbol{\Sigma}_{\varphi\varphi}$~\cite{Colas:2021llj}. We have thus found an exact solution to the problem at hand (namely compute the reduced state of the system), to which we will now compare effective methods in order to test their robustness. 

As explained in \Sec{sec:intro}, one of the main physical effects driven by the interaction with an environment is decoherence, namely the transition from a pure quantum state into a statistical mixture. The loss of quantum coherence can be measured with the so-called purity parameter $\gamma(\eta) \equiv \Tr\left(\widehat{\rho}_{\mathrm{red}}^2\right)$ which measures the amount of quantum entanglement between the system and the environment. Pure states correspond to $\gamma=1$, and mixed states have $\gamma<1$ (with $\gamma=0$ corresponding to a maximally mixed state). The amount by which $\chi$ decoheres $\varphi$ is given by
	\begin{eqnarray}\label{eq:gamma}
		\gamma(\eta)  = \frac{1}{4}\det\left[\boldmathsymbol{\Sigma}_{\varphi\varphi}(\eta)\right]^{-1}\, ,
	\end{eqnarray}
the expression being valid for any Gaussian state~\cite{Colas:2021llj}. In the absence of interactions between the system and the environment, $\det\boldmathsymbol{\Sigma}_{\varphi\varphi}= 1/4$ so $\gamma= 1$. Otherwise, the system is said to have decohered when $\gamma\ll 1$. 
		
	\subsection{Effective description: the TCL${}_2$ master equation}\label{subsec:CME}
	
We now turn our attention to the TCL${}_2$ master equation~\eqref{eq:TCL2expdev}. We remind that it is formulated in the interaction picture, where the interaction Hamiltonian reads $\widetilde{\mathcal{H}}_{ \mathrm{int}}(\eta) =  a^2(\eta) \widetilde{v}_{\varphi}(\eta)\widetilde{v}_{\chi}(\eta)$. The TCL${}_2$ master equation~\eqref{eq:TCL2exp} thus takes the form
 	\begin{align}\label{eq:TCL2CL}
 		\frac{\dd \widetilde{\rho}_{\mathrm{red}}}{\dd \eta} = - \lambda^4 a^2(\eta) \int_{\eta_0}^{\eta} \dd \eta' a^2(\eta') \bigg\{&\Rea\left[\mathcal{K}^{>}(\eta,\eta')\right] \left[\widetilde{v}_{\varphi}(\eta),\left[\widetilde{v}_{\varphi}(\eta'), \widetilde{\rho}_{\mathrm{red}} (\eta) \right]\right] \nonumber \\ 
 		+ i &\Ima\left[\mathcal{K}^{>}(\eta,\eta')\right] \left[\widetilde{v}_{\varphi}(\eta),\left\{\widetilde{v}_{\varphi}(\eta'), \widetilde{\rho}_{\mathrm{red}} (\eta) \right\}\right]\bigg\} ,
 	\end{align}
where the memory kernel is given by
 	\begin{align}
	\label{eq:TCK:kernel}
 		\mathcal{K}^{>}(\eta,\eta') \equiv \mathrm{Tr}_{\mathrm{E}}\left[\widehat{v}_{\chi} (\eta) \widehat{v}_{\chi} (\eta')\widehat{\rho}_{\mathrm{E}}\right] 
 	\end{align}
and we recall that $\widehat{\rho}_{\mathrm{E}}$ corresponds to the state of the environment in the absence of interactions with the system [a derivation of \Eq{eq:TCL2CL} following microphysical considerations is also presented in \App{app:TCL2:microphysical:derivation}]. Since $\widehat{v}_{\chi}(\eta) \widehat{v}_{\chi} (\eta')$ is not hermitian for $\eta \neq \eta'$, the kernel $\mathcal{K}^{>}(\eta,\eta')$ is complex and can be evaluated as follows. In the interaction picture, operators evolve with the free Hamiltonian, so one can use the results obtained in \Sec{subsec:exact} in the uncoupled basis. More precisely, a similar mode-function decomposition as in \Eq{eq:modefctdecomp} can be introduced,
 	\begin{align}
	\label{eq:mode:decomp:interaction}
 		\widetilde{v}_i (\eta) = v_i(\eta) \widehat{a}_i + v^{*}_i(\eta) \widehat{a}^{\dag}_i
 	\end{align}
where $i=\varphi,\chi$, and an analogous expression for $\widetilde{p}_i(\eta)$. The mode functions are still given by \Eqs{eq:modefctl}-\eqref{eq:modefctph}, where $m_\uell$ and $m_{\mathrm{h}}$ are simply replaced with $m$ and $M$. This leads to
 	\begin{align}
	\label{eq:K:mode:function}
 		\mathcal{K}^{>}(\eta,\eta') &= v_{\chi} (\eta) v^{*}_{\chi} (\eta').
 	\end{align}

	\subsubsection*{Interpretating the master equation}

	While the above form~\eqref{eq:TCL2CL} of the cosmological master equation is compact, it makes the connection with quantum Brownian motion~\cite{Caldeira:1982iu, Hu:1993qa, PhysRevD.45.2843, PhysRevD.53.2012, Huang:2022hru} less apparent. A form that is easier to interpret can be obtained by expressing all operators at the same time. This can be achieved by inverting the mode-function expansion to yield $\widehat{a}_\varphi$ and $\widehat{a}_\varphi^\dagger$ in terms of $\widetilde{v}_\varphi(\eta)$ and $\widetilde{p}_\varphi(\eta)$. Inserting those expressions in \Eq{eq:mode:decomp:interaction} evaluated at time $\eta'$ leads to	
	\bea
	\label{eq:v:etap:v:eta}
	\widetilde{v}_{\varphi} (\eta')= -2\Imag{p_{\varphi} (\eta) v^{*}_{\varphi} (\eta')} \widetilde{v}_{\varphi} (\eta) +2 \Imag{v_{\varphi} (\eta) v^{*}_{\varphi} (\eta')}\widetilde{p}_{\varphi} (\eta)\, .
	\eea
Here we have used that $\Imag{ v_{\varphi} (\eta)p^*_{\varphi} (\eta)}=-1/2$, which comes from the canonical commutation relation $[\widetilde{v}_{\varphi}(\eta),\widetilde{p}_{\varphi}(\eta)]=1$. Plugging \Eq{eq:v:etap:v:eta} into \Eq{eq:TCL2CL}, one finds
\bea\label{eq:METCL2}
		\frac{\dd \widetilde{\rho}_{\mathrm{red}}}{\dd \eta}  =&-i\Big[\overbrace{ \frac{1}{2}\widetilde{\bs{z}}_{i} (\eta) \bs{\Delta}_{ij}(\eta) \widetilde{\bs{z}}_{j} (\eta)}^{\widetilde{H}^{\mathrm{(LS)}}(\eta)},\widetilde{\rho}_{\mathrm{red}}(\eta)\Big]- \frac{1}{2}\ds\sum_{i,j}\boldsymbol{D}_{ij}(\eta)\left[\widetilde{\bs{z}}_{i} (\eta),\left[\widetilde{\bs{z}}_{j} (\eta),\widetilde{\rho}_{\mathrm{red}}(\eta)\right]\right]  \\ &- \frac{i}{2} \bs{\Delta}_{12}(\eta) \ds\sum_{i,j}\boldsymbol{\omega}_{ij}\left[\widetilde{\bs{z}}_{i} (\eta),\left\{\widetilde{\bs{z}}_{j} (\eta),\widetilde{\rho}_{\mathrm{red}}(\eta)\right\}\right] ,
\eea
which defines the ``Lamb-shift'' Hamiltonian $\widetilde{H}^{\mathrm{(LS)}}$ (see below), where $\boldsymbol{\omega}=\left(\begin{array}{cc}
			0 & 1 \\
			-1 & 0
		\end{array}\right)$, and where we have used the canonical commutation relation again.	In this expression, $\widetilde{\bs{z}} (\eta) \equiv\left(\widetilde{v}_{\varphi} (\eta), \widetilde{p}_{\varphi} (\eta)\right)^{\mathrm{T}}$ and the two-by-two matrices $\bs{D}$ and $\bs{\Delta}$ are given by
	\begin{align}
		\boldsymbol{D}_{11}(\eta)&= -4 \lambda^4 a^2(\eta) \int_{\eta_0}^{\eta}\dd \eta'  a^2(\eta') \Imag{p_{\varphi} (\eta) v^{*}_{\varphi} (\eta')}
		\Real{v_{\chi} (\eta) v^{*}_{\chi} (\eta') } ,\label{eq:Dphiphi}\\
		\boldsymbol{D}_{12}(\eta)= \boldsymbol{D}_{21}(\eta) &=2 \lambda^4 a^2(\eta) \int_{\eta_0}^{\eta}\dd \eta'  a^2(\eta')\Imag{v_{\varphi} (\eta) v^{*}_{\varphi} (\eta')}\Real{v_{\chi} (\eta) v^{*}_{\chi} (\eta') }, \label{eq:Dphipi} \\
		\boldsymbol{D}_{22}(\eta) &= 0 ,\label{eq:D22}
	\end{align}
	and
	\begin{align}
		\bs{\Delta}_{11}(\eta)&= -4 \lambda^4 a^2(\eta)  \int_{\eta_0}^{\eta}\dd \eta'  a^2(\eta')\Imag{p_{\varphi} (\eta) v^{*}_{\varphi} (\eta')}\Imag{v_{\chi} (\eta) v^{*}_{\chi} (\eta') } ,\label{eq:Fphiphi}\\
		\bs{\Delta}_{12}(\eta)= \bs{\Delta}_{21}(\eta)&= 2 \lambda^4 a^2(\eta) \int_{\eta_0}^{\eta}\dd \eta'  a^2(\eta')\Imag{v_{\varphi} (\eta) v^{*}_{\varphi} (\eta')}\Imag{v_{\chi} (\eta) v^{*}_{\chi} (\eta') } ,\label{eq:Fphipi} \\ 
		\bs{\Delta}_{22}(\eta) &=0\, . \label{eq:Delta22}
	\end{align}
The corresponding equation in the Schr\"odinger picture can be obtained using the fact that operators are mapped between the two pictures with the free Hamiltonian of the system, see \Eq{eq:rho:Hint}, and one finds\footnote{Here we use that since $\boldsymbol{D}$ is symmetric, $\bs{\omega}$ is anti-symmetric given the canonical commutation relations between phase-space variables $[\widehat{\boldsymbol{z}}_i,\widehat{\boldsymbol{z}}_j]=\bs{w}_{ij}$, one has
\bea
\boldsymbol{D}_{ij} \left[\widehat{\boldsymbol{z}}_i,\left[\widehat{\boldsymbol{z}}_j,\widehat{\rho}_{\mathrm{red}}\right]\right] &= \boldsymbol{D}_{ij} \left(-2 \widehat{\boldsymbol{z}}_i\widehat{\rho}_{\mathrm{red}} \widehat{\boldsymbol{z}}_j  +  \left\{\widehat{\boldsymbol{z}}_j \widehat{\boldsymbol{z}}_i, \widehat{\rho}_{\mathrm{red}} \right\}\right), \\
        \label{eq:result2} \bs{\omega}_{ij} \left[\widehat{\boldsymbol{z}}_i,\left\{\widehat{\boldsymbol{z}}_j,\widehat{\rho}_{\mathrm{red}}\right\}\right] &= \bs{\omega}_{ij} \left(2 \widehat{\boldsymbol{z}}_i\widehat{\rho}_{\mathrm{red}} \widehat{\boldsymbol{z}}_j  - \left\{\widehat{\boldsymbol{z}}_j \widehat{\boldsymbol{z}}_i, \widehat{\rho}_{\mathrm{red}} \right\}\right).
        \eea
}
	\begin{align}\label{eq:METCL2fin}
	\frac{\dd \widehat{\rho}_{\mathrm{red}}}{\dd \eta}  &=-i\left[\widehat{H}^{(\varphi)}(\eta) + \widehat{H}^{\mathrm{(LS)}}(\eta),\widehat{\rho}_{\mathrm{red}}(\eta)\right]+ \ds\sum_{i,j}\bs{\mathcal{D}}_{ij}(\eta)\left[\widehat{\bs{z}}_{i}\widehat{\rho}_{\mathrm{red}}(\eta) \widehat{\bs{z}}_{j}-\frac{1}{2}\left\{\widehat{\bs{z}}_{j}\widehat{\bs{z}}_{i},\widehat{\rho}_{\mathrm{red}}(\eta)\right\}\right] .
	\end{align}
In this expression, the dissipator matrix is defined as 
	\begin{align}\label{eq:dissipator}
		\bs{\mathcal{D}}(\eta) \equiv \bs{D}(\eta) - i \bs{\Delta}_{12}(\eta) \bs{\omega} = \begin{pmatrix}
			\bs{D}_{11}(\eta) & \bs{D}_{12}(\eta) - i \bs{\Delta}_{12}(\eta) \\ 
			\bs{D}_{12}(\eta) + i \bs{\Delta}_{12}(\eta) & 0   
		\end{pmatrix} .
	\end{align} 
One can see that \Eq{eq:METCL2fin} has the same form as the Lindblad equation~\eqref{eq:GKSL}, with the crucial difference that the dissipator matrix $\bs{\mathcal{D}}(\eta)$ is not positive semi-definite in the present case.\footnote{If the dynamical map generated by \Eq{eq:METCL2fin} were Markovian in the sense introduced in footnote~\ref{footnote:Markovian}, \ie if it described a semi-group evolution, then according to Lindblad theorem \cite{Lindblad:1975ef} the fact that its dissipator is not semi-definite positive would imply that it is not Completely Positive and Trace Preserving (CPTP). However, \Eq{eq:METCL2fin} belongs to the class of so-called ``Gaussian master equations'', which were shown to be CPTP in \Refs{Ferialdi_2016,Di_si_2014} (and thus map a quantum state to another proper quantum state). The contrapositive of Lindblad's theorem thus imposes that our master equation is non-Markovian~{\cite{Breuer:2002pc, 2008JPhA...41q5304W, RevModPhys.88.021002, Moustos:2016lol, Shandera:2017qkg, Nicacio:2022djs, Prudhoe:2022pte, Spaventa:2022hks, Chruscinski:2022hvy}}.  
\label{footnote:Markovian:versus:CPTP}} 
It is also worth stressing that \Eq{eq:METCL2} has the same form as the master equation obtained by Hu, Paz and Zhang in their seminal paper \cite{PhysRevD.45.2843} and that describes quantum Brownian motion.
The first term in the right-hand side of \Eq{eq:METCL2} provides a unitary contribution, which renormalises the energy levels of the system due to the interaction with the environment \cite{Brasil_2013, Manzano_2020, Breuer:2002pc}. This is why it is often referred to as the Lamb--shift Hamiltonian. In our case, it reads
	\begin{align}\label{eq:HLSexp}
		\widehat{H}^{(\varphi)}(\eta) + \widehat{H}^{\mathrm{(LS)}}(\eta) &= \frac{1}{2} \left[\widehat{p}_{\varphi}\widehat{p}_{\varphi} + \left(k^2 + m^2a^2 + \bs{\Delta}_{11} \right) \widehat{v}_{\varphi} \widehat{v}_{\varphi} +\left(\frac{a'}{a}+\bs{\Delta}_{12}\right)  \left\{\widehat{v}_{\varphi}, \widehat{p}_{\varphi}\right\} \right].
	\end{align}
One can thus see that $\bs{\Delta}_{11}$ renormalises the mass of the field $\varphi$, while $\bs{\Delta}_{12}$ renormalises the comoving Hubble parameter. Note that, in the context of effective-field theoretic calculations, these contributions are usually re-absorbed in an effective speed of sound $\cs^2$~\cite{Cheung:2007st, Baumann:2011su, Garcia-Saenz:2018vqf}.
The second and the third terms in \Eq{eq:METCL2} are of a different nature, since they capture the non-unitary evolution of the system and thus cannot be described by an effectively local Lagrangian. This is due to dissipation and decoherence, which respectively correspond to the imaginary and the real part of the dissipator matrix in \Eq{eq:METCL2fin}.\footnote{The fact that the real and the imaginary part of the memory kernel lead to distinct physical effects is also encountered in the influence-functional approach~\cite{Hu:1993qa, Lombardo:2004fr, Jackson:2010cw, Jackson:2012qp, Boyanovsky:2015jen, Boyanovsky:2018fxl, Boyanovsky:2018soy, Burrage:2018pyg, Burrage:2019szw, Pinol:2020cdp, Choudhury:2022ati, Kading:2022jjl}, of which the master equation is the dynamical generator~\cite{2002quant.ph..9153B,Boyanovsky:2015xoa}. Indeed, in the influence functional description, $\Ima\left[\mathcal{K}^{>}(\eta,\eta')\right]$ is related to the retarded and advanced Green's function of the environment and can be interpreted as a dissipation kernel, while $\Rea\left[\mathcal{K}^{>}(\eta,\eta')\right]$ is related to the Keldysh-Green's function~\cite{Calzetta:2008iqa, Kamenev_2009} and can be interpreted as a noise kernel~\cite{Hsiang:2021vgx}.}

Finally, in phase space, the TCL${}_2$ master equation takes the form of a Fokker-Planck equation for the reduced Wigner function $W_{\mathrm{red}}$. The latter is defined by the Wigner-Weyl transform of the reduced density matrix \cite{doi:10.1119/1.2957889}, and provides a quantum analogue of a phase-space quasi probability distribution. In \App{app:phase:space}, we derive general results on the phase-space representation of the TCL${}_2$ master equation. In particular, we find that performing the Wigner-Weyl transform of \Eq{eq:METCL2} leads to	\begin{align}\label{eq:Wignerredeom}
	\frac{\dd  W_{\mathrm{red}}}{\dd \eta}=\left\{\widetilde{H}^{(\varphi)} + \widetilde{H}^{\mathrm{(LS)}} ,W_{\mathrm{red}}\right\} + \bs{\Delta}_{12} \ds\sum_{i}\frac{\partial}{\partial \boldsymbol{z}_i}\left(\boldsymbol{z}_iW_{\mathrm{red}}\right) -\frac{1}{2}\ds\sum_{i,j}\left[\bs{\omega} \boldsymbol{D} \bs{\omega}\right]_{ij} \frac{\partial^2 W_{\mathrm{red}}}{\partial \boldsymbol{z}_i\partial \boldsymbol{z}_j}\, ,	
	\end{align}
where brackets correspond to Poisson brackets (not to be confused with the anti-commutator). Only the term involving $\widetilde{H}^{(\varphi)} + \widetilde{H}^{\mathrm{(LS)}}$ is unitary, as mentioned above. The second term, proportional to $\bs{\Delta}_{12}$, is dissipative: it is a drift (or friction) term that accounts for the energy transfer from the system into the environment~\cite{Hollowood:2017bil}. Finally, the term proportional to $\bs{\omega} \boldsymbol{D} \bs{\omega}$ corresponds to diffusion and leads to decoherence. One can show that this equation admits Gaussian solutions, hence the reduced state of the system is still Gaussian in TCL.
\subsection{Transport equations}
As mentioned above, the state being Gaussian, it is fully characterised by its covariance matrix. Since the initial covariance matrix is the same in all  approaches (TCL${}_2$, exact, SPT) a first strategy to benchmark the cosmological master equation consists in comparing the equation of motion for the covariance of the system, usually refereed to as the transport equations. 
\subsubsection*{TCL${}_2$ transport equation}
In the TCL approach, the transport equations can be obtained by differentiating \Eq{eq:covdef} with respect to time in the Schr\"odinger picture, and using \Eq{eq:METCL2fin} to evaluate $\dd\widehat{\rho}_{\mathrm{red}}/\dd\eta$. This gives
			\begin{align}\label{eq:TCLeom}
				\frac{\dd \bs{\Sigma}_{\mathrm{TCL}}}{\dd \eta}&=\boldsymbol{\omega}\left(\bs{H}^{(\varphi)}+\bs{\Delta}\right) \bs{\Sigma}_{\mathrm{TCL}}-\bs{\Sigma}_{\mathrm{TCL}} \left(\bs{H}^{(\varphi)}+\bs{\Delta}\right) \boldsymbol{\omega}- \bs{\omega D \omega} -2\bs{\Delta}_{12}\bs{\Sigma}_{\mathrm{TCL}} \, ,
			\end{align}
where $\bs{D}$ and $\bs{\Delta}$ were introduced in \Eqs{eq:Dphiphi}-\eqref{eq:Delta22}. The first two terms correspond to the unitary evolution, which as stressed above receives an additional contribution from the Lamb-shift Hamiltonian. The last two terms respectively correspond to the diffusion (a source term proportional to $\bs{D}$) and the dissipation (a damping term proportional to $\bs{\Delta}_{12}$).
\subsubsection*{Exact transport equation}
In the exact approach presented in \Sec{subsec:exact}, the transport equations for the full system-plus-environment setup can be obtained by differentiating \Eq{eq:covdef} with respect to time in the Heisenberg picture, and using the Heisenberg equations to evaluate $\dd\widehat{\bs{z}}/\dd\eta$. The Hamiltonian~\eqref{eq:Hamiltonian:k:s} being quadratic, one finds
			\bea
			\frac{\dd\boldsymbol{\Sigma}}{\dd \eta}=\boldsymbol{\Omega H \Sigma}-\boldsymbol{\Sigma H \Omega},
			\eea
where $\bs{H}$ was defined in \Eq{eq:Hmat} and $\boldsymbol{\Omega}$ is a four-by-four block-diagonal matrix where each $2\times2$ block on the diagonal is $\boldsymbol{\omega}$. 
						
Using blockwise multiplication we can split the above into a set of coupled differential equations for the covariance of the system ($\boldmathsymbol{\Sigma}_{\varphi\varphi}$), of the environment ($\boldmathsymbol{\Sigma}_{\chi\chi}$), and for their cross-covariance ($\boldmathsymbol{\Sigma}_{\varphi\chi}$). Using \Eqs{eq:Hvarphimat}, it reads
			\begin{align}
				\frac{\dd\boldmathsymbol{\Sigma}_{\varphi\varphi}}{\dd \eta}&=\boldsymbol{\omega H}^{(\varphi)} \boldmathsymbol{\Sigma}_{\varphi\varphi}-\boldmathsymbol{\Sigma}_{\varphi\varphi} \boldsymbol{H}^{(\varphi)} \boldsymbol{\omega}+\boldsymbol{\omega V} \boldmathsymbol{\Sigma}_{\varphi\chi}^{\mathrm{T}}-\boldmathsymbol{\Sigma}_{\varphi\chi}\boldsymbol{V} \boldsymbol{\omega}, \label{eq:covSdiffexact}\\
				\frac{\dd\boldmathsymbol{\Sigma}_{\chi\chi}}{\dd \eta}&=\boldsymbol{\omega H}^{(\chi)} \boldmathsymbol{\Sigma}_{\chi\chi}-\boldmathsymbol{\Sigma}_{\chi\chi} \boldsymbol{H}^{(\chi)} \boldsymbol{\omega}+\boldsymbol{\omega V}^{\mathrm{T}}\boldmathsymbol{\Sigma}_{\varphi\chi}-\boldmathsymbol{\Sigma}_{\varphi\chi}^{\mathrm{T}}\boldsymbol{V} \boldsymbol{\omega}, \\
				\frac{\dd\boldmathsymbol{\Sigma}_{\varphi\chi}}{\dd \eta}&=\boldsymbol{\omega H}^{(\varphi)} \boldmathsymbol{\Sigma}_{\varphi\chi}-\boldmathsymbol{\Sigma}_{\varphi\chi} \boldsymbol{H}^{(\chi)} \boldsymbol{\omega}+\boldsymbol{\omega V \Sigma}_{\chi \chi}-\boldmathsymbol{\Sigma}_{\varphi\varphi}\boldsymbol{V\omega}.
			\end{align}
Note that these transport equations can also be obtained in the phase-space representation (\ie using Wigner functions), as explained in \App{app:phase:space}. In the present case, a first integral of the above system can be easily constructed, since we know that, in spite of having three covariance matrices ($\boldmathsymbol{\Sigma}_{\varphi\varphi}$, $\boldmathsymbol{\Sigma}_{\chi\chi}$ and $\boldmathsymbol{\Sigma}_{\varphi\chi}$), only two combinations are independent (namely $\boldmathsymbol{\Sigma}_{\uell}$ and $\boldmathsymbol{\Sigma}_{\mathrm{h}}$). More precisely, from \Eq{eq:exactCovphiphi} one can show that $\boldmathsymbol{\Sigma}_{\varphi\chi} = \boldmathsymbol{\Sigma}_{\varphi\chi}^{\mathrm{T}} = \tan(2\theta)(\boldmathsymbol{\Sigma}_{\varphi\varphi} - \boldmathsymbol{\Sigma}_{\chi\chi})/2.$ 
Focusing on the dynamics of the reduced system, \Eq{eq:covSdiffexact} can thus be written as
			\begin{align}\label{eq:transportexact}
				\frac{\dd\boldmathsymbol{\Sigma}_{\varphi\varphi}}{\dd \eta}&=\boldsymbol{\omega}\left(\bs{H}^{(\varphi)}+\bs{\Delta}_{\mathrm{ex}}\right) \boldmathsymbol{\Sigma}_{\varphi\varphi}-\boldmathsymbol{\Sigma}_{\varphi\varphi} \left(\bs{H}^{(\varphi)}+\bs{\Delta}_{\mathrm{ex}}\right) \boldsymbol{\omega} - \bs{\omega} \boldsymbol{D}_{\mathrm{ex}}\bs{\omega} \, ,
			\end{align}
where
			\begin{align}\label{eq:LSex}
				\bs{\Delta}_{\mathrm{ex}} \equiv -\frac{\lambda^2}{M^2 - m^2} \bs{V}
				\quad\text{and}\quad
				\boldsymbol{D}_{\mathrm{ex}} \equiv  -\frac{\lambda^2}{M^2 - m^2} \left( \bs{\omega} \boldmathsymbol{\Sigma}_{\chi\chi} \bs{V} -\bs{V}  \boldmathsymbol{\Sigma}_{\chi\chi}  \bs{\omega}\right).
			\end{align}
The reason why we write the exact transport equation in this form is to allow for an easy comparison with its TCL${}_2$ counterpart~\eqref{eq:TCLeom}. This suggests to interpret $\bs{\Delta}_{\mathrm{ex}}$ as a Lamb-shift contribution in the exact approach, and $\boldsymbol{D}_{\mathrm{ex}}$ as a diffusion matrix. From \Eq{eq:LSex}, the only non-vanishing entries of those matrices are given by
\begin{align}
\label{eq:Delta:ex:comp}
\bs{\Delta}_{\mathrm{ex},11} =& -\frac{\lambda^4 a^2}{M^2-m^2}\, ,\\	
\bs{D}_{\mathrm{ex},11} =	&-2 \frac{\lambda^4 a^2}{M^2-m^2}\boldmathsymbol{\Sigma}_{\chi\chi,12}\, ,\quad\quad
\bs{D}_{\mathrm{ex},12} =	 \frac{\lambda^4 a^2}{M^2-m^2}\boldmathsymbol{\Sigma}_{\chi\chi,11}\, .
\label{eq:D:ex:comp}
\end{align}
Note that, in the asymptotic past, when $a\to 0$, the above coefficients vanish, which confirms that the two fields become effectively uncoupled and that Bunch-Davies initial conditions can be safely set, see footnote~\ref{footnote:BunchDavies}.
\subsubsection*{SPT transport equation}
In the perturbative approach introduced in \Sec{sec:inin}, at leading order, the transport equation is simply given by the exact transport equation,
 \Eq{eq:transportexact}, where the right-hand side is truncated at order $\lambda^4$:
\begin{align}\label{eq:SPTeom}
\frac{\dd \bs{\Sigma}_{\mathrm{SPT}}}{\dd \eta}&=\boldsymbol{\omega}\bs{H}^{(\varphi)}\bs{\Sigma}_{\mathrm{SPT}}-\bs{\Sigma}_{\mathrm{SPT}}\bs{H}^{(\varphi)}\boldsymbol{\omega} 
+\boldsymbol{\omega}\boldsymbol{\Delta}_{\mathrm{ex}}\boldsymbol{\Sigma}_{\varphi\varphi}^{\mathrm{free}}-\boldsymbol{\Sigma}_{\varphi\varphi}^{\mathrm{free}}\boldsymbol{\Delta}_{\mathrm{ex}}\boldsymbol{\omega}
- \bs{\omega} \bs{D}_{\mathrm{SPT}} \bs{\omega} \, .
\end{align}
Here, $\boldsymbol{\Sigma}_{\varphi\varphi}^{\mathrm{free}}$ corresponds to $\boldsymbol{\Sigma}_{\varphi\varphi}$ evaluated in the free theory and is given by 	the second part of \Eq{eq:Sigma:l-h} with the mode functions $v_\varphi$ and $p_\varphi$. Similarly, $\bs{D}_{\mathrm{SPT}}$ is given by \Eq{eq:D:ex:comp} where $\boldsymbol{\Sigma}_{\chi\chi}$ is replaced with $\boldsymbol{\Sigma}_{\chi\chi}^{\mathrm{free}}$, which is given by the second part of \Eq{eq:Sigma:l-h} with the mode functions $v_\chi$ and $p_\chi$. Note that $\bs{\Delta}_{\mathrm{ex}}$ does not need to be expanded since it is already of order $\lambda^4$, see \Eq{eq:Delta:ex:comp}. 
			
Even though the covariance matrix in SPT can be obtained by integrating the above transport equation, in the present situation an exact solution to the full theory is known, so it can also be obtained by expanding \Eq{eq:exactCovphiphi} in $\lambda$. Here, not only $\theta^2=\lambda^4/(m^2-M^2)^2+\order{\lambda^8}$ needs to be expanded, see \Eq{eq:thetamix}, but also $m_\uell^2=m^2-\lambda^4/(M^2-m^2)+\order{\lambda^8}$ and $m^2_\mathrm{h}=M^2+\lambda^4/(M^2-m^2)+\order{\lambda^8}$ in $\bs{\Sigma}_{\mathrm{h}}$ and $\bs{\Sigma}_\uell$, see \Eqs{eq:meff1}-\eqref{eq:meff2}. On the numerical results presented below, we have checked that these two approaches coincide.
\subsection{Spurious terms}
The TCL${}_2$ coefficients are expressed as integrals between $\eta_0$ and $\eta$, see \Eqs{eq:Dphiphi}-\eqref{eq:Delta22}, where $\eta_0\rightarrow-\infty$ if Bunch-Davies initial conditions are chosen. Formally, they can be written as
\bea
\label{eq:primF:def}
\boldmathsymbol{D}_{11} = F_{\boldmathsymbol{D}_{11}}\left(\eta,\eta\right)-F_{\boldmathsymbol{D}_{11}}\left(\eta,\eta_0\right),
\eea 
where $ F_{\boldmathsymbol{D}_{11}}(\eta,\cdot)$ is the primitive of the integrand appearing in \Eq{eq:Dphiphi}, which itself depends on $\eta$, and with similar notations for the other TCL${}_2$ coefficients. The $F$ functions are derived explicitly in \App{app:TCL2coef}, where it is shown that the integrals \eqref{eq:Dphiphi}-\eqref{eq:Delta22} can be performed analytically and involve products of four Hankel functions. The second term in \Eq{eq:primF:def}, the one of the form $F(\eta,\eta_0)$, features several properties that we now describe and that will lead us to dub it ``spurious''.

First, the spurious terms involve the initial time $\eta_0$, which implies that they carry explicit dependence on the initial conditions. If the environment memory kernel~\eqref{eq:TCK:kernel} is sufficiently peaked around $\eta'=\eta$, that is if the integrands in \Eqs{eq:Dphiphi}-\eqref{eq:Delta22} are much smaller around $\eta'=\eta_0$ than around $\eta'=\eta$, then this contribution should be suppressed compared to the non-spurious one. This is similar to the Lindbladian limit discussed in \Sec{sec:Lindblad}. Whether or not this is the case can be verified explicitly in the super-Hubble regime (\ie at late time, $-k\eta\ll 1$) where the $F$ functions take simple forms. The expansion in the limit $-k\eta\ll 1$ is performed in \App{sec:TCLcoeff:superH}, where it is shown that the spurious terms dominate for all coefficients. More precisely, $F_{\bs{D}_{11}}(\eta,\eta)\propto (-k\eta)^{-2}$ while $F_{\bs{D}_{11}}(\eta,\eta_0)\propto (-k\eta)^{-7/2}$, $F_{\bs{D}_{12}}(\eta,\eta)\propto (-k\eta)^{-1}$ while $F_{\bs{D}_{11}}(\eta,\eta_0)\propto (-k\eta)^{-5/2}$, $F_{\bs{\Delta}_{11}}(\eta,\eta)\propto (-k\eta)^{-2}$ while $F_{\bs{\Delta}_{11}}(\eta,\eta_0)\propto (-k\eta)^{-7/2}$, and  $F_{\bs{\Delta}_{12}}(\eta,\eta)$ vanishes while $F_{\bs{\Delta}_{12}}(\eta,\eta_0)\propto (-k\eta)^{-5/2}$. Let us stress that the late-time domination of the spurious terms is strongly related to having a dynamical background. This is the first indication we encounter that applying the master-equation program to cosmology may not be as straightforward as in other situations. 

Second, in \App{sec:TCLcoeff:exact}, we notice that, using various identities satisfied by the Hankel functions, the expressions for the non-spurious contributions can be vastly simplified. More precisely, after a lengthy though straightforward calculation we find that 
\bea
\label{eq:TCLcoeff:crucial:non-spurious_simp}
F_{\bs{D}_{11}}(\eta,\eta)=&\bs{D}_{\mathrm{SPT},11}(\eta)\, ,\quad
F_{\bs{D}_{12}}(\eta,\eta)=\bs{D}_{\mathrm{SPT},12}(\eta)\, ,\\
 F_{\bs{\Delta}_{11}}(\eta,\eta)=&\bs{\Delta}_{\mathrm{ex},11}(\eta)\, ,\quad \ \ 
 F_{\bs{\Delta}_{12}}(\eta,\eta)=0\, .
 \eea
Let us now recall the result obtained in \Sec{sec:inin}, namely the fact that the perturbative version of TCL is strictly equivalent to SPT. This implies that $\bs{\Sigma}_{\mathrm{TCL}}=\bs{\Sigma}_{\mathrm{SPT}}+\order{\lambda^8}$, where $\bs{\Sigma}_{\mathrm{SPT}}$ only contains terms of order $\lambda^0$ (namely $\bs{\Sigma}_{\varphi\varphi}^{\mathrm{free}}$) and $\lambda^4$. As a consequence, the right-hand sides of \Eqs{eq:TCLeom} and~\eqref{eq:SPTeom} coincide at order $\lambda^4$. The terms of order $\lambda^0$ are trivially identical, and for the terms of order $\lambda^4$ one obtains (recalling that both $\bs{D}$ and $\bs{\Delta}$ are of order $\lambda^4$)
\bea
\bs{\omega}\bs{\Delta}\bs{\Sigma}_{\varphi\varphi}^{\mathrm{free}} - \bs{\Sigma}_{\varphi\varphi}^{\mathrm{free}}\bs{\Delta}\bs{\omega} - \bs{\omega} \bs{D}\bs{\omega}-2\bs{\Delta}_{12}\bs{\Sigma}_{\mathrm{SPT}} = \bs{\omega} \bs{\Delta}_{\mathrm{ex}}\bs{\Sigma}_{\varphi\varphi}^{\mathrm{free}}-\bs{\Sigma}_{\varphi\varphi}^{\mathrm{free}}\bs{\Delta}_{\mathrm{ex}}\bs{\omega}-\bs{\omega}\bs{D}_{\mathrm{SPT}}\bs{\omega}\, .
\eea
Each term in the left-hand side can be decomposed into a non-spurious part and a spurious part, see \Eq{eq:primF:def}. An important remark is that, thanks to \Eq{eq:TCLcoeff:crucial:non-spurious_simp}, the non-spurious part exactly coincides with the right-hand side, hence the spurious contributions cancel out. We have therefore proven that the spurious terms are absent from the perturbative limit of TCL and only arise at higher order. 
This is obviously consistent with the fact that, at leading order, TCL coincides with the exact theory, which is not plagued by any spurious contribution.

Third, we have checked that if one includes the spurious terms when solving the TCL transport equation~\eqref{eq:TCLeom}, then the result quickly blows up. This is due to the late-time divergences of the spurious contributions mentioned above. On the contrary, as we will see below, if one removes them, then the result is remarkably well-behaved. 

To summarise, spurious terms cancel out at leading order in the interaction strength, and at higher order, the fact that they carry an explicit dependence on the initial time,  combined with their late-time divergent behaviour, indicates that they cannot be resummed. This leads us to conclude that, for the simple model we have considered here, resummation cannot be efficiently performed with the standard master-equation program.

However, this may be due to the over-simplicity of that particular model, which contains a single degree of freedom in the environment. As we further argue in \Sec{sec:discuss}, if ``larger'' environment are considered, initial-time dependent terms may be parametrically suppressed. In order to gain insight on such situations, in what follows we analyse the consequences of removing the spurious terms ``by hand''.

If spurious contributions are removed, \Eq{eq:TCLcoeff:crucial:non-spurious_simp} indicates that $\bs{D}=\bs{D}_{\mathrm{SPT}}$ and that $\bs{\Delta}=\bs{\Delta}_{\mathrm{ex}}$. The $\bs{\Delta}$ matrix is perfectly captured by TCL since $\bs{\Delta}_{\mathrm{ex}}$ only contains contributions proportional to $\lambda^4$, see \Eq{eq:LSex}. In particular, there is no damping term, \ie $\bs{\Delta}_{12}=0$ (note however that this is due to the specifics of the interaction we consider, which is such that $\bs{V}_{12}=0$). The diffusive part, \ie the one driven by $\bs{D}$, is however only partly contained in TCL, where $\bs{D}=\bs{D}_{\mathrm{SPT}}$, whereas $\bs{D}_{\mathrm{ex}}$ contains terms of higher-order in $\lambda$. We therefore expect spurious-free TCL to lie somewhere between SPT and the exact theory, which we now further investigate.

	\section{Non-perturbative resummation}\label{sec:sol}
In \Sec{sec:inin}, we have shown that the TCL master equation reduces to standard perturbation theory when solved at leading order in the interaction strength. In \App{sec:app:SPT:TCL} this equivalence is shown explicitly for the toy model introduced in \Sec{sec:model}. However, the TCL master equation can also be treated as a \textit{bona fide} dynamical map for the quantum state of the system, and solved as it is. In that case, its ability to resum secular effects has been investigated in various contexts~\cite{Burgess:2015ajz, Burgess:2014eoa, Boyanovsky:2015tba, Hollowood:2017bil, Brahma:2021mng}, and we now want to study how late-time resummation proceeds in the (spurious-free) cosmological Caldeira-Leggett model. 
\subsection{Power spectra}\label{subsec:pert}
As mentioned above, both in the exact and TCL descriptions, the state of the system remains Gaussian, hence it is fully characterised by its covariance matrix, \ie by its power spectra. This is why we first compare these setups at the level of their power spectra. If the cosmological Caldeira-Leggett model were to describe cosmological perturbations, note that the configuration-configuration power spectrum would be directly related to cosmological observables, such as the CMB temperature anisotropies. 

The power spectra in the exact theory are given by \Eq{eq:exactCovphiphi}, and as explained above, by expanding these formulas at first order in $\lambda^4$ one obtains their SPT counterpart. In the TCL setup, the power spectra can be obtained by solving the transport equation~\eqref{eq:TCLeom}. In the model under consideration, there is no damping term, $ \bs{\Delta}_{12}=0$, but in general it can be absorbed by introducing 
		\begin{align}
		\label{eq:damping}
			\boldsymbol{\sigma}_{\mathrm{TCL}}\equiv \ee^{\Gamma(\eta,\eta_0)} \boldsymbol{\Sigma}_{\mathrm{TCL}}
			\quad\quad\text{with}\quad\quad\Gamma(\eta,\eta_0)\equiv 2\int^{\eta}_{\eta_0}\dd \eta' \bs{\Delta}_{12}(\eta')\, ,
		\end{align}
		which is solution of a damping-free transport equation, namely
		\begin{align}
		\label{eq:transport:rescaled}
			\frac{\dd \bs{\sigma}_{\mathrm{TCL}}}{\dd \eta}&=\boldsymbol{\omega}\left(\bs{H}^{(\varphi)}+\bs{\Delta}\right) \bs{\sigma}_{\mathrm{TCL}}-\bs{\sigma}_{\mathrm{TCL}} \left(\bs{H}^{(\varphi)}+\bs{\Delta}\right) \boldsymbol{\omega} - \ee^{\Gamma(\eta,\eta_0)}\bs{\omega D \omega} \, .
		\end{align}
This equation can be seen as a homogeneous part, describing unitary evolution, and a source term, describing diffusion. The homogeneous part is generated by the Hamiltonian $H^{(\varphi)} + H^{(\mathrm{LS})}$, and by denoting $\boldsymbol{g}_{\mathrm{LS}}(\eta,\eta_0)$ the associated Green's matrix, the solution of \Eq{eq:transport:rescaled} reads
		\begin{align}
			\bs{\sigma}_{\mathrm{TCL}} (\eta ) = \boldsymbol{g}_{\mathrm{LS}}(\eta,\eta_0) \bs{\sigma}_{\mathrm{TCL}}(\eta_0) \boldsymbol{g}^{\mathrm{T}}_{\mathrm{LS}}(\eta,\eta_0) - \int_{\eta_0}^{\eta} \dd \eta' \ee^{\Gamma(\eta',\eta_0)}   \boldsymbol{g}_{\mathrm{LS}}(\eta,\eta')  \left[\bs{\omega} \bs{D}(\eta') \bs{\omega}\right]\boldsymbol{g}^{\mathrm{T}}_{\mathrm{LS}}(\eta,\eta') .
		\end{align}
Note that $\boldsymbol{g}_{\mathrm{LS}}$ is obtained from the Lamb-shift corrected mode functions
\bea
\boldsymbol{g}_{\mathrm{LS}}(\eta,\eta') = 2 \begin{pmatrix}
\Ima\left[v_{\mathrm{LS}}(\eta) p^*_{\mathrm{LS}}(\eta')\right] & -  \Ima\left[v_{\mathrm{LS}}(\eta) v^*_{\mathrm{LS}}(\eta')\right]\\
\Ima\left[p_{\mathrm{LS}}(\eta) p^*_{\mathrm{LS}}(\eta')\right] & -\Ima\left[p_{\mathrm{LS}}(\eta) v^*_{\mathrm{LS}}(\eta')\right]
\end{pmatrix},
\eea
where $v_{\mathrm{LS}}$ is the solution of $v_{\mathrm{LS}}''+\omega_{\mathrm{LS}}^2 v_{\mathrm{LS}}=0$ where $\omega_{\mathrm{LS}}^2 =k^2+m^2 a^2 + \bs{\Delta}_{11} - \bs{\Delta}_{12}'+\bs{\Delta}_{12}^2-2 \bs{\Delta}_{12}a'/a$, see \Eq{eq:HLSexp}, with Bunch-Davies initial conditions, and $p_{\mathrm{LS}}=v_{\mathrm{LS}}'-(a'/a) v_{\mathrm{LS}}$ as in \Eq{eq:momenta:def}.\footnote{In the present case, since $\bs{\Delta}_{12}=0$ and $\bs{\Delta}_{11}=\bs{\Delta}_{\mathrm{ex},11}$, where $\bs{\Delta}_{\mathrm{ex},11}$ is given in \Eq{eq:Delta:ex:comp}, one has $\omega_{\mathrm{LS}}^2=k^2+[m^2-\lambda^4/(M^2-m^2)]a^2$. This implies that $v_{\mathrm{LS}}$ and $p_{\mathrm{LS}}$ can be expressed in terms of Hankel functions as in \Eqs{eq:modefctl} and \eqref{eq:modefctpl}, with $\nu$ replaced by $\nu_{\mathrm{LS}}=\frac{3}{2}\sqrt{1-\left(\frac{2 m_{\mathrm{LS}}}{3H}\right)^2}$ where $m_{\mathrm{LS}}^2=m^2-\lambda^4/(M^2-m^2)$. This is consistent with effective-field theoretic approaches where the masses of light scalar fields are renormalised by heavy fields with contributions $\mathcal{O}(\lambda^4/M^2)$ \cite{Burgess:2007pt}.\label{footnote:nuLS}}
This leads to
		\begin{align}\label{eq:finsol}
			\bs{\Sigma}_{\mathrm{TCL}} (\eta) =& \ee^{-\Gamma(\eta,\eta_0)} \boldsymbol{g}_{\mathrm{LS}}(\eta,\eta_0) \bs{\Sigma}_{\mathrm{TCL}}(\eta_0) \boldsymbol{g}^{\mathrm{T}}_{\mathrm{LS}}(\eta,\eta_0) 
			\nonumber \\ &
			- \int_{\eta_0}^{\eta} \dd \eta' \ee^{-\Gamma(\eta,\eta')}   \boldsymbol{g}_{\mathrm{LS}}(\eta,\eta')  \left[\bs{\omega} \bs{D}(\eta') \bs{\omega}\right]\boldsymbol{g}^{\mathrm{T}}_{\mathrm{LS}}(\eta,\eta').
		\end{align}
In practice, this integral is computed numerically from a large negative value of $\eta_0$ (sufficiently large that we check the result does not depend on $\eta_0$). 
\subsubsection*{Growth rate}	
\begin{figure}
			\centering
			\includegraphics[width=0.8\textwidth]{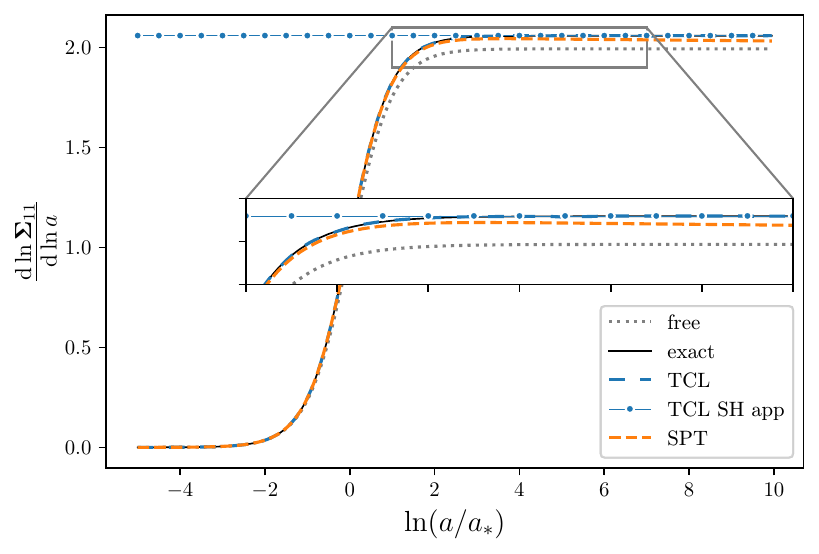}
			\caption{Growth rate of the configuration-configuration power spectrum as a function of time, labeled with the scale factor ($a_*=k/H$ corresponds to the time of Hubble exit, \ie~when $\eta_*=-1/k$). The result is displayed in the free (grey), exact (black), TCL${}_2$ (blue) and SPT (orange) theories. The blue dotted line corresponds to the super-Hubble expansion for TCL, see \Eq{eq:TCL:superH:exp:Cov}, which leads to the growth rate $\dd\ln\bs{\Sigma}_{\mathrm{TCL},11}/\dd\ln a = 2\nu_{\mathrm{LS}}-1$. The parameters are set to $\lambda=H$, $m=H/10$ and $M=\sqrt{10}H$.}
			\label{fig:GR}
\end{figure} 	
First we compare in \Fig{fig:GR} the growth rate of the configuration-configuration power spectrum, $\dd\ln\bs{\Sigma}_{11}/\dd\ln a$. The result is given in the free theory (\ie setting $\lambda=0$, grey line), in the exact theory (black line), in TCL (blue line) and in SPT (orange line). The difference between these different setups becomes more pronounced at late time, on which the inset zooms in. One can see that TCL provides an excellent approximation, better than SPT, which itself is closer to the exact result than the free theory.  

The behaviour of the TCL covariance matrix can be further understood by investigating the super-Hubble (\ie late time, $-k\eta\ll 1$) limit of the transport equation~\eqref{eq:TCLeom}. In this regime, an expansion of the coefficients can be found in \App{sec:TCLcoeff:superH}. By inserting power-law ansatz for the entries of the covariance matrix, one finds that the diffusion term becomes negligible at large scales, and that
\bea
\label{eq:TCL:superH:exp:Cov}
\bs{\Sigma}_{\mathrm{TCL},11} \propto a^{2\nu_{\mathrm{LS}}-1}\, ,\quad
\bs{\Sigma}_{\mathrm{TCL},12} \propto a^{2\nu_{\mathrm{LS}}}\, ,\quad
\bs{\Sigma}_{\mathrm{TCL},22} \propto a^{2\nu_{\mathrm{LS}}+1}\, ,
\eea
where $\nu_{\mathrm{LS}}$ was introduced in footnote~\ref{footnote:nuLS}. The corresponding growth rate, $2\nu_{\mathrm{LS}}-1$, is displayed in \Fig{fig:GR} with the dotted blue line, and one can check that it asymptotes the TCL result at late time indeed. 

In the exact theory, the term involving $\bs{\Sigma}_\uell$ dominates over the one involving $\bs{\Sigma}_{\mathrm{h}}$ in \Eq{eq:exactCovphiphi}, so the growth rate is given by $2\nu_\uell-1$, where $\nu_\uell$ is given below \Eq{eq:dyn1}. It is worth stressing that by expanding $\nu_\uell$ at leading order in $\lambda^4$, one recovers $\nu_{\mathrm{LS}}$ [namely $m_\uell^2=m_{\mathrm{LS}}^2+\order{\lambda^8}$]. As a consequence,  TCL correctly reproduces the growth rate at first order in $\lambda^4$. 

Although this may seem as a perturbative result, let us stress that the resummed non-perturbative feature lies in \Eq{eq:TCL:superH:exp:Cov}. Indeed, in SPT,  expanding $\bs{\Sigma}_{\varphi\varphi}$ at leading order in $\lambda^4$ leads to 
\bea
\label{eq:growth:SPT}
\bs{\Sigma}_{\mathrm{SPT},11}\propto a^{2\nu_\varphi-1}\left[1+ \frac{\lambda^4}{H^2\nu_\varphi\left(M^2-m^2\right)}\ln a\right]
\eea
 at late time, where $\nu_\varphi=\frac{3}{2}\sqrt{1-\left(\frac{2m}{3H}\right)^2}$. This matches \Eq{eq:TCL:superH:exp:Cov} at leading order in $\lambda^4$, but \Eq{eq:TCL:superH:exp:Cov} contains all higher-order terms in $\lambda^{4}$ that allow the logs to be resumed. In particular, \Eq{eq:growth:SPT} implies that at late time, the growth rate in SPT approaches the one of the free theory, $2\nu_\varphi-1$, while as stated above the growth rate of TCL incorporates the first correction in $\lambda^4$. 
 \subsubsection*{Relative deviation to the exact result}
 \begin{figure}
			\centering
			\includegraphics[width=0.8\textwidth]{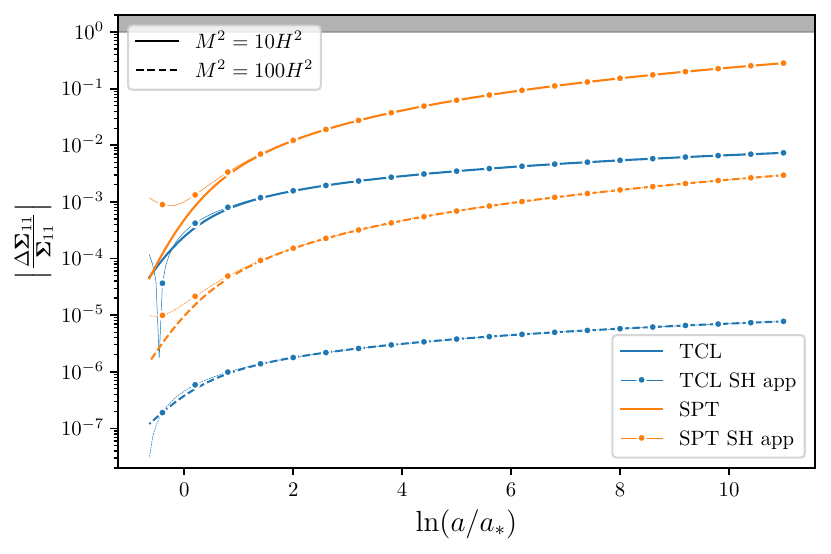}
			\caption{Relative error in the configuration-configuration power spectrum in TCL${}_2$ ($\vert \bs{\Sigma}_{\mathrm{TCL},11}-\bs{\Sigma}_{\varphi\varphi,11}\vert/\bs{\Sigma}_{\varphi\varphi,11}$, blue lines) and SPT ($\vert \bs{\Sigma}_{\mathrm{SPT},11}-\bs{\Sigma}_{\varphi\varphi,11}\vert/\bs{\Sigma}_{\varphi\varphi,11}$, orange lines). The result is displayed as a function of time, labeled with the scale factor, and for $M^2=10H^2$ (solid lines) and  $M^2=100H^2$ (dashed lines). The dotted lines correspond to the super-Hubble formula~\eqref{eq:TCL:superH:exp:Cov}, which indeed provide a good fit at late time. The parameters are taken as $m^2 = 10^{-4} H^2$ and $\lambda^2 = H^2$. The grey-shaded area is where the error is larger than $100\%$.}
			\label{fig:zvar}
		\end{figure} 
The performance reached by TCL or SPT is given by the relative deviation of their covariance matrices to the exact result. This is displayed in \Fig{fig:zvar} for $m^2 = 10^{-4} H^2$ and $\lambda^2 = H^2$, which purposely corresponds to a large coupling. One can check that TCL is always more accurate than SPT, and that the difference in accuracy becomes more pronounced at larger $M$. This can be understood as follows. In the super-Hubble regime, TCL behaves according to \Eq{eq:TCL:superH:exp:Cov}, which is super-imposed in \Fig{fig:zvar} and indeed provides a good fit. It leads to
\bea
\label{eq:TCL:relat:error:superH}
\frac{\left\vert\bs{\Sigma}_{\mathrm{TCL},11}-\bs{\Sigma}_{\varphi\varphi,11}\right\vert}{\bs{\Sigma}_{\varphi\varphi,11}} \simeq a^{2\left(\nu_{\mathrm{LS}}-\nu_\uell\right)}-1 = \frac{\lambda^8 \ln(a)}{\nu_\varphi H^2\left(M^2-m^2\right)^3}+\order{\lambda^{12}}\, .
\eea
The last result is expanded at leading order in $\lambda$ (hence in $\ln a$), which provides a good approximation as long as the relative error is much smaller than one, as in \Fig{fig:zvar}. In SPT, \Eq{eq:growth:SPT} gives rise to
\bea
\label{eq:SPT:relat:error:superH}
\frac{\left\vert\bs{\Sigma}_{\mathrm{SPT},11}-\bs{\Sigma}_{\varphi\varphi,11}\right\vert}{\bs{\Sigma}_{\varphi\varphi,11}} \simeq \frac{\lambda^8\ln^2(a)}{2\left(M^2-m^2\right)^2 H^4 \nu_\varphi^2}+\order{\lambda^{12}}
\eea
at late time. There are two main differences between \Eqs{eq:TCL:relat:error:superH} and~\eqref{eq:SPT:relat:error:superH}. First, when the environment is heavy, $M\gg H$, the relative error in TCL decays as $\lambda^8/M^{6}$ while it is suppressed by $\lambda^8/M^4$ in SPT. This explains why, when going from $M^2=10H^2$ to $M^2=100H^2$ in \Fig{fig:zvar}, the relative error decreases by a factor $10^3$ in TCL and by a factor $10^2$ in SPT. This indicates that, although both results become more accurate as the environment is heavier, the gain in accuracy is much stronger for TCL. Second, the relative error in SPT increases as $\ln^2(a)$ at late time, while it only increases as $\ln(a)$ in TCL. This is why in \Fig{fig:zvar}, the difference in accuracy between these two approaches becomes even larger as time proceeds.

\begin{figure}
			\centering
			\includegraphics[width=0.8\textwidth]{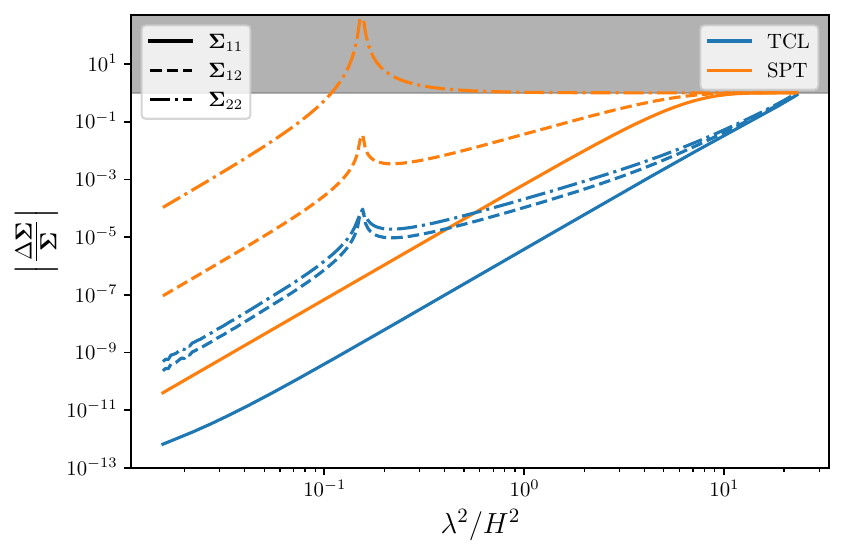}
			\caption{Relative error in all entries of the covariance matrix of the system. The blue lines correspond to the TCL${}_2$ result $\vert \bs{\Sigma}_{\mathrm{TCL},ij}-\bs{\Sigma}_{\varphi\varphi,ij}\vert /\vert \bs{\Sigma}_{\varphi\varphi,ij}\vert$, while the orange lines correspond to standard perturbation theory (SPT) $\vert \bs{\Sigma}_{\varphi\varphi,ij}^{(2)}-\bs{\Sigma}_{\varphi\varphi,ij}\vert /\vert \bs{\Sigma}_{\varphi\varphi,ij}\vert$.
			Different line styles correspond to different entries of the covariance matrix, and the parameters of the model are chosen as $m^2 = 10^{-4} H^2$, $M^2 = 10^{2} H^2$ and $a/a_* =\ee^5$. The grey-shaded area is where the error is larger than $100\%$. The peaky features correspond to where the exact power spectrum $\bs{\Sigma}_{\varphi\varphi,12}$ vanishes and $\bs{\Sigma}_{\varphi\varphi,22}$ goes through a local minimum.
			}
			\label{fig:TCLinin}
\end{figure} 
Finally, in \Fig{fig:TCLinin} we display the relative error for all power spectra (\ie all entries of the covariance matrix), as a function of the interaction strength $\lambda$. When $\lambda$ is small, the relative error scales as $\lambda^8$ for both SPT and TCL, in agreement with the fact that both methods match the exact result at order $\lambda^4$ [see \Sec{sec:inin}, see also \Eqs{eq:TCL:relat:error:superH} and~\eqref{eq:SPT:relat:error:superH}].
One can also see that both in TCL and in SPT, the reconstruction of the configuration-configuration power spectrum is better than for the configuration-momentum power spectrum, which is itself better than the momentum-momentum power spectrum. In TCL, all power spectra are accurately computed up to large values of $\lambda$. For instance, even when $\lambda/H=1$, the relative error is smaller than $10^{-4}$ for all power spectra. In SPT however, the momentum-momentum power spectrum is already out of control for such values of $\lambda$. Indeed, the correlators involving the momentum are given with less precision in SPT, and the perturbative expansion breaks down for the momentum-momentum power spectrum much sooner than for the configuration-configuration power spectrum. This will be of prime importance below, since those correlators play an essential role in the process of decoherence.
		\subsection{Decoherence}
		\label{sec:decoherence}
\begin{figure}
			\centering
				\includegraphics[width=.8\textwidth]{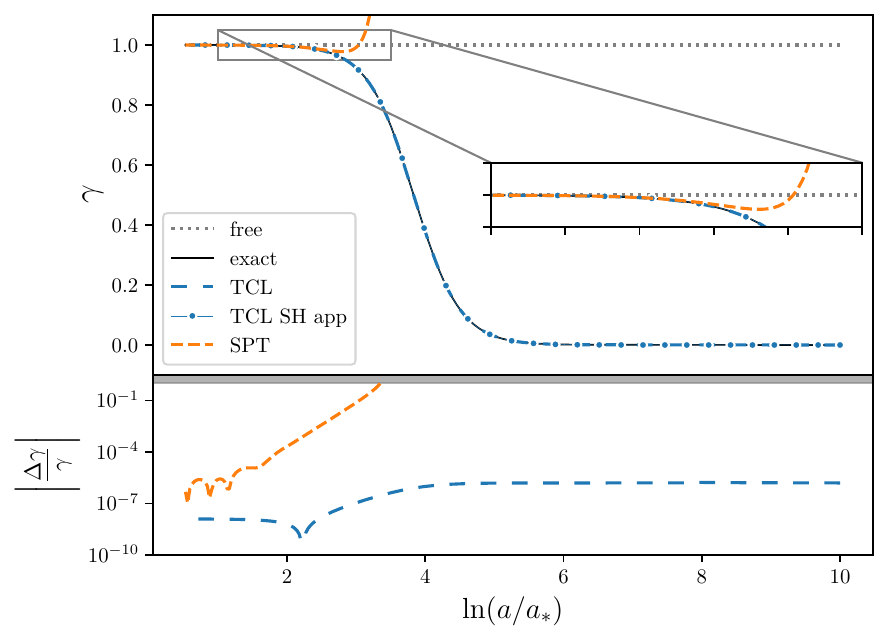}
			\caption{Purity parameter~\eqref{eq:gamma} in the free (grey), exact (black), TCL${}_2$ (blue), and SPT (orange) approaches as a function of time. The dotted line corresponds to the super-Hubble expansion of the TCL result, see \Eq{eq:purity:TCL:superH}. The lower panel shows the error of TCL${}_2$ and SPT relative to the exact result. The parameters are set to $m^2 = 10^{-4} H^2$ and $M^2 = 10^{2} H^2$ with $\lambda^2 = 10^{-1} H^2$. The perturbative result rapidly diverges, while TCL${}_2$ accurately predicts the amount of decoherence even in the fully decohered regime.}\label{fig:DecoTCLininzvar}
\end{figure} 
We turn our attention to decoherence that we measure using the purity whose expression for Gaussian states is given by \Eq{eq:gamma}. The result is displayed in \Fig{fig:DecoTCLininzvar}. As time proceeds, the system entangles with its environment, decoherence occurs (\ie $\gamma$ decreases away from $1$), and the system becomes maximally mixed soon after Hubble-crossing for the parameters used in the figure. 

The lower panel displays the error relative to the exact result. One can see that, when time proceeds, the SPT result quickly diverges. So perturbation theory is only able to describe quasi pure states, for which $1-\gamma\ll 1$, and breaks down when decoherence proceeds. The reason for the weak performance of SPT is that the purity parameter is driven by the so-called cosmological decaying mode, which is encoded in the power spectra involving the momentum. Around \Fig{fig:TCLinin} we saw that those are precisely the correlators that SPT predicts with the least accuracy.  On the contrary, TCL${}_2$ remarkably describes the full decoherence process, and is able to approximate the full quantum state even in the strongly decohered regime. The relative error freezes to a tiny value at large scales (here of the order $10^{-6}$), which is a manifestation of the resummation occurring in TCL. 

The simplest way to access the late-time behaviour of the purity is to derive an equation of motion for $\det(\bs{\Sigma}_{\mathrm{TCL}})$ from the transport equation~\eqref{eq:TCLeom}, namely
\bea
\label{eq:detSigma:eom}
\frac{\dd\det(\bs{\Sigma}_{\mathrm{TCL}})}{\dd\eta}=\bs{D}_{11}\bs{\Sigma}_{\mathrm{TCL},11}+2\bs{D}_{12}\bs{\Sigma}_{\mathrm{TCL},12}\
-4\bs{\Delta}_{12}\det(\bs{\Sigma}_{\mathrm{TCL}}).
\eea
All unitary contributions (\ie those involving $\bs{H}^{(\varphi)}$ and $\bs{\Delta}_{11}$) have cancelled out (indeed, only non-unitary contributions can change the purity). This implies that diffusion, controlled by $\bs{D}$, is crucial in the process of decoherence (since $\bs{\Delta}_{12}=0$ in the present case). It contrasts with \Sec{subsec:pert} where we had found that $\bs{D}$ gives negligible corrections to the power spectra on large scales -- those negligible corrections are precisely the ones driving decoherence. The results of \App{sec:TCLcoeff:superH} together with \Eq{eq:TCL:superH:exp:Cov} indicate that the two first terms of \Eq{eq:detSigma:eom} are of the same order $a^{2\nu_{\mathrm{LS}}-1}$ at late time. In this limit \Eq{eq:detSigma:eom} can be integrated, and one obtains
\bea
\label{eq:purity:TCL:superH}
\det(\bs{\Sigma}_{\mathrm{TCL}}) \simeq \frac{1}{4}+\frac{2^{2\nu_{\mathrm{LS}}-3}}{\pi}\Gamma^2\left(\nu_{\mathrm{LS}}\right)\left(\frac{\lambda}{H}\right)^4\left(\frac{H}{M}\right)^3\left(\frac{a}{a_*}\right)^{2\nu_{\mathrm{LS}}}\, ,
\eea
where the prefactors in \Eq{eq:TCL:superH:exp:Cov} have been set by neglecting diffusion (alternatively, they can be set by asymptotic matching at Hubble crossing and this gives a very similar result) and we have neglected contributions exponentially suppressed by $M/H$ to reach a concise expression (they be easily kept but do not bring any particular insight). The purity $\gamma_{\mathrm{TCL}}=1/(4 \det\bs{\Sigma}_{\mathrm{TCL}})$ obtained from this expression is displayed in \Fig{fig:DecoTCLininzvar} with the dotted line. One can check that it provides an excellent approximation to the full TCL result, hence to the exact result too. 

The above formula~\eqref{eq:purity:TCL:superH} also allows us to study under which conditions decoherence occurs for the model at hand. It is non perturbative in $\lambda$ since one should recall that $\nu_{\mathrm{LS}}$ depends on $\lambda$, see footnote~\ref{footnote:nuLS}, although the rate of decoherence is mostly proportional to $(\lambda/H)^4$. Similarly, although $\nu_{\mathrm{LS}}$ depends on $M$, decoherence occurs at a rate mostly proportional to $(H/M)^3$, so it is slower for heavier environments. Finally, it is very efficient on super-Hubble scales, since it scales as $a^{2\nu_{\mathrm{LS}}}\sim a^3$, so roughly as the spatial volume, as often encountered~\cite{Burgess:2006jn,Martin:2018zbe}. For instance, for scales of astrophysical interest today that are such that $a/a_*\sim \ee^{50}$ at the end of inflation, for $M/H=100$ and $m/H=10^{-2}$, one finds that decoherence proceeds during inflation as soon as $\lambda/H>10^{-15}$, a very small value indeed.

\section{Conclusion}
\label{sec:discuss}
Let us now summarise our main results and open up a few prospects. In this work, we have investigated how the master-equation program can be implemented in cosmology. To this end, we have used a toy model where two scalar fields are linearly coupled and evolve on a de-Sitter background. It has the advantage of being exactly solvable, an ``integrable system'', in which the performance of effective methods can be assessed and compared with more traditional, perturbative techniques.  

We have derived the second-order Time-ConvolutionLess (TCL) equation in this setup, which is a master equation for the reduced density matrix of the system (here the lighter field), and which features the memory kernel of the environment (here the heavier field). It possesses three contributions: a unitary ``Lamb-shift'' term (renormalisation of the bare Hamiltonian), a dissipation term (energy exchange with the environment) and a diffusive term (driving the quantum decoherence process). They can all be expressed in terms of integrals ranging from the initial time to the time at which the master equation is written.

Usually, the memory kernel is sufficiently peaked around the coincident configuration that these integrals are dominated by their upper bound, hence they carry negligible dependence on the initial time. This is the case if the relaxation time of the environment around its stationary configuration is small compared to the time scale over which the evolution of the system is tracked. This is the so-called Markovian, or Lindbladian limit. In the present case however, due to the presence of a dynamical background, there is no such thing as a stationary configuration for the environment, which strongly departs from being a thermal bath. In practice we find that these integrals carry a non-negligible dependence on the initial time, through a set of terms that we have dubbed ``spurious''.

We have then shown that these spurious terms cancel out when the TCL equation is solved perturbatively in the coupling constant, \ie they are absent from the perturbative version of the theory. This is consistent with the fact that the perturbative solution to the TCL equation is strictly equivalent to standard perturbation theory (such as the in-in formalism for instance).
When solving the TCL equation non-perturbatively however, they lead to unphysical diverging behaviours, which clearly signals their problematic nature.

However, if one removes them ``by hand'' (which does not necessarily makes the dynamics Markovian, see footnote~\ref{footnote:Markovian:versus:CPTP}), one finds that the TCL equation provides an excellent approximation to the full theory: it successfully reproduces all power spectra up to large values of the interaction strength, and it tracks the amount of decoherence very accurately, including at late time when the system is in a strongly mixed state. This is due to an explicit resummation of logarithmic terms (\ie of powers of $\ln a$, where $a$ is the scale factor of the universe), and in \App{app:growingdecaying} we show that this resummation is more efficient than the late-time resummation technique proposed in \Refa{Boyanovsky:2015tba}.
The incorporation of these late-time secular effects makes TCL vastly superior to perturbative methods. Although we have found that it does not require particularly heavy environment, the advantage of TCL compared to perturbative methods is even more pronounced when the mass $M$ of the environmental field is larger than the Hubble scale $H$, since the relative error of the former scales as $(H/M)^6$ while it scales as $(H/M)^4$ for the latter. 

\begin{center}
To summarise, we have found that the master-equation program can be successfully applied in cosmological backgrounds, \emph{provided spurious terms are suppressed}. 
\end{center}

The presence of the spurious terms may be related to the simplicity of our toy model, where only one field is contained in the environment, which can therefore not be considered as a proper reservoir. If multiple fields were present indeed, all with different masses, thus oscillating at different frequencies, the memory kernel would be suppressed away from the coincident limit through the accumulation of random phases~\cite{Breuer:2002pc,RevModPhys.88.021002} (technically, the memory kernel would involve some Fourier transform of the mass distribution of the environmental fields, which may be peaked if that distribution is sufficiently broad). This mechanism was studied \eg in the context of black-hole physics in \Refa{Burgess:2021luo}. Another possibility would be to consider non-linear interactions between the two fields, which would imply that one Fourier mode in the system couples to all Fourier modes in the environment, hence making the number of environmental degrees of freedom to which the system couples infinite. One could also consider situations in which non-linearities only arise within the environmental sector,\footnote{Let us note that the presence of non-linearities, even if confined to the environmental sector, would leave an imprint on the non-Gaussian statistics of the system~\cite{Chen:2009zp, Assassi:2013gxa}.} as in quasi-single field models~\cite{Chen:2009zp, Pi:2012gf}. The same mechanism of random phase addition would presumably occur in those cases, which would also lead to a suppression of the spurious terms. Whether or not that suppression is enough should be the subject of further investigations. Another, maybe more adventurous question, is whether or not one can design an improved master equation, where the removal of spurious contributions is automatically taken care of. Indeed, our results show that master equations free from spurious terms are extremely powerful at deriving reliable predictions for cosmology, and perform much better than perturbative methods. We plan to address these issues in future works.

\section{Acknowledgments}	

It is a pleasure to thank Suddhasattwa Brahma, Cliff Burgess, Denis Comelli, Jaime Calder\'on-Figueroa, Bei-Lok Hu, Christian K\"ading, Greg Kaplanek and Amaury Micheli for interesting discussions.

	\appendix	
	
	\section{Microphysical derivation of the TCL${}_2$ master equation}
	\label{app:TCL2:microphysical:derivation}
	In this appendix, we present an alternative derivation of the TCL${}_2$ master equation~\eqref{eq:TCL2CL} in the curved-space Caldeira-Leggett model, which does not rely on the cumulant expansion of the Nakajima-Zwanzig equation. We start from the Liouville--Von-Neumann equation in the interaction picture~\eqref{eq:LVN}, namely
	\bea \label{eq:rhotildeeom}
	\frac{\dd \widetilde{\rho}}{\dd \eta}=
	-i \lambda^2 \left[\widetilde{\mathcal{H}}_{\mathrm{int}}(\eta),\widetilde{\rho}(\eta)\right]\, .
	\eea
As noted in \Eq{eq:formal}, it can be solved formally as 
	\begin{align}
		\widetilde{\rho}(\eta)  &= \widetilde{\rho}(\eta_0) -i \lambda^2 \int_{\eta_0}^{\eta}\dd \eta' \left[\widetilde{\mathcal{H}}_{\mathrm{int}}(\eta'),\widetilde{\rho}(\eta')\right].
	\end{align}
	Inserting this expression into \Eq{eq:rhotildeeom}, one obtains
	\begin{align}
	\label{eq:TCL:app:interm1}
		\frac{\dd \widetilde{\rho}}{\dd \eta}  = -i \lambda^2 \left[\widetilde{\mathcal{H}}_{\mathrm{int}}(\eta),\widetilde{\rho}(\eta_0)\right] - \lambda^4 \int_{\eta_0}^{\eta}\dd \eta'\left[\widetilde{\mathcal{H}}_{\mathrm{int}}(\eta),\left[\widetilde{\mathcal{H}}_{\mathrm{int}}(\eta'),\widetilde{\rho}(\eta')\right] \right] + \mathcal{O}(\lambda^6)\, .
	\end{align}
	This procedure could be iterated to obtain higher-order nested commutators, controlled by higher powers of the interaction strength. If the coupling constant $\lambda$ is small (Born approximation), one may stop at order $\mathcal{O}(\lambda^4)$ where the first non-unitary effects appear. 
	
Our next task is to turn \Eq{eq:TCL:app:interm1} into an ordinary differential equation that is local in time for the reduced density matrix $\widetilde{\rho}_{\mathrm{red}}(\eta)$. By tracing \Eq{eq:TCL:app:interm1} over the environmental degrees of freedom, one finds
	\begin{align}
		\frac{\dd \widetilde{\rho}_{\mathrm{red}}}{\dd \eta}  \simeq -i \lambda^2 \mathrm{Tr}_{\mathrm{E}}\left[\widetilde{\mathcal{H}}_{\mathrm{int}}(\eta),\widetilde{\rho}(\eta_0)\right] - \lambda^4 \int_{\eta_0}^{\eta}\dd \eta'\mathrm{Tr}_{\mathrm{E}}\left[\widetilde{\mathcal{H}}_{\mathrm{int}}(\eta),\left[\widetilde{\mathcal{H}}_{\mathrm{int}}(\eta'),\widetilde{\rho}(\eta')\right] \right] .
	\end{align}  
	In the interaction picture, the deviation of $\widetilde{\rho}$ from its initial configuration is necessarily controlled by some positive power $p$ of the interaction strength, 
	\begin{align}\label{eq:factorisation}
		\widetilde{\rho}(\eta) = \widetilde{\rho}_{\mathrm{red}}(\eta_0) \otimes \widetilde{\rho}_{\mathrm{E}}(\eta_0) + \lambda^{p} \widetilde{\rho}_{\mathrm{correl}}(\eta) 
	\end{align}
	where $\mathrm{Tr}_{\mathrm{E}}(\widetilde{\rho}_{\mathrm{correl}}) = \mathrm{Tr}_{\mathrm{S}}(\widetilde{\rho}_{\mathrm{correl}}) = 0$. Consequently,
	\bea\label{eq:exp1}
		\frac{\dd \widetilde{\rho}_{\mathrm{red}}}{\dd \eta}  =& -i \lambda^2 \mathrm{Tr}_{\mathrm{E}}\left[\widetilde{\mathcal{H}}_{\mathrm{int}}(\eta),\widetilde{\rho}_{\mathrm{red}}(\eta_0) \otimes \widetilde{\rho}_{\mathrm{E}}(\eta_0)\right]  \\
		&-i \lambda^{p+2} \mathrm{Tr}_{\mathrm{E}}\left[\widetilde{\mathcal{H}}_{\mathrm{int}}(\eta),\widetilde{\rho}_{\mathrm{correl}}(\eta_0)\right] \\
		&- \lambda^4 \int_{\eta_0}^{\eta}\dd \eta'\mathrm{Tr}_{\mathrm{E}}\left[\widetilde{\mathcal{H}}_{\mathrm{int}}(\eta),\left[\widetilde{\mathcal{H}}_{\mathrm{int}}(\eta'),\widetilde{\rho}_{\mathrm{red}}(\eta') \otimes \widetilde{\rho}_{\mathrm{E}}(\eta')\right] \right] \\ 
		&- \lambda^{p+4} \int_{\eta_0}^{\eta}\dd \eta'\mathrm{Tr}_{\mathrm{E}}\left[\widetilde{\mathcal{H}}_{\mathrm{int}}(\eta),\left[\widetilde{\mathcal{H}}_{\mathrm{int}}(\eta'),\widetilde{\rho}_{\mathrm{correl}}(\eta') \right] \right] .
\eea	
	Which term dominates depends on the value of $p$, which can be determined as follows. Let us first recall that $\widetilde{\mathcal{H}}_{\mathrm{int}}(\eta)$ was given above \Eq{eq:TCL2CL} and reads 
	\begin{align}
	\label{eq:app:Hint}
		\widetilde{\mathcal{H}}_{ \mathrm{int}}(\eta) &=  a^2(\eta) \widetilde{v}_{\varphi}(\eta)\widetilde{v}_{\chi}(\eta)\, ,
	\end{align}
	which leads to
	\begin{eqnarray}
		\mathrm{Tr}_{\mathrm{E}}\left[\widetilde{\mathcal{H}}_{\mathrm{int}}(\eta), \widetilde{\rho}_{\mathrm{red}}(\eta_0) \otimes \widetilde{\rho}_{\mathrm{E}}(\eta_0)\right] = a^2(\eta)\left[\widetilde{v}_{\varphi}(\eta), \widetilde{\rho}_{\mathrm{red}}(\eta_0) \right] \mathrm{Tr}_{\mathrm{E}}\left[\widetilde{v}_{\chi}(\eta) \widetilde{\rho}_{\mathrm{E}}(\eta_0)\right].
	\end{eqnarray}
	Note that 
	\begin{align}
		\mathrm{Tr}_{\mathrm{E}}\left[\widetilde{v}_{\chi}(\eta) \widetilde{\rho}_{\mathrm{E}}(\eta_0)\right] = \Big<\widetilde{v}_{\chi} (\eta-\eta_0)\Big>
	\end{align}	
	which is the mean value of the environment field operator. Such a mean value can always be absorbed in a redefinition of the field, such that the first term in the right-hand side of \Eq{eq:exp1} vanishes. 
	From \Eq{eq:factorisation}, $\dd \widetilde{\rho}_{\mathrm{red}}/\dd \eta$ necessarily contains a term of order $\mathcal{O}\left(\lambda^{p}\right) $, so the only possibility is that $p=4$. As a consequence, the terms of order  $\mathcal{O}\left(\lambda^{p+2}\right) $ and $\mathcal{O}\left(\lambda^{p+4}\right) $ can be neglected, and one finds
	\begin{align}\label{eq:almostTCL2}
		\frac{\dd \widetilde{\rho}_{\mathrm{red}}}{\dd \eta}  = 
		- \lambda^4 \int_{\eta_0}^{\eta}\dd \eta'\mathrm{Tr}_{\mathrm{E}}\left[\widetilde{\mathcal{H}}_{\mathrm{int}}(\eta),\left[\widetilde{\mathcal{H}}_{\mathrm{int}}(\eta'),\widetilde{\rho}_{\mathrm{red}}(\eta') \otimes \widetilde{\rho}_{\mathrm{E}}(\eta')\right] \right] .
	\end{align}  
At leading order in $\lambda$, one can safely replace $\widetilde{\rho}_{\mathrm{red}}(\eta')$ by $\widetilde{\rho}_{\mathrm{red}}(\eta)$ and $\widetilde{\rho}_{\mathrm{E}}(\eta')$ by $\widehat{\rho}_{\mathrm{E}}$ in \Eq{eq:almostTCL2}. This leads to a manifestly time-local equation, namely
        \begin{align}\label{eq:TCL2app}
		\frac{\dd \widetilde{\rho}_{\mathrm{red}}}{\dd \eta}  = 
		- \lambda^4 \int_{\eta_0}^{\eta}\dd \eta'\mathrm{Tr}_{\mathrm{E}}\left[\widetilde{\mathcal{H}}_{\mathrm{int}}(\eta),\left[\widetilde{\mathcal{H}}_{\mathrm{int}}(\eta'),\widetilde{\rho}_{\mathrm{red}}(\eta) \otimes \widehat{\rho}_{\mathrm{E}}\right] \right] 
	\end{align}  
	which is consistent with \Eq{eq:TCL2exp}. 
Replacing $\mathcal{H}_{\mathrm{int}}$ by \Eq{eq:app:Hint} and expanding the commutators yields the result
	\begin{align}\label{eq:preLindblad}
		\frac{\dd \widetilde{\rho}_{\mathrm{red}}}{\dd \eta} =& - \lambda^4 a^2(\eta)\int_{\eta_0}^{\eta}\dd \eta'  a^2(\eta') \nonumber \\
		\Bigg\{&\left[\widetilde{v}_{\varphi} (\eta) \widetilde{v}_{\varphi} (\eta') \widetilde{\rho}_{\mathrm{red}}(\eta) -  \widetilde{v}_{\varphi} (\eta') \widetilde{\rho}_{\mathrm{red}}(\eta) \widetilde{v}_{\varphi} (\eta) \right]\mathcal{K}^{>}(\eta,\eta') \nonumber\\
		-&\left[\widetilde{v}_{\varphi} (\eta)  \widetilde{\rho}_{\mathrm{red}}(\eta) \widetilde{v}_{\varphi} (\eta') -   \widetilde{\rho}_{\mathrm{red}}(\eta) \widetilde{v}_{\varphi} (\eta') \widetilde{v}_{\varphi} (\eta) \right]\mathcal{K}^{<}(\eta,\eta')
		\Bigg\}, 
	\end{align} 
	where the memory kernels are defined as in \Eq{eq:TCK:kernel}, namely
	\begin{align}
		\mathcal{K}^{>}(\eta,\eta') \equiv \mathrm{Tr}_{\mathrm{E}}\left[\widehat{v}_{\chi} (\eta) \widehat{v}_{\chi} (\eta')\widehat{\rho}_{\mathrm{E}}\right],
		 \label{eq:Kplus}\\
		\mathcal{K}^{<}(\eta,\eta') \equiv \mathrm{Tr}_{\mathrm{E}}\left[\widehat{v}_{\chi} (\eta') \widehat{v}_{\chi} (\eta)\widehat{\rho}_{\mathrm{E}}\right] . \label{eq:Kminus}
	\end{align}
As in \Eq{eq:K:mode:function}, they can be expressed in terms of the mode functions
	\begin{align}
	\label{eq:kernel:mode:function}
		\mathcal{K}^{>}(\eta,\eta') &= v_{\chi} (\eta) v^{*}_{\chi} (\eta') \\ 
		\mathcal{K}^{<}(\eta,\eta') &= v^{*}_{\chi} (\eta) v_{\chi}(\eta')  = \mathcal{K}^{>*}(\eta,\eta'),
	\end{align}
	and the master equation reads
	\begin{align}\label{eq:CME}
		\frac{\dd \widetilde{\rho}_{\mathrm{red}}}{\dd \eta} =& - \lambda^4 a^2(\eta)\int_{\eta_0}^{\eta}\dd \eta'  a^2(\eta') \nonumber \\
		\Big\{&\left[\widetilde{v}_{\varphi} (\eta) \widetilde{v}_{\varphi} (\eta') \widetilde{\rho}_{\mathrm{red}}(\eta) -  \widetilde{v}_{\varphi} (\eta') \widetilde{\rho}_{\mathrm{red}}(\eta) \widetilde{v}_{\varphi} (\eta) \right]\mathcal{K}^{>}(\eta,\eta') \nonumber\\
		-&\left[\widetilde{v}_{\varphi} (\eta)  \widetilde{\rho}_{\mathrm{red}}(\eta) \widetilde{v}_{\varphi} (\eta') -   \widetilde{\rho}_{\mathrm{red}}(\eta) \widetilde{v}_{\varphi} (\eta') \widetilde{v}_{\varphi} (\eta) \right]\mathcal{K}^{>*}(\eta,\eta') 
		\Big\}.
	\end{align} 
Expanding $\mathcal{K}^>$ into its real and imaginary part, one recovers \Eq{eq:TCL2CL}.

			\section{Phase-space representation of the TCL${}_2$ master equation}\label{app:phase:space}
	An alternative representation of the quantum state is given in the phase-space by the Wigner function (see \Refa{doi:10.1119/1.2957889} for a brief introduction). For Gaussian states, the Wigner function takes the simple form of a multivariate Gaussian \cite{PhysRevA.36.3868}, which makes it particularly convenient to work with.
	
The Wigner function is defined as the inverse Weyl transform of the density matrix. For a generic quantum operator $\widehat{O}$, 	the inverse Wigner-Weyl transform reads 
	\begin{align}
		W_{\widehat{O}}(v_\varphi,p_\varphi) = 2 \int_{-\infty}^{\infty} \dd y \ee^{-2ip_\varphi y}\left<v_\varphi+y\right| \widehat{O} \left| v_\varphi-y \right>
	\end{align}
and is a function of the phase-space variables $v_\varphi$ and $p_\varphi$. The above formula is written in the configuration representation, it can also be written in the momentum representation,
	\begin{align}
		W_{\widehat{O}}(v_\varphi, p_\varphi) = 2 \int_{-\infty}^{\infty} \dd k \ee^{2ikv_\varphi}\left<p_\varphi+k\right| \widehat{O} \left| p_\varphi-k \right> .
	\end{align}
In this way, commutators of quantum operators are mapped to the Poisson brackets of their phase-space representations. Indeed, using the above formulas, one finds
		\begin{align}
			W_{\left[\widehat{v}_{\varphi},  \widehat{O}\right]} &= i \frac{\partial}{\partial p_\varphi}  W_O ~~~~\text{and}~~~~
			W_{\left[\widehat{p}_{\varphi},  \widehat{O}\right]} = - i \frac{\partial}{\partial v_\varphi}  W_{O}  \, ,\\ 
			W_{\left\{\widehat{v}_{\varphi},  \widehat{O}\right\}} &= 2 v_\varphi W_{O} ~~~~\text{and}~~~~
			W_{\left\{\widehat{p}_{\varphi},  \widehat{O}\right\}} = 2 p_\varphi W_{O}\, . 
		\end{align}
This leads to 
		\begin{align}
			 i \bs{\omega}_{ij}W_{\left[\widehat{\boldsymbol{z}}_j, \widehat{O}\right]} &= \frac{\partial W_{O}}{\partial \boldsymbol{z}_i} \label{eq:result3}\, , \\ 
			 \frac{1}{2} W_{\left\{\widehat{\boldsymbol{z}}_i, \widehat{O}\right\}} &= \boldsymbol{z}_i W_{O}\, ,\label{eq:result4}
		\end{align} 
where we have introduced the phase-space vector $\boldsymbol{z} = \left(v_\varphi,p_\varphi \right)^{\mathrm{T}}$. 

These relations can be used to compute the inverse Weyl transform of the TCL${}_2$ master equation~\eqref{eq:METCL2}. Using that $\bs{\omega}$ is antisymmetric, one finds
		\begin{align}\label{eq:Wignerredeomre}
	\frac{\dd  W_{\mathrm{red}}}{\dd \eta}=\left\{\widetilde{H}_0 + \widetilde{H}^{\mathrm{(LS)}} ,W_{\mathrm{red}}\right\} + \bs{\Delta}_{12} \ds\sum_{i}\frac{\partial}{\partial \boldsymbol{z}_i}\left(\boldsymbol{z}_iW_{\mathrm{red}}\right) -\frac{1}{2}\ds\sum_{i,j}\left[\bs{\omega} \boldsymbol{D} \bs{\omega}\right]_{ij} \frac{\partial^2 W_{\mathrm{red}}}{\partial \boldsymbol{z}_i\partial \boldsymbol{z}_j}\, ,	
	\end{align}
	where $W_{\mathrm{red}}=W_{\widehat{\rho}_{\mathrm{red}}}$ is the reduced Wigner function, \ie the inverse Wigner-Weyl transform of the reduced density matrix $\widehat{\rho}_{\mathrm{red}}$. The curly brackets now represent Poisson's brackets, not to be confused with the anticommutators for quantum operators. This coincides with \Eq{eq:Wignerredeom}.	
	
The first term in \Eq{eq:Wignerredeomre} corresponds to the free evolution dressed by the Lamb-shift Hamiltonian. This part of the equation only captures unitary/time-reversible evolution. The second term is a damping term reading as a total derivative and the last term is a diffusion term. These last two terms can be combined into a single second-order differential operator involving the dissipator matrix defined in \Eq{eq:dissipator}, and they induce a non-unitary evolution.
	
	Let us finally mention that the TCL${}_2$ transport equation can be simply obtained from \Eq{eq:Wignerredeomre} using the Gaussianity of the state. Indeed, the state being Gaussian, the reduced Wigner function is given by 
	\bea
\label{eq:Wred:Gaussian}
	W_{\mathrm{red}}=&\sqrt{\frac{1}{4\pi^2\det\bs{\Sigma}_{\mathrm{TCL}}}}\exp\left[-\frac{1}{2}\ds\sum_{i,j}\boldsymbol{z}_i\left(\bs{\Sigma}_{\mathrm{TCL}}\right)^{-1}_{ij}\boldsymbol{z}_j\right],
	\eea
	where $\bs{\Sigma}_{\mathrm{TCL}}$ is the covariance of the reduced system. Upon inserting \Eq{eq:Wred:Gaussian} into \Eq{eq:Wignerredeomre}, one obtains
		\begin{align}\label{eq:TCLeomre}
				\frac{\dd \bs{\Sigma}_{\mathrm{TCL}}}{\dd \eta}&=\boldsymbol{\omega}\left(\bs{H}^{(\varphi)}+\bs{\Delta}\right) \bs{\Sigma}_{\mathrm{TCL}}-\bs{\Sigma}_{\mathrm{TCL}} \left(\bs{H}^{(\varphi)}+\bs{\Delta}\right) \boldsymbol{\omega}- \bs{\omega D \omega} -2\bs{\Delta}_{12}\bs{\Sigma}_{\mathrm{TCL}} \, ,
			\end{align} 
which indeed coincides with \Eq{eq:TCLeom}.

		\section{Coefficients of the transport equation for TCL${}_2$}\label{app:TCL2coef}
	In this appendix, we work out the coefficients of the transport equation for TCL${}_2$, defined in  \Eqs{eq:Dphiphi}, \eqref{eq:Dphipi}, \eqref{eq:Fphiphi} and \eqref{eq:Fphipi}. They involve  the scale factor, which in a de-Sitter universe is given by $a=k/(Hz)$, as well as the mode functions
	\begin{align}
    \label{eq:mode:phi}	v_{\varphi}(\eta) =& \frac{1}{2}\sqrt{\frac{\pi z}{k}} \ee^{i\frac{\pi}{2}\left(\nu_\varphi+\frac{1}{2}\right)} H_{\nu_\varphi}^{(1)}(z) 
    \, ,\\ 
    p_{\varphi}(\eta) =&-\frac{1}{2}\sqrt{\frac{k \pi}{z}} \ee^{i\frac{\pi}{2}\left(\nu_\varphi+\frac{1}{2}\right)} \left[ \left(\nu_\varphi+\frac{3}{2}\right) H_{\nu_\varphi}^{(1)}(z)-zH_{\nu_\varphi+1}^{(1)}(z) \right]  \, , \\
	 \label{eq:mode:chi}
   v_{\chi}(\eta) =&\frac{1}{2}\sqrt{\frac{\pi z}{k}} \ee^{-\frac{\pi}{2}\mu_\chi+i\frac{\pi}{4}} H_{i\mu_\chi}^{(1)}(z) 
   \, ,\\ 
   p_{\chi}(\eta) =&-\frac{1}{2}\sqrt{\frac{k \pi}{z}}\ee^{-\frac{\pi}{2}\mu_\chi+i\frac{\pi}{4}}\left[ \left(i\mu_\chi+\frac{3}{2}\right) H_{i\mu_\chi}^{(1)}(z)-zH_{i\mu_\chi+1}^{(1)}(z) \right]  \, .
	\end{align}
Here, we recall that $z = - k \eta$, $H^{(1)}_{\nu}$ is the Hankel function of the first kind and of order $\nu$ and 
\bea
\label{eq:nuvarphi:muchi}
 \nu_\varphi= \frac{3}{2}\sqrt{1 - \left(\frac{2m}{3H}\right)^2}\quad\text{and}\quad \mu_\chi= \frac{3}{2}\sqrt{ \left(\frac{2M}{3H}\right)^2-1}\, .
\eea
\subsection{Exact results}
\label{sec:TCLcoeff:exact}
In order to perform the integrals involved in \Eqs{eq:Dphiphi}-\eqref{eq:Fphipi}, we will make use of the formula
	\bea
	\label{eq:Int:Bessel}
	\int \mathcal{C}_{\nu_1}(Az) \mathcal{D}_{\nu_2}(Az) \frac{\dd z}{z} = \frac{\mathcal{C}_{{\nu_1}}(Az)\mathcal{D}_{\nu_2}(Az)}{{\nu_1}+\nu_2} + \frac{Az}{{\nu_1}^2-\nu_2^2} \left[\mathcal{C}_{\nu_1}(Az)\mathcal{D}_{\nu_2+1}(Az) - \mathcal{C}_{{\nu_1}+1}(Az)\mathcal{D}_{\nu_2}(Az)\right],
	\eea
	see Eq.~(10.22.6) of \Refa{NIST:DLMF},
	where $\mathcal{C}_{\nu_1}$ and $\mathcal{D}_{\nu_2}$ are any of the Bessel functions, and $A$ is a fixed arbitrary parameter. Anticipating the computation, let us finally define
	\begin{align}
		\label{eq:Ffct}		F_{\nu,\mu}(z) &\equiv \frac{H^{(2)}_{\nu}(z) H^{(1)}_{i\mu}(z)}{\nu+i\mu} + \frac{z}{\nu^2+\mu^2} \left[H^{(2)}_{\nu}(z)H^{(1)}_{i\mu+1}(z) - H^{(2)}_{\nu+1}(z)H^{(1)}_{i\mu}(z)\right] ,\\
		\label{eq:Gfct}		G_{\nu,\mu}(z) &\equiv \frac{H^{(2)}_{\nu}(z) H^{(2)}_{-i\mu}(z)}{\nu-i\mu} + \frac{z}{\nu^2+\mu^2} \left[H^{(2)}_{\nu}(z)H^{(2)}_{-i\mu+1}(z) - H^{(2)}_{\nu+1}(z)H^{(2)}_{-i\mu}(z)\right],
	\end{align}
in terms of which it will be convenient to express our results.	
	\subsubsection*{$\bs{D}_{11}$ coefficient}
	We start with {$\bs{D}_{11}$ defined in \Eq{eq:Dphiphi}, namely
	\bea
	\bs{D}_{11}(\eta) &= -4 \lambda^4 a^2(\eta)\int_{\eta_0}^{\eta} \dd \eta' a^2(\eta') \Ima\left[p_{\varphi} (\eta) v^{*}_{\varphi} (\eta')\right] \Rea\left[v_{\chi} (\eta) v^{*}_{\chi} (\eta')\right] .
	\eea
	Expanding the real and the imaginary parts, and replacing $a = k / (Hz)$, it is given by
	\bea
	\bs{D}_{11}(z) = i \frac{k^3}{z^2}\frac{\lambda^4}{H^4} \int_z^{z_0} \frac{\dd z'}{(z')^2} \left[p_{\varphi}(z) v_{\varphi}^*(z') - p_{\varphi}^*(z) v_{\varphi}(z')\right] \left[v_{\chi}(z) v_{\chi}^*(z')+v_{\chi}^*(z) v_{\chi}(z')\right]\, .
	\eea
We thus have four terms,
	\bea
	\bs{D}_{11}(z) = &i
	\frac{k^3}{z^2}\frac{\lambda^4}{H^4} p_{\varphi}(z) v_{\chi}(z) \int_z^{z_0} \frac{\dd z'}{(z')^2} v_{\varphi}^*(z')v_{\chi}^*(z')\\
	&+i\frac{k^3}{z^2}\frac{\lambda^4}{H^4} p_{\varphi}(z) v_{\chi}^*(z) \int_z^{z_0} \frac{\dd z'}{(z')^2} v_{\varphi}^*(z')v_{\chi}(z')\\
	&-i\frac{k^3}{z^2}\frac{\lambda^4}{H^4} p_{\varphi}^*(z) v_{\chi}(z) \int_z^{z_0} \frac{\dd z'}{(z')^2} v_{\varphi}(z')v_{\chi}^*(z')\\
	&-i\frac{k^3}{z^2}\frac{\lambda^4}{H^4} p_{\varphi}^*(z) v_{\chi}^*(z) \int_z^{z_0} \frac{\dd z'}{(z')^2} v_{\varphi}(z')v_{\chi}(z')\, ,
	\eea
	which can be re-organised as
	\bea\label{eq:D11analytic}
	\bs{D}_{11}(z) = & 2 \frac{k^3}{z^2}\frac{\lambda^4}{H^4} \Ima\left[p_{\varphi}(z) v_{\chi}(z) \int^z_{z_0} \frac{\dd z'}{(z')^2} v_{\varphi}^*(z')v_{\chi}^*(z')+p_{\varphi}(z) v_{\chi}^*(z) \int^z_{z_0} \frac{\dd z'}{(z')^2} v_{\varphi}^*(z')v_{\chi}(z')\right]\, .
	\eea
	Therefore, we have two integrals to compute. Making use of \Eq{eq:Int:Bessel}, they are given by
	\bea
	\int^z_{z_0} \frac{\dd z'}{(z')^2} v_{\varphi}^*(z')v_{\chi}^*(z')=& - i \frac{\pi}{4k}\ee^{-\frac{\pi}{2}\mu_\chi-i\frac{\pi}{2}\nu_\varphi} \left[G_{\nu_\varphi,\mu_\chi}(z)-G_{\nu_\varphi,\mu_\chi}(z_0)\right] ,\\
	\int^z_{z_0} \frac{\dd z'}{(z')^2} v_{\varphi}^*(z')v_{\chi}(z')=& \frac{\pi}{4k}\ee^{-\frac{\pi}{2}\mu_\chi-i\frac{\pi}{2}\nu_\varphi} \left[F_{\nu_\varphi,\mu_\chi}(z)-F_{\nu_\varphi,\mu_\chi}(z_0)\right].
	\eea
	We conclude that $\bs{D}_{11}$ can be written as 
	\bea
	\label{eq:FD11:def}
    \boldmathsymbol{D}_{11}(z) = F_{\boldmathsymbol{D}_{11}}\left(z,z\right)-F_{\boldmathsymbol{D}_{11}}\left(z,z_0\right),
    \eea 
    where 
	\begin{align}\label{eq:FD11}
	    F_{\boldmathsymbol{D}_{11}}\left(z_1,z_2\right) =& \frac{\pi}{2} \frac{k^2}{z^2}\frac{\lambda^4}{H^4} \ee^{-\frac{\pi}{2}\mu_\chi}\Ima\Bigg[- i p_{\varphi}(z_1) v_{\chi}(z_1)   G_{\nu_\varphi,\mu_\chi}(z_2) \ee^{-i\frac{\pi}{2}\nu_\varphi} \nonumber \\
	    &+ p_{\varphi}(z_1) v_{\chi}^*(z_1) F_{\nu_\varphi,\mu_\chi}(z_2) \ee^{-i\frac{\pi}{2}\nu_\varphi} \Bigg].
	\end{align}
It is worth noting that in the case where $z_1=z_2$, this function can be further simplified by making repeated use of the Wronskian identity (see Eq.~(10.5.5) of \Refa{NIST:DLMF}), namely
\bea
\label{eq:Wronskian}
H_{\nu+1}^{(1)}(z)H_\nu^{(2)}(z)- H_{\nu}^{(1)}(z)H_{\nu+1}^{(2)}(z) = -\frac{4i}{\pi z}\, .
\eea	
Recalling that $[H_{\nu}^{(1)}(z)]^*=H_{\nu^*}^{(2)}(z)$ and $[H_{\nu}^{(2)}(z)]^*=H_{\nu^*}^{(1)}(z)$, after a tedious but straightforward calculation it leads to 
\bea
 F_{\boldmathsymbol{D}_{11}}\left(z,z\right) = -\frac{2}{\nu_\varphi^2+\mu_\chi^2} \left(\frac{k}{z}\right)\left(\frac{\lambda}{H}\right)^4\Rea\left[v_\chi(z)p_\chi^*(z)\right]\, .
\eea
Given that \Eq{eq:nuvarphi:muchi} leads to $\nu_\varphi^2+\mu_\chi^2=(M^2-m^2)/H^2$, and since $\boldmathsymbol{\Sigma}^{\mathrm{free}}_{\chi\chi,12}(z) =\Rea [ v_\chi(z)p^*_\chi(z)] $, this can be rewritten as 
\bea
F_{\boldmathsymbol{D}_{11}}\left(z,z\right) = -2 \frac{\lambda^4a^2}{M^2-m^2}\boldmathsymbol{\Sigma}^{\mathrm{free}}_{\chi\chi,12}(z)\, ,
\eea
where we have also used that $a=-k/(Hz)$ in a de-Sitter universe. This corresponds to $\boldmathsymbol{D}_{11}$ in the exact theory when evaluated at leading order in the interaction strength, see \Eq{eq:D:ex:comp}. In other words, we have shown that
\bea
F_{\boldmathsymbol{D}_{11}}\left(z,z\right) = \boldmathsymbol{D}_{\mathrm{SPT},11}(z)\, .
\eea
	\subsubsection*{$\bs{D}_{12}$ coefficient}
	The other coefficients can be computed similarly. For $\bs{D}_{12}$ defined in \Eqs{eq:Dphipi}, one has
	\bea
		\bs{D}_{12}(\eta) =& 2 \lambda^4 a^2(\eta)\int_{\eta_0}^{\eta} \dd \eta' a^2(\eta') \Ima\left[v_{\varphi} (\eta) v^{*}_{\varphi} (\eta')\right] \Rea\left[v_{\chi} (\eta) v^{*}_{\chi} (\eta')\right]\\
	 =& \frac{i}{2}\frac{k^3}{z^2}\frac{\lambda^4}{H^4} \int_{z_0}^z \frac{\dd z'}{(z')^2} \left[v_{\varphi}(z) v_{\varphi}^*(z')-v_{\varphi}^*(z) v_{\varphi}(z')\right] \left[v_{\chi}(z) v_{\chi}^*(z')+v_{\chi}^*(z) v_{\chi}(z') \right] .
	\eea
This leads to
	\bea\label{eq:D12exp}
    \boldmathsymbol{D}_{12}(z) = F_{\boldmathsymbol{D}_{12}}\left(z,z\right)-F_{\boldmathsymbol{D}_{12}}\left(z,z_0\right),
    \eea 
    where 
    \begin{align}\label{eq:FD12}
	    F_{\boldmathsymbol{D}_{12}}\left(z_1,z_2\right) &= - \frac{\pi}{4} \frac{k^2}{z^2}\frac{\lambda^4}{H^4} \ee^{-\frac{\pi}{2}\mu_\chi}\Ima\Bigg[- i v_{\varphi}(z_1) v_{\chi}(z_1)   G_{\nu_\varphi,\mu_\chi}(z_2) \ee^{-i\frac{\pi}{2}\nu_\varphi} \nonumber \\
	    +& v_{\varphi}(z_1) v_{\chi}^*(z_1) F_{\nu_\varphi,\mu_\chi}(z_2) \ee^{-i\frac{\pi}{2}\nu_\varphi} \Bigg].
	\end{align}
As for $F_{\boldmathsymbol{D}_{11}}$, this expression can be simplified in the coincident configuration $z_1=z_2$ by repeatedly using the Wronskian identity~\eqref{eq:Wronskian}, and one finds
\bea
F_{\boldmathsymbol{D}_{12}}\left(z,z\right) &= \left(\frac{k}{z}\right)^2\left(\frac{\lambda}{H}\right)^4\frac{\left\vert v_\chi(z)\right\vert^2}{\mu_\chi^2+\nu_\varphi^2}\, .
\eea
Using again that $\nu_\varphi^2+\mu_\chi^2=(M^2-m^2)/H^2$, this can be written as 
\bea
F_{\boldmathsymbol{D}_{12}}\left(z,z\right) &= \frac{\lambda^4a^2}{M^2-m^2}\boldmathsymbol{\Sigma}^{\mathrm{free}}_{\chi\chi,11}(z)=\boldmathsymbol{D}_{\mathrm{SPT},12}(z)\, ,
\eea 
where we recognise the leading-order contribution in $\boldmathsymbol{D}_{\mathrm{ex},12}$, see \Eq{eq:D:ex:comp}.
	\subsubsection*{$\bs{\Delta}_{11}$ coefficient}
	For $\bs{\Delta}_{11}$ defined in \Eqs{eq:Fphiphi}, one has
	\bea
	\bs{\Delta}_{11}(\eta) =& - 4 \lambda^4 a^2(\eta)\int_{\eta_0}^{\eta} \dd \eta' a^2(\eta') \Ima\left[p_{\varphi} (\eta) v^{*}_{\varphi} (\eta')\right] \Ima\left[v_{\chi} (\eta) v^{*}_{\chi} (\eta')\right]\\
	=& -\frac{k^3}{z^2}\frac{\lambda^4}{H^4} \int_{z_0}^z \frac{\dd z'}{(z')^2}\left[p_{\varphi}(z) v_{\varphi}^*(z')-p_{\varphi}^*(z) v_{\varphi}(z')\right] \left[v_{\chi}(z) v_{\chi}^*(z')-v_{\chi}^*(z) v_{\chi}(z')\right] .
	\eea
This leads to
	\bea
    \boldmathsymbol{\Delta}_{11}(z) = F_{\boldmathsymbol{\Delta}_{11}}\left(z,z\right)-F_{\boldmathsymbol{\Delta}_{11}}\left(z,z_0\right),
    \eea 
    where 
    \begin{align}\label{eq:FDelta11}
	    F_{\boldmathsymbol{\Delta}_{11}}\left(z_1,z_2\right) &= - \frac{\pi}{2} \frac{k^2}{z^2}\frac{\lambda^4}{H^4} \ee^{-\frac{\pi}{2}\mu_\chi}\Rea\Bigg[- i p_{\varphi}(z_1) v_{\chi}(z_1)   G_{\nu_\varphi,\mu_\chi}(z_2) \ee^{-i\frac{\pi}{2}\nu_\varphi} \nonumber \\
	    -& p_{\varphi}(z_1) v_{\chi}^*(z_1) F_{\nu_\varphi,\mu_\chi}(z_2) \ee^{-i\frac{\pi}{2}\nu_\varphi} \Bigg].
	\end{align}
This expression can be simplified when $z_1=z_2$ using the Wronskian identity~\eqref{eq:Wronskian}, if one further uses two additional properties of the Hankel functions. The first one is the recurrence relation (see Eq.~(10.6.1) of \Refa{NIST:DLMF})
\bea
H_{\nu-1}^{(2)}(z)+H_{\nu+1}^{(2)}(z)=\frac{2\nu}{z}H_{\nu}^{(2)}\, ,
\eea
and the second one is the inversion formula
\bea
H_{-\nu}^{(1)}(z)=\ee^{i\pi\nu} H_\nu^{(1)}(z)\, ,\quad\quad H_{-\nu}^{(2)}(z)=\ee^{-i\pi\nu} H_\nu^{(2)}(z)\, .
\eea
After a tedious but straightforward calculation, this leads to
\bea
F_{\boldmathsymbol{\Delta}_{11}}\left(z,z\right) = - \left(\frac{k}{z}\right)^2 \left(\frac{\lambda}{H}\right)^4\frac{1}{\nu_\varphi^2+\mu_\chi^2}=\boldmathsymbol{\Delta}_{\mathrm{ex},11}\, ,
\eea
where we have recognised $\boldmathsymbol{\Delta}_{\mathrm{ex},11}$, see \Eq{eq:Delta:ex:comp}, using again that $\nu_\varphi^2+\mu_\chi^2=(M^2-m^2)/H^2$. Note that, here, the agreement between $F_{\boldmathsymbol{\Delta}_{11}}(z,z) $ and $\boldmathsymbol{\Delta}_{\mathrm{ex},11}$ is valid at all orders, given that $\boldmathsymbol{\Delta}_{\mathrm{ex},11}$ only contains terms of order $\lambda^4$.
	\subsubsection*{$\bs{\Delta}_{12}$ coefficient}
	Finally, for $\bs{\Delta}_{12}$ defined in \Eqs{eq:Fphipi}, one has
	\bea
	\bs{\Delta}_{12}(\eta) =&2 \lambda^4 a^2(\eta)\int_{\eta_0}^{\eta} \dd \eta' a^2(\eta') \Ima\left[v_{\varphi} (\eta) v^{*}_{\varphi} (\eta')\right] \Ima\left[v_{\chi} (\eta) v^{*}_{\chi} (\eta')\right]\\
		 =& \frac{1}{2}\frac{k^3}{z^2}\frac{\lambda^4}{H^4}  \int_{z_0}^z \frac{\dd z'}{(z')^2}\left[v_{\varphi}(z) v_{\varphi}^*(z')-v_{\varphi}^*(z) v_{\varphi}(z')\right] \left[v_{\chi}(z) v_{\chi}^*(z')-v_{\chi}^*(z) v_{\chi}(z')\right] ,
	\eea
and this leads to
	\bea\label{eq:Delta12exp}
    \bs{\Delta}_{12}(z) = F_{\bs{\Delta}_{12}}\left(z,z\right)-F_{\bs{\Delta}_{12}}\left(z,z_0\right),
    \eea 
    where 
    \begin{align}\label{eq:FDelta12}
	    F_{\bs{\Delta}_{12}}\left(z_1,z_2\right) &= \frac{\pi}{4} \frac{k^2}{z^2}\frac{\lambda^4}{H^4} \ee^{-\frac{\pi}{2}\mu_\chi}\Rea\Bigg[- i v_{\varphi}(z_1) v_{\chi}(z_1)   G_{\nu_\varphi,\mu_\chi}(z_2) \ee^{-i\frac{\pi}{2}\nu_\varphi} \nonumber \\
	    -& v_{\varphi}(z_1) v_{\chi}^*(z_1) F_{\nu_\varphi,\mu_\chi}(z_2) \ee^{-i\frac{\pi}{2}\nu_\varphi} \Bigg].
	\end{align}
This expression can be simplified when $z_1=z_2$ using the Wronskian identity~\eqref{eq:Wronskian}, and we find that it vanishes,
\bea
\label{eq:Delta12:non:spurious}
F_{\bs{\Delta}_{12}}\left(z,z\right)=0\, .
\eea
In particular, it implies that $F_{\bs{\Delta}_{12}}(z,z)=\bs{\Delta}_{\mathrm{ex},12}(z)$ and that, as for $F_{\bs{\Delta}_{11}}(z,z)$, this is valid at all orders in $\lambda^4$ given that both quantities identically vanish. 
	\subsection{Sub-Hubble limit}
In order to gain analytic insight, let us expand the coefficients derived above in the sub-Hubble ($z\gg 1$) and super-Hubble ($z\ll 1$) limits. In the sub-Hubble limit, one can use the asymptotic expansion
\begin{align}
        H^{(1)}_{\nu}(z)	=& \sqrt{\frac{2}{\pi z}}    \ee^{-iz-i\frac{\pi}{2}\nu-i\frac{\pi}{4}} \sum_{k=0}^{\infty} a_k(\nu) \left(\frac{i}{z}\right)^k\, ,\\
        H^{(2)}_{\nu}(z)	=& \sqrt{\frac{2}{\pi z}}    \ee^{iz+i\frac{\pi}{2}\nu+i\frac{\pi}{4}} \sum_{k=0}^{\infty} a_k(\nu) \left(\frac{-i}{z}\right)^k\, ,
\end{align}
see Eq.~(10.17.5) of \Refa{NIST:DLMF}, with
	\begin{align}
	    a_k(\nu) = \frac{\left( \frac{1}{2} - \nu \right)_k \left( \frac{1}{2} + \nu \right)_k}{(-2)^k k!}
	\end{align}
	where the parenthesis with lower index indicate the Pochhammer's symbol, \ie $(x)_k=\Gamma(x+k)/\Gamma(x)$. Inserting these formulas into \Eqs{eq:Ffct} and \eqref{eq:Gfct}, one obtains
	\begin{align}	\label{eq:expFsubH}
	F_{\nu,\mu}(z) =&  - \frac{\ee^{\frac{\pi}{2}(\mu + i \nu)}}{\pi} \left(\frac{4i}{\nu^2+\mu^2} + \frac{2}{z} - i \frac{\nu^2 + \mu^2}{2z^2} \right)+ \mathcal{O}(z^{-3})\, ,\\
	\label{eq:expGsubH}
	G_{\nu,\mu}(z) = & - \frac{\ee^{\frac{\pi}{2}(\mu + i \nu)}}{\pi} \frac{\ee^{-2iz}}{z^2} + \mathcal{O}(z^{-3})\, .
	\end{align}
	Note that $F_{\nu,\mu}(z)$ is non vanishing in the sub-Hubble regime. Let us now expand \Eqs{eq:FD11}, \eqref{eq:FD12}, \eqref{eq:FDelta11} and \eqref{eq:FDelta12} in the limit $z_1,z_2\gg 1$. At leading order, one obtains
	\begin{align}
	    F_{\boldmathsymbol{D}_{11}}\left(z_1,z_2\right)\simeq &\frac{k^2 \lambda^4}{2H^4 z_1^3}\left(\frac{2}{\nu_\varphi^2 + \mu_\chi^2}-1+\frac{z_1}{z_2}\right) \, ,\\
	    F_{\boldmathsymbol{D}_{12}}\left(z_1,z_2\right)\simeq& \frac{k \lambda^4}{2H^4\left(\nu_\varphi^2 + \mu_\chi^2\right) z_1^2}\, ,\\
  	   F_{\boldmathsymbol{\Delta}_{11}}\left(z_1,z_2\right)\simeq & -\frac{k^2 \lambda^4}{H^4\left(\nu_\varphi^2 + \mu_\chi^2\right) z_1^2}\, ,\\
	   F_{\boldmathsymbol{\Delta}_{12}}\left(z_1,z_2\right)\simeq & \frac{k \lambda^4(z_1-z_2)}{4H^4z_1^3 z_2}\, .
	\end{align}
	\subsection{Super-Hubble limit}
	\label{sec:TCLcoeff:superH}
	To organise the super-Hubble expansion, we introduce the quantities
	\begin{align}
		\label{eq:app:SHcoef1}		\alpha_{\nu}(z) &\equiv \frac{1+ i \cot \pi \nu}{\Gamma(1+\nu)}\left(\frac{z}{2}\right)^{\nu - \frac{3}{2}} ,\quad\quad
				\beta_{\nu}(z) \equiv \frac{-i}{\sin \pi\nu} \frac{1}{\Gamma(1-\nu)}\left(\frac{z}{2}\right)^{ \frac{3}{2}-\nu },\\
		\label{eq:app:SHcoef2}		\gamma_{\mu}(z) &\equiv \frac{1+  \coth \pi \mu}{\Gamma(1+i\mu)}\left(\frac{z}{2}\right)^{i\mu},\quad\quad~~\,
				\delta_{\mu}(z) \equiv  \frac{-1}{\sinh \pi\mu} \frac{1}{\Gamma(1-i\mu)}\left(\frac{z}{2}\right)^{ -i\mu },
			\end{align}	
 together with the function
			\begin{align}
		\label{eq:SHf}		f_x(z) \equiv& \sum_{k=0}^\infty \frac{(-1)^k \left(\frac{z}{2}\right)^{2k}}{k! \left(x+1\right)_k}\\
		= &1 - \frac{\left(\frac{z}{2}\right)^2}{x+1} +\frac{\left(\frac{z}{2}\right)^4}{2(x+1)(x+2)} + \mathcal{O}(z^6) 
			\end{align}
			such that
			\begin{align}
		\label{eq:SHHnu}		H^{(1)}_{\nu}(z) &= \alpha_{\nu}(z) f_{\nu}(z)\left(\frac{z}{2}\right)^{\frac{3}{2}} + \beta_{\nu}(z) f_{-\nu}(z)\left(\frac{z}{2}\right)^{-\frac{3}{2}}\, , \\
		\label{eq:SHHmu}		H^{(1)}_{i \mu}(z) &= \gamma_{\mu}(z) f_{i\mu}(z) + \delta_{\mu}(z) f_{-i\mu}(z)\, ,
			\end{align}
	and
	\begin{align}
\label{eq:SHHnu+1}		H^{(1)}_{\nu+1}(z) &= \frac{\alpha_{\nu}(z)}{\nu+1} f_{\nu+1}(z)\left(\frac{z}{2}\right)^{\frac{5}{2}} + \nu \beta_{\nu}(z) f_{-\nu-1}(z)\left(\frac{z}{2}\right)^{-\frac{5}{2}} \, ,\\
\label{eq:SHHmu+1}		H^{(1)}_{i \mu+1}(z) &= \frac{\gamma_{\mu}(z)}{i \mu + 1} f_{i\mu+1}(z)\frac{z}{2} + i \mu \delta_{\mu}(z) f_{-i\mu-1}(z)\left(\frac{z}{2}\right)^{-1}\, ,
	\end{align}	
	see Eqs.~(10.2.2), (10.4.7) and (10.4.8) of \Refa{NIST:DLMF}. This allows one to expand \Eqs{eq:Ffct} and \eqref{eq:Gfct}, and one finds
	\begin{align}	\label{eq:expF}
	F_{\nu,\mu}(z)=&- 2\sqrt{2}z^{-3/2} \beta^{*}_{\nu}\left[\frac{(\nu+i\mu) \gamma_\mu +(\nu-i\mu)\delta_\mu}{\nu^2+\mu^2} \right] + \frac{z^{1/2}}{\sqrt{2}}\beta^{*}_{\nu}\left[\frac{(1-i\mu) \gamma_\mu +(1+i\mu)\delta_\mu}{\left(1+\mu^2\right)\left(\nu-1\right)} \right] \nonumber\\ & +\frac{z^{3/2}}{2\sqrt{2}}\alpha^{*}_{\nu}\left[\frac{(\nu-i\mu) \gamma_\mu +(\nu+i\mu)\delta_\mu}{\nu^2+\mu^2} \right]+\mathcal{O}(z^{5/2})\, .
	\end{align}
	and
	\begin{align}	\label{eq:expG}
	G_{\nu,\mu}(z)=& - 2\sqrt{2}z^{-3/2} \beta^{*}_{\nu} \left[\frac{(\nu-i\mu) \gamma^{*}_\mu +(\nu+i\mu)\delta^{*}_\mu}{\nu^2+\mu^2} \right] +  \frac{z^{1/2}}{\sqrt{2}}\beta^{*}_{\nu}\left[\frac{(1+i\mu) \gamma^{*}_\mu +(1-i\mu)\delta^{*}_\mu}{\left(1+\mu^2\right)\left(\nu-1\right)} \right] \nonumber\\ & +\frac{z^{3/2}}{2\sqrt{2}}\alpha^{*}_{\nu}\left[\frac{(\nu+i\mu) \gamma^{*}_\mu +(\nu-i\mu)\delta^{*}_\mu}{\nu^2+\mu^2} \right]+\mathcal{O}(z^{5/2})\, .
	\end{align}
Hereafter, to lighten the notation, we have dropped the explicit $z$-dependence of $\alpha_\nu$, $\beta_\nu$, $\gamma_\mu$ and $\delta_\mu$. This is because, since $\nu_\varphi$ is close to $3/2$ in practice, see \Eq{eq:nuvarphi:muchi}, this does not affect the power counting in $z$, see \Eqs{eq:app:SHcoef1}-\eqref{eq:app:SHcoef2}.

	It is worth noting that the terms of orders $z^{-3/2}$ and $z^{1/2}$ cancel out in $F^{*}_{\nu,\mu}(z)+ G_{\nu,\mu}(z)$ since $\beta_\nu$ is pure imaginary, see \Eq{eq:app:SHcoef2}. One indeed has
	\bea
	F^{*}_{\nu,\mu}(z)+ G_{\nu,\mu}(z) =\frac{1}{\nu^2+\mu^2}\frac{z^{3/2}}{\sqrt{2}}\Rea\left( \alpha_{\nu}\right)\left[(\nu+i\mu) \gamma^{*}_\mu +(\nu-i\mu)\delta^{*}_\mu \right]+\mathcal{O}(z^{5/2}).
	\eea
	Let us now expand the coefficients of the transport equation in the super-Hubble limit, \ie when $z\ll 1$ (but keeping $z_0$ arbitrary).
	\subsubsection*{$\bs{D}_{11}$ coefficient}
	For $\bs{D}_{11}$, one finds
	\bea
	\bs{D}_{11} =  & z^{-7/2}  S_{\bs{D}_{11}}^{(-7/2)} (z,z_0)
	+ \frac{\pi}{2} \frac{\ee^{-\pi \mu_{\chi}} }{\nu_{\varphi}^2 + \mu_{\chi}^2} \frac{\lambda^4}{H^4}  \left[\frac{3}{2} \left|\gamma_{\mu_{\chi}}+\delta_{\mu_{\chi}}\right|^2+2\mu_{\chi} \Ima \left(\gamma^*_{\mu_{\chi}} \delta_{\mu_{\chi}}\right) \right]\frac{k^2}{z^2} \\ &
	+z^{-3/2} S_{\bs{D}_{11}}^{(-3/2)}  (z,z_0)\, ,
	\eea
where
	\begin{align}
		\label{eq:spurD11}	S_{\bs{D}_{11}}^{(-7/2)}(z,z_0) = -&
		\frac{\pi^2k^2}{2\sqrt{2}} \left(\nu_\varphi - \frac{3}{2}\right) \frac{\lambda^4}{H^4} \Ima\left\lbrace \beta_{\nu_\varphi} \left[F_{\nu_\varphi,\mu_\chi}^*\left(z_0\right)+G_{\nu_\varphi,\mu_\chi}\left(z_0\right)\right] \left(\gamma_{\mu_{\chi}}+\delta_{\mu_{\chi}}\right) \right\rbrace \ee^{-\pi \mu_{\chi}} 
	\end{align}
	and 	\begin{align}\label{eq:ho1}
	   S_{\bs{D}_{11}}^{(-3/2)} (z,z_0)&= - \frac{\pi^2}{16\sqrt{2}}\frac{1}{(1+\mu_\chi^2)(\nu_\varphi-1)}\frac{\lambda^4}{H^4} k^2 \Ima\Bigg[ \beta_{\nu_\varphi} \left[F_{\nu_\varphi,\mu_\chi}^*\left(z_0\right)+G_{\nu_\varphi,\mu_\chi}\left(z_0\right)\right]\nonumber \\ 
	    \bigg(&\gamma_{\mu_{\chi}}(\mu_\chi+i)\left\{\mu_\chi(-7+2\nu_\varphi)+i\left[10 + \nu_\varphi (-7+2\nu_\varphi)\right]\right\}  \\ 
	    +&\delta_{\mu_{\chi}}(\mu_\chi-i)\left\{\mu_\chi(-7+2\nu_\varphi)-i\left[10 + \nu_\varphi (-7+2\nu_\varphi)\right]\right\}\bigg) \Bigg] \ee^{-\pi \mu_{\chi}} \nonumber \\ &+\frac{\pi^2}{16\sqrt{2}}\left(\frac{3}{2}+\nu_\varphi \right)\frac{\lambda^4}{H^4} \frac{k^2}{z^{1/2}}  
	    \Ima \bigg\{ \left[- \alpha^{*}_{\nu_\varphi} F_{\nu_\varphi,\mu_\chi}^*\left(z_0\right)+\alpha_{\nu_\varphi} G_{\nu_\varphi,\mu_\chi}\left(z_0\right)\right] \nonumber \\  &\left(\gamma_{\mu_{\chi}}+\delta_{\mu_{\chi}}\right) \bigg\}\ee^{-\pi \mu_{\chi}}
	    +\mathcal{O}(z^{1/2}) \nonumber
	\end{align}
	are spurious contributions, \ie they arise from the term $F_{\bs{D}_{11}}(z,z_0)$ in \Eq{eq:FD11:def}.
	\subsubsection*{$\bs{D}_{12}$ coefficient}
	One finds
	\bea
	\bs{D}_{12} = &z^{-5/2} S_{\bs{D}_{12}}^{(-5/2)}(z,z_0)
	 +
	  \frac{\pi}{4} \frac{\left|\gamma_{\mu_{\chi}}+\delta_{\mu_{\chi}}\right|^2}{\nu_{\varphi}^2 + \mu_{\chi}^2} \frac{\lambda^4}{H^4} \ee^{-\pi \mu_{\chi}}  \frac{k}{z}  + z^{-1/2} S_{\bs{D}_{12}}^{(-1/2)}(z,z_0)\, ,
	\eea
where
	\begin{align}
		\label{eq:spurD12}	S_{\bs{D}_{12}}^{(-5/2)}(z,z_0) = &
		\frac{\pi^2}{4\sqrt{2}} \frac{\lambda^4}{H^4} k  \Ima\left\lbrace \beta_{\nu_\varphi} \left[F_{\nu_\varphi,\mu_\chi}^*\left(z_0\right)+G_{\nu_\varphi,\mu_\chi}\left(z_0\right)\right] \left(\gamma_{\mu_{\chi}}+\delta_{\mu_{\chi}}\right) \right\rbrace \ee^{-\pi \mu_{\chi}} 
	\end{align}
	and 
	\bea\label{eq:ho2}
	   S_{\bs{D}_{12}}^{(-1/2)}(z,z_0)=&  \frac{\pi^2 k}{16\sqrt{2}}\frac{\ee^{-\pi \mu_{\chi}}}{(1+\mu_\chi^2)(\nu_\varphi-1)}\frac{\lambda^4}{H^4} \Ima\Bigg\lbrace \beta_{\nu_\varphi} \left[F_{\nu_\varphi,\mu_\chi}^*\left(z_0\right)+G_{\nu_\varphi,\mu_\chi}\left(z_0\right)\right] \\ 
	    &\left[(1+\mu_\chi^2)\left(\gamma_{\mu_{\chi}}+\delta_{\mu_{\chi}}\right) + (1-i\mu_\chi)(1-\nu_\varphi)\gamma_{\mu_{\chi}}  + (1+i\mu_\chi)(1-\nu_\varphi)\delta_{\mu_{\chi}}\right]\Bigg\rbrace 
	\eea
	are again spurious contributions.
	\subsubsection*{$\bs{\Delta}_{11}$ coefficient}
	For $\bs{\Delta}_{11}$, we have
	\bea
	\label{eq:app:Delta11:superH}
	\bs{\Delta}_{11} = & z^{-7/2} S_{\bs{\Delta}_{11}}^{(-7/2)}(z,z_0)  
	-	 \frac{\pi}{2} \frac{\mu_{\chi}\ee^{-\pi \mu_{\chi}}}{\nu_{\varphi}^2 + \mu_{\chi}^2} \frac{\lambda^4}{H^4}   \left(\left|\gamma_{\mu_{\chi}}\right|^2 - \left| \delta_{\mu_{\chi}}\right|^2\right)  \frac{k^2}{z^2}+ z^{-3/2} S_{\bs{\Delta}_{11}}^{(-3/2)}(z,z_0)\, ,
	\eea
where	
	\begin{align}
		\label{eq:spurDelta11}	S_{\bs{\Delta}_{11}}^{(-7/2)}(z,z_0) = &
		\frac{\pi^2 k^2}{2\sqrt{2}} \left(\nu_\varphi - \frac{3}{2}\right) \frac{\lambda^4}{H^4}   \Rea\left\lbrace \beta_{\nu_\varphi} \left[F_{\nu_\varphi,\mu_\chi}^*\left(z_0\right)+G_{\nu_\varphi,\mu_\chi}\left(z_0\right)\right] \left(\gamma_{\mu_{\chi}}+\delta_{\mu_{\chi}}\right) \right\rbrace \ee^{-\pi \mu_{\chi}} 
	\end{align}
	and 
		\bea\label{eq:ho3}
	    S_{\bs{\Delta}_{11}}^{(-3/2)}(z,z_0)  =& \frac{\pi^2 k^2}{16\sqrt{2}}\frac{1}{(1+\mu_\chi^2)(\nu_\varphi-1)}\frac{\lambda^4}{H^4}  \Rea\Bigg[ \beta_{\nu_\varphi} \left[F_{\nu_\varphi,\mu_\chi}^*\left(z_0\right)+G_{\nu_\varphi,\mu_\chi}\left(z_0\right)\right] \\ 
	    \bigg(&\gamma_{\mu_{\chi}}(\mu_\chi+i)\left\{\mu_\chi(-7+2\nu_\varphi)+i\left[10 + \nu_\varphi (-7+2\nu_\varphi)\right]\right\}  \\ 
	    +&\delta_{\mu_{\chi}}(\mu_\chi-i)\left\{\mu_\chi(-7+2\nu_\varphi)-i\left[10 + \nu_\varphi (-7+2\nu_\varphi)\right]\right\}\bigg) \Bigg] \ee^{-\pi \mu_{\chi}} \\ &-\frac{\pi^2}{16\sqrt{2}}\left(\frac{3}{2}+\nu_\varphi \right)\frac{\lambda^4}{H^4} \frac{k^2}{z^{1/2}}  
	    \Rea \bigg\{ \left[- \alpha^{*}_{\nu_\varphi} F_{\nu_\varphi,\mu_\chi}^*\left(z_0\right)+\alpha_{\nu_\varphi} G_{\nu_\varphi,\mu_\chi}\left(z_0\right)\right]  \\  &\left(\gamma_{\mu_{\chi}}+\delta_{\mu_{\chi}}\right) \bigg\}\ee^{-\pi \mu_{\chi}}
	\eea
	are spurious contributions. It is also worth noting that, in \Eq{eq:app:Delta11:superH}, one can simplify 
	\bea
	\left|\gamma_{\mu_{\chi}}\right|^2 - \left| \delta_{\mu_{\chi}}\right|^2 = 2\frac{\ee^{\pi\mu}}{\pi\mu}\, .
	\eea
	\subsubsection*{$\bs{\Delta}_{12}$ coefficient}
	Finally, for $\bs{\Delta}_{12}$, one obtains
	\bea
	\bs{\Delta}_{12}(z) = & z^{-5/2} S_{\bs{\Delta}_{12}}^{(-5/2)}\left(z,z_0\right) + z^{-1/2} S_{\bs{\Delta}_{12}}^{(-1/2)}\left(z,z_0\right)
	\eea
	which only contains spurious terms as shown in \Eq{eq:Delta12:non:spurious}, given by
	\begin{align}
		\label{eq:spurDelta12}		S_{\bs{\Delta}_{12}}^{(-5/2)}\left(z,z_0\right)=  -
		\frac{\pi^2 k}{4\sqrt{2}} \frac{\lambda^4}{H^4}  \Rea\left\lbrace \beta_{\nu_\varphi} \left[F_{\nu_\varphi,\mu_\chi}^*\left(z_0\right)+G_{\nu_\varphi,\mu_\chi}\left(z_0\right)\right] \left(\gamma_{\mu_{\chi}}+\delta_{\mu_{\chi}}\right) \right\rbrace \ee^{-\pi \mu_{\chi}} 
	\end{align}
	and 
	\bea
	    S_{\bs{\Delta}_{12}}^{(-1/2)}\left(z,z_0\right) &= -\frac{\pi^2 k}{16\sqrt{2}}\frac{\ee^{-\pi \mu_{\chi}}}{(1+\mu_\chi^2)(\nu_\varphi-1)}\frac{\lambda^4}{H^4} \Rea\Big\lbrace \beta_{\nu_\varphi} \left[F_{\nu_\varphi,\mu_\chi}^*\left(z_0\right)+G_{\nu_\varphi,\mu_\chi}\left(z_0\right)\right] \\ 
	    &~~~~\left[(1+\mu_\chi^2)\left(\gamma_{\mu_{\chi}}+\delta_{\mu_{\chi}}\right) + (1-i\mu_\chi)(1-\nu_\varphi)\gamma_{\mu_{\chi}}  + (1+i\mu_\chi)(1-\nu_\varphi)\delta_{\mu_{\chi}}\right]\Big\rbrace  \, .
	\eea

	\section{Comparison between TCL and perturbation theory in the curved-space Caldeira-Leggett model}
	\label{sec:app:SPT:TCL}
In this appendix, we compare Standard Perturbation Theory (SPT) to the perturbative solutions of the TCL master equation, in the context of the curved-space Caldeira-Leggett model introduced in \Sec{sec:model}. This will allow us to exhibit a concrete manifestation of the generic statement proven in \Sec{sec:inin}, that TCL${}_n$ solved perturbatively at order $n$ coincides with SPT${}_n$.
	\subsection{Perturbation theory}
The two-field system detailed in \Sec{subsec:exact} being linear, the field operators admit a decomposition of the form
        		\begin{align}
        			\widehat{v}_{\varphi}(\eta) &= v_{\varphi\varphi}(\eta) \widehat{a}_{\varphi} +  v^{*}_{\varphi\varphi}(\eta) \widehat{a}^{\dag}_{\varphi} + v_{\varphi\chi}(\eta) \widehat{a}_{\chi} +  v^{*}_{\varphi\chi}(\eta) \widehat{a}^{\dag}_{\chi}\, , \\ 
        			\widehat{v}_{\chi}(\eta) &= v_{\chi\varphi}(\eta) \widehat{a}_{\varphi} +  v^{*}_{\chi\varphi}(\eta) \widehat{a}^{\dag}_{\varphi} + v_{\chi\chi}(\eta) \widehat{a}_{\chi} +  v^{*}_{\chi\chi}(\eta) \widehat{a}^{\dag}_{\chi} \, ,
        		\end{align}
        where $(\widehat{a}_{\varphi};\widehat{a}^{\dag}_{\varphi})$ and $(\widehat{a}_{\chi};\widehat{a}^{\dag}_{\chi})$ are the creation and annihilation operators of the $\varphi$ and $\chi$ quanta respectively. This generalises the decomposition~\eqref{eq:modefctdecomp} to the case where fields interact and exchange quanta. A similar decomposition can be introduced for the momenta operators $\widehat{p}_{\varphi}$ and $\widehat{p}_{\chi}$, where the Hamiltonian~\eqref{eq:Hmat}-\eqref{eq:Hvarphimat} gives the mode functions 
        \bea
        \label{eq:mode:function:generalised:momentum}
        p_{ij}(\eta) = v_{ij}'(\eta)-\frac{a'}{a} v_{ij}(\eta)
        \eea
        for $i,j\in\{\varphi,\chi\}$. Using Heisenberg's equations, one finds that the mode functions evolve according to
        \bea
        \label{eq:eom:vij}
        v_{ij}''+\omega_i^2(\eta) v_{ij}= -\lambda^2 a^2(\eta) v_{\bar{i}j}\, ,
        \eea
        where we have introduced $\omega^2_{\varphi}(\eta) \equiv k^2 + m^2 a^2(\eta)-a''/a$ and $\omega^2_{\chi}(\eta) \equiv k^2 + M^2 a^2(\eta)-a''/a$,	and where $\bar{i}=\chi$ when $i=\varphi$ and $\bar{i}=\varphi$ when $i=\chi$. This constitutes a set of coupled differential equations, where the coupling is mediated by $\lambda^2$. It can thus be solved perturbatively in $\lambda$.  		
        		\begin{itemize}
        			\item \textbf{Zeroth order}: The right-hand side of \Eq{eq:eom:vij} vanishes, hence the uncoupled dynamics is recovered, namely $v_{ii}^{(0)}(\eta)=v_i(\eta)$ and $v_{i\bar{i}}^{(0)}(\eta)=0$, where $v_{\varphi}$ and $v_{\chi}$ are the free-field mode functions [\ie they are given by \Eq{eq:modefctl} if one replaces $\nu_\uell$ by $\nu_\varphi$ and $\mu_{\mathrm{h}}$ by $\mu_{\chi}$]. One also has $p_{ii}^{(0)}(\eta)=p_i(\eta)$ and $p_{i\bar{i}}^{(0)}(\eta)=0$.
        			\item \textbf{First order}: At first order, the right-hand side of \Eq{eq:eom:vij} needs to be replaced with the zeroth-order solution. This does not change the diagonal mode functions $v_{ii}^{(1)}(\eta)=v_{ii}^{(0)}(\eta)$ and $p_{ii}^{(1)}(\eta)=p_{ii}^{(0)}(\eta)$, while the cross mode functions now obey $v^{(1)\prime\prime}_{i\bar{i}}+\omega_i^2 v^{(1)}_{i \bar{i}} = \lambda^2 a^2 v_{\bar{i}}$. Using the Green's functions of the homogeneous (hence uncoupled) system of differential equation, $g_i(\eta,\eta')=2 \Imag{v_{i} (\eta) v^{*}_{i} (\eta')} $, this gives rise to
        			\begin{align}
        			\label{eq:v1:sol}
        				v^{(1)}_{i\bar{i}}(\eta) =  - 2 \lambda^2 \int_{\eta_0}^{\eta} \dd \eta_1 a^2(\eta_1) \Imag{v_{i} (\eta) v^{*}_{i} (\eta_1)} v_{\bar{i}} (\eta_1) .			\end{align}
Using \Eq{eq:mode:function:generalised:momentum}, this leads to
\bea
\label{eq:p1:sol}
p^{(1)}_{i\bar{i}}(\eta) =  - 2 \lambda^2 \int_{\eta_0}^{\eta} \dd \eta_1 a^2(\eta_1) \Imag{p_{i} (\eta) v^{*}_{i} (\eta_1)} v_{\bar{i}} (\eta_1) .			\eea
        			\item \textbf{Second order}: At second order,  \Eq{eq:eom:vij} is sourced by the first-order solution, so the diagonal mode functions obey $v_{ii}^{(2)\prime\prime}+\omega_i^2 v_{ii}^{(2)}=-\lambda^2 a^2 v_{\bar{i} i}^{(1)}$. Using again the homogeneous Green functions, together with \Eq{eq:v1:sol}, this gives rise to 
        				\begin{align}
        				v^{(2)}_{ii}(\eta) = & v_{i} (\eta) + 4 \lambda^4 \int_{\eta_0}^{\eta} \dd \eta_1  a^2(\eta_1) \int_{\eta_0}^{\eta_1} \dd \eta_2  a^2(\eta_2) 
        				\Imag{v_{i} (\eta) v^{*}_{i} (\eta_1)} \Imag{v_{\bar{i}} (\eta_1) v^{*}_{\bar{i}} (\eta_2)} v_{i} (\eta_2).
        			\end{align}
Using \Eq{eq:mode:function:generalised:momentum}, this leads to
\bea
p^{(2)}_{ii}(\eta) = & p_{i} (\eta) + 4 \lambda^4 \int_{\eta_0}^{\eta} \dd \eta_1  a^2(\eta_1) \int_{\eta_0}^{\eta_1} \dd \eta_2  a^2(\eta_2) 
        				\Imag{p_{i} (\eta) v^{*}_{i} (\eta_1)} \Imag{v_{\bar{i}} (\eta_1) v^{*}_{\bar{i}} (\eta_2)} v_{i} (\eta_2).
\eea
        		One may also compute the cross mode functions, and carry on the expansion, but that would lead to subdominant corrections to the power spectra.
        		\end{itemize}
The covariance matrix can be computed using \Eq{eq:covdef}, and one finds
        		\begin{align}
        			\label{eq:covinin}		\bs{\Sigma}_{\varphi \varphi}(\eta) &= \begin{pmatrix}
        				\left|v_{\varphi\varphi} (\eta) \right|^2 + \left|v_{\varphi\chi}(\eta) \right|^2 & \Rea  \left[v_{\varphi\varphi}(\eta) p^{*}_{\varphi\varphi} (\eta)\right] + \Rea  \left[v_{\varphi\chi}(\eta) p^{*}_{\varphi\chi}(\eta) \right] \\ 
        				\Rea  \left[v_{\varphi\varphi}(\eta) p^{*}_{\varphi\varphi} (\eta)\right] + \Rea  \left[v_{\varphi\chi}(\eta) p^{*}_{\varphi\chi} (\eta)\right]	& \left|p_{\varphi\varphi}(\eta) \right|^2 + \left|p_{\varphi\chi} (\eta)\right|^2
        			\end{pmatrix}, \\ 
        			\bs{\Sigma}_{\chi \chi}(\eta) &= \begin{pmatrix}
        				\left|v_{\chi\chi}(\eta) \right|^2 + \left|v_{\chi\varphi} (\eta)\right|^2 & \Rea  \left[v_{\chi\chi} (\eta)p^{*}_{\chi\chi} (\eta)\right] + \Rea  \left[v_{\chi\varphi}(\eta) p^{*}_{\chi\varphi} (\eta)\right] \\
        				\Rea  \left[v_{\chi\chi} (\eta)p^{*}_{\chi\chi} (\eta)\right] + \Rea  \left[v_{\chi\varphi} (\eta)p^{*}_{\chi\varphi}(\eta) \right] & \left|p_{\chi\chi} (\eta)\right|^2 + \left|p_{\chi\varphi} (\eta)\right|^2
        			\end{pmatrix}, \\ 
        			\label{eq:crosscov}		\bs{\Sigma}_{ \varphi\chi}(\eta) &= \begin{pmatrix}
        				\Rea  \left[v_{\varphi\varphi} (\eta) v^{*}_{\chi\varphi} (\eta)\right] + \Rea  \left[v_{\chi\chi} (\eta) v^{*}_{\varphi\chi} (\eta)\right]  & \Rea  \left[v_{\varphi\varphi}(\eta) p^{*}_{\chi\varphi}(\eta) \right] + \Rea  \left[p_{\chi\chi}(\eta) v^{*}_{\varphi\chi}(\eta) \right] \\
        				\Rea  \left[p_{\varphi\varphi} (\eta) v^{*}_{\chi\varphi} (\eta)\right] + \Rea  \left[v_{\chi\chi}(\eta) p^{*}_{\varphi\chi}(\eta) \right] & \Rea  \left[p_{\varphi\varphi}(\eta) p^{*}_{\chi\varphi}(\eta) \right] + \Rea  \left[p_{\chi\chi}(\eta) p^{*}_{\varphi\chi} (\eta)\right]
        			\end{pmatrix}.
        		\end{align} 
        By inserting the mode functions obtained above into these expressions, one obtains the first perturbative corrections to the power spectra. For the configuration-configuration power spectrum of the $\varphi$ field, one finds	
		\begin{align}
			\boldmathsymbol{\Sigma}^{(2)}_{\varphi\varphi,11}(\eta) = \left|v^{(0)}_{\varphi\varphi} (\eta) \right|^2 + \left|v^{(1)}_{\varphi\chi}(\eta) \right|^2 + 2 \Rea \left[v^{(2-0)}_{\varphi\varphi} (\eta) v^{(0)*}_{\varphi\varphi} (\eta)\right],
\end{align}
where we have introduced the short-hand notation $v^{(2-0)}_{\varphi\varphi} (\eta)=v^{(2)}_{\varphi\varphi} (\eta)-v^{(0)}_{\varphi\varphi} (\eta)$, which selects the terms of order $\lambda^2$ in $v^{(2)}_{\varphi\varphi} (\eta)$. This gives rise to
\bea
\label{eq:P11:SPT:2}
				\boldmathsymbol{\Sigma}^{(2)}_{\varphi\varphi,11}(\eta)	=& \left|v_{\varphi} (\eta)\right|^2 + 4 \lambda^4 \left|\int_{\eta_0}^{\eta} \dd \eta_1 a^2(\eta_1) \Imag{v_{\varphi} (\eta) v^{*}_{\varphi} (\eta_1)} v_{\chi} (\eta_1)\right|^2 \\ 
				+ 8 \lambda^4 &\Rea  \left\lbrace v_{\varphi} (\eta) \int_{\eta_0}^{\eta}\dd \eta_1  a^2(\eta_1)  \int_{\eta_0}^{\eta_1}\dd \eta_2  a^2(\eta_2)\Imag{v_{\varphi} (\eta) v^{*}_{\varphi} (\eta_1)} \Imag{v_{\chi} (\eta_1) v^{*}_{\chi} (\eta_2)}  v^{*}_{\varphi} (\eta_2)\right\rbrace.
\eea
For the configuration-momentum power spectrum, one obtains
\bea
			\boldmathsymbol{\Sigma}^{(2)}_{\varphi\varphi,12}(\eta) =&	\Rea  \left[v^{(0)}_{\varphi\varphi}(\eta) p^{(0)*}_{\varphi\varphi} (\eta)\right] + \Rea  \left[v^{(1)}_{\varphi\chi}(\eta) p^{(1)*}_{\varphi\chi}(\eta) \right] 
			 + \Rea  \left[v^{(0)}_{\varphi\varphi}(\eta) p^{(2-0)*}_{\varphi\varphi} (\eta) + v^{(2-0)}_{\varphi\varphi}(\eta) p^{(0)*}_{\varphi\varphi} (\eta)\right],
\eea
namely
\bea
\label{eq:P12:SPT:2}
\boldmathsymbol{\Sigma}^{(2)}_{\varphi\varphi,12}(\eta) =&	\Rea  \left[v_{\varphi}(\eta) p^{*}_{\varphi} (\eta)\right]\\
+4\lambda^4 &\int_{\eta_0}^\eta \dd\eta' a^2(\eta')\Ima\left[v_\varphi(\eta) v_\varphi^*(\eta')\right]v_\chi(\eta')
\int_{\eta_0}^\eta \dd\eta'' a^2(\eta'')\Ima\left[p_\varphi(\eta) v_\varphi^*(\eta'')\right]v_\chi(\eta'')\\
+4\lambda^4& \Rea\left\lbrace v_\varphi(\eta) \int_{\eta_0}^{\eta} \dd \eta_1  a^2(\eta_1) \int_{\eta_0}^{\eta_1} \dd \eta_2  a^2(\eta_2) 
        				\Imag{p_{\varphi} (\eta) v^{*}_{\varphi} (\eta_1)} \Imag{v_{\chi} (\eta_1) v^{*}_{\chi} (\eta_2)} v_{\varphi}^* (\eta_2) \right\rbrace\\
+4\lambda^4& \Rea\left\lbrace p_\varphi(\eta) \int_{\eta_0}^{\eta} \dd \eta_1  a^2(\eta_1) \int_{\eta_0}^{\eta_1} \dd \eta_2  a^2(\eta_2) 
        				\Imag{v_{\varphi} (\eta) v^{*}_{\varphi} (\eta_1)} \Imag{v_{\chi} (\eta_1) v^{*}_{\chi} (\eta_2)} v_{\varphi}^* (\eta_2) \right\rbrace .
\eea
Finally, for the momentum-momentum power spectrum, one has
\bea
			\boldmathsymbol{\Sigma}^{(2)}_{\varphi\varphi,22}(\eta) = \left|p^{(0)}_{\varphi\varphi} (\eta) \right|^2 + \left|p^{(1)}_{\varphi\chi}(\eta) \right|^2 + 2 \Rea \left[p^{(2-0)}_{\varphi\varphi} (\eta) p^{(0)*}_{\varphi\varphi} (\eta)\right], 
\eea
which leads to
\bea
\label{eq:P22:SPT:2}
			\boldmathsymbol{\Sigma}^{(2)}_{\varphi\varphi,22}(\eta) = & \left|p_{\varphi} (\eta)\right|^2 + 4 \lambda^4 \left|\int_{\eta_0}^{\eta} \dd \eta_1 a^2(\eta_1) \Imag{p_{\varphi} (\eta) v^{*}_{\varphi} (\eta_1)} v_{\chi} (\eta_1)\right|^2 \\ 
				+ 8 \lambda^4 &\Rea  \left\lbrace  p_{\varphi} (\eta) \int_{\eta_0}^{\eta}\dd \eta_1  a^2(\eta_1)  \int_{\eta_0}^{\eta_1}\dd \eta_2  a^2(\eta_2)\Imag{p_{\varphi} (\eta) v^{*}_{\varphi} (\eta_1)} \Imag{v_{\chi} (\eta_1) v^{*}_{\chi} (\eta_2)}  v^{*}_{\varphi} (\eta_2)\right\rbrace.
\eea
        \subsection{Perturbative solution of TCL}\label{app:TCL2}
Let us start with the TCL${}_2$ master equation written in the form 
		\begin{align}\label{eq:CMEbis}
			\frac{\dd \widetilde{\rho}_{\mathrm{red}}}{\dd \eta} =& - \lambda^4 a^2(\eta)\int_{\eta_0}^{\eta}\dd \eta'  a^2(\eta') \nonumber \\
			&\Big\{\left[\widetilde{v}_{\varphi} (\eta) \widetilde{v}_{\varphi} (\eta') \widetilde{\rho}_{\mathrm{red}}(\eta) -  \widetilde{v}_{\varphi} (\eta') \widetilde{\rho}_{\mathrm{red}}(\eta) \widetilde{v}_{\varphi} (\eta) \right]\mathcal{K}^{>}(\eta,\eta') \nonumber\\
			&-\left[\widetilde{v}_{\varphi} (\eta)  \widetilde{\rho}_{\mathrm{red}}(\eta) \widetilde{v}_{\varphi} (\eta') -   \widetilde{\rho}_{\mathrm{red}}(\eta) \widetilde{v}_{\varphi} (\eta') \widetilde{v}_{\varphi} (\eta) \right]\mathcal{K}^{>*}(\eta,\eta')
			\Big\} .
		\end{align} 
This equation was obtained in \Eq{eq:CME} from microphysical considerations and is just a convenient rewriting of \Eq{eq:TCL2CL}. We want to solve it at order $\lambda^4$, \ie drop all contributions of higher order. Since the right-hand side is already proportional to $\lambda^4$, this implies that it can be evaluated in the free theory, where $\widetilde{\rho}_{\mathrm{red}}(\eta) \simeq \widetilde{\rho}_{\mathrm{red}}(\eta_0)$. One can thus integrate \Eq{eq:CMEbis}, which leads to 
\begin{align}\label{eq:rhoredapprox}
			\widetilde{\rho}_{\mathrm{red}}^{(2)} (\eta) =& \widetilde{\rho}_{\mathrm{red}} (\eta_0) - \lambda^4 \int_{\eta_0}^{\eta}\dd \eta'  a^2(\eta') \int_{\eta_0}^{\eta'}\dd \eta''  a^2(\eta'') \nonumber \\
			&\Big\{\left[\widetilde{v}_{\varphi} (\eta') \widetilde{v}_{\varphi} (\eta'')\widetilde{\rho}_{\mathrm{red}} (\eta_0) -  \widetilde{v}_{\varphi} (\eta'')\widetilde{\rho}_{\mathrm{red}} (\eta_0) \widetilde{v}_{\varphi} (\eta') \right]v_{\chi} (\eta') v^{*}_{\chi} (\eta'') \nonumber\\
			&-\left[\widetilde{v}_{\varphi} (\eta') \widetilde{\rho}_{\mathrm{red}} (\eta_0) \widetilde{v}_{\varphi} (\eta'') -  \widetilde{\rho}_{\mathrm{red}} (\eta_0) \widetilde{v}_{\varphi} (\eta'') \widetilde{v}_{\varphi} (\eta') \right]v^{*}_{\chi} (\eta') v_{\chi} (\eta'') 
			\Big\},
		\end{align} 
where we have used that the memory kernels are related to the free mode functions via \Eq{eq:kernel:mode:function}. 			

Let us now compute the entries of the covariance matrix using this expression for $\widetilde{\rho}_{\mathrm{red}}^{(2)}$. The configuration-configuration power spectrum reads
		\begin{align}
			\bs{\Sigma}^{(2)}_{\mathrm{TCL},11} (\eta) = \mathrm{Tr}\left[\widetilde{v}_{\varphi} (\eta) \widetilde{v}_{\varphi} (\eta)\widetilde{\rho}_{\mathrm{red}}^{(2)} (\eta)\right],
		\end{align}
		that is 
\bea 
\label{eq:computPk}
			\bs{\Sigma}^{(2)}_{\mathrm{TCL},11} (\eta) = & \mathrm{Tr}\left[\widetilde{v}_{\varphi} (\eta) \widetilde{v}_{\varphi} (\eta)\widetilde{\rho}_{\mathrm{red}} (\eta_0)\right] 
			 \\ &
			- \lambda^4 \int_{\eta_0}^{\eta}\dd \eta'  a^2(\eta')  v_{\chi} (\eta') \int_{\eta_0}^{\eta'}\dd \eta''  a^2(\eta'') v^{*}_{\chi} (\eta'')
			 \\ &
			\left\lbrace \mathrm{Tr}\left[\widetilde{v}_{\varphi} (\eta) \widetilde{v}_{\varphi} (\eta)\widetilde{v}_{\varphi} (\eta') \widetilde{v}_{\varphi} (\eta'')\widetilde{\rho}_{\mathrm{red}} (\eta_0)\right] - \mathrm{Tr}\left[\widetilde{v}_{\varphi} (\eta) \widetilde{v}_{\varphi} (\eta) \widetilde{v}_{\varphi} (\eta'')\widetilde{\rho}_{\mathrm{red}} (\eta_0) \widetilde{v}_{\varphi} (\eta')\right]\right\rbrace
			 \\ &
			+ \lambda^4 \int_{\eta_0}^{\eta}\dd \eta'  a^2(\eta')  v^{*}_{\chi} (\eta') \int_{\eta_0}^{\eta'}\dd \eta''  a^2(\eta'') v_{\chi} (\eta'')
			 \\ &
			\left\lbrace \mathrm{Tr}\left[\widetilde{v}_{\varphi} (\eta) \widetilde{v}_{\varphi} (\eta)\widetilde{v}_{\varphi} (\eta') \widetilde{\rho}_{\mathrm{red}} (\eta_0) \widetilde{v}_{\varphi} (\eta'') \right] - \mathrm{Tr}\left[\widetilde{v}_{\varphi} (\eta) \widetilde{v}_{\varphi} (\eta)  \widetilde{\rho}_{\mathrm{red}} (\eta_0) \widetilde{v}_{\varphi} (\eta'') \widetilde{v}_{\varphi} (\eta')\right]\right\rbrace .
\eea
Since the initial state is the Bunch-Davies vacuum, $\widetilde{\rho}_{\mathrm{red}} (\eta_0)=\left|\cancel{0}\right> \left<\cancel{0}\right|$, using the mode-function decomposition~\eqref{eq:mode:decomp:interaction} one obtains
\bea
\label{eq:Puupert}
			\bs{\Sigma}^{(2)}_{\mathrm{TCL},11} (\eta) =& \left|v_{\varphi} (\eta)\right|^2  \\ &
			- 4 \lambda^4 \Rea  \left[  v^2_{\varphi} (\eta)  \int_{\eta_0}^{\eta}\dd \eta'  a^2(\eta') v^{*}_{\varphi} (\eta') v_{\chi} (\eta') \int_{\eta_0}^{\eta'}\dd \eta''  a^2(\eta'') v^{*}_{\varphi} (\eta'') v^{*}_{\chi} (\eta'')   \right. \\ & \left.
			- \left|v_{\varphi}(\eta) \right|^2  \int_{\eta_0}^{\eta}\dd \eta'  a^2(\eta') v_{\varphi} (\eta') v_{\chi} (\eta') \int_{\eta_0}^{\eta'}\dd \eta''  a^2(\eta'') v^{*}_{\varphi} (\eta'') v^{*}_{\chi} (\eta'') \right].
\eea
		This expression matches \Eq{eq:P11:SPT:2}, as can be shown by expanding the real and imaginary parts and relabeling the integration domain.
Following the same method, one finds
\bea
\label{eq:Puppert}
			\bs{\Sigma}^{(2)}_{\mathrm{TCL},12} (\eta) 	=& \Real{v_{\varphi} (\eta)p^{*}_{\varphi} (\eta)}  \\ 
			& - 4 \lambda^4 \Rea  \Bigg\{v_{\varphi} (\eta)p_{\varphi} (\eta) \int_{\eta_0}^{\eta}\dd \eta'  a^2(\eta') v^{*}_{\varphi} (\eta') v_{\chi} (\eta') \int_{\eta_0}^{\eta'}\dd \eta''  a^2(\eta'') v^{*}_{\varphi} (\eta'') v^{*}_{\chi} \left(\eta''\right)  \\
			& - \Real{v_{\varphi} (\eta)p^{*}_{\varphi} (\eta)} \int_{\eta_0}^{\eta}\dd \eta'  a^2(\eta') v_{\varphi} (\eta') v_{\chi} (\eta') \int_{\eta_0}^{\eta'}\dd \eta''  a^2(\eta'') v^{*}_{\varphi} (\eta'') v^{*}_{\chi} (\eta'') \Bigg\},
\eea
which can be shown to match \Eq{eq:P12:SPT:2}, and
\bea\label{eq:Ppppert}
			\bs{\Sigma}^{(2)}_{\mathrm{TCL},22} (\eta) 	= & \left|p_{\varphi} (\eta)\right|^2  \\
			& - 4 \lambda^4 \Rea  \Bigg[p^2_{\varphi} (\eta) \int_{\eta_0}^{\eta}\dd \eta'  a^2(\eta') v^{*}_{\varphi} (\eta') v_{\chi} (\eta') \int_{\eta_0}^{\eta'}\dd \eta''  a^2(\eta'') v^{*}_{\varphi} (\eta'') v^{*}_{\chi} (\eta'') \\
			& - \left|p_{\varphi}(\eta) \right|^2 \int_{\eta_0}^{\eta}\dd \eta'  a^2(\eta') v_{\varphi} (\eta') v_{\chi} (\eta') \int_{\eta_0}^{\eta'}\dd \eta''  a^2(\eta'') v^{*}_{\varphi} (\eta'') v^{*}_{\chi} (\eta'') \Bigg], 
\eea
which can be shown to match \Eq{eq:P22:SPT:2}.

	\section{Comparison with other late-time resummation techniques}\label{app:growingdecaying}
	In this section, we compare TCL with the late-time resummation technique proposed in \Refa{Boyanovsky:2015tba} and also studied in \Refa{Brahma:2021mng}. The idea is to keep track of the growing mode only, in order to simplify the analysis in the late-time limit. As we will make clear, the method also implicitly performs an additional layer of approximation compared to TCL, which makes it less efficient. 
	
    The starting point is to rewrite the free mode function
    \bea
    v_\varphi(z) =  \frac{1}{2}\sqrt{\frac{\pi z}{k}} \ee^{i\frac{\pi}{2}\left(\nu_{\varphi}+\frac{1}{2}\right)} H_{\nu_{\varphi}}^{(1)}(z) ,
    \eea
    where we recall that $z=-k\eta$, as (see Eq.~(10.4.3) of \Refa{NIST:DLMF})
    \bea
    v_\varphi(z) = \ee^{i\frac{\pi}{2}\left(\nu_{\varphi}+\frac{1}{2}\right)} \frac{v_-(z) + i v_+(z)}{\sqrt{2}}
    \eea
    where
	\begin{eqnarray}
		\label{eq:modefct+}	v_{+}(z) = \sqrt{\frac{\pi z}{2 k}} Y_{\nu_{\varphi}}(z)~~~~\text{and}~~~~
		\label{eq:modefct-}	v_{-}(z) = \sqrt{\frac{\pi z}{2 k}} J_{\nu_{\varphi}}(z) 
	\end{eqnarray}
are real functions. Here, $J_{\nu}$ and $Y_{\nu}$ are the Bessel functions of the first and second kind respectively, and of order $\nu$. The reason why this decomposition is convenient is that $v_-$ corresponds to the cosmological ``decaying mode'' [\ie $v_-(\eta)$ decreases on super-Hubble scales], while $v_+$ stands for the growing mode. Let us recall that the heavy-field mode function cannot be divided into a growing mode and a decaying mode, since both modes oscillate with similar amplitude on super-Hubble scales.

In the interaction picture, where operators evolve as in the free theory, the mode-function expansion~\eqref{eq:mode:decomp:interaction} of the field operators can then be written as
	\begin{align}
		\widetilde{v}_{\varphi}(\eta) &= v_{\varphi} (\eta) \widehat{a}_{\varphi} + v^{*}_{\varphi} (\eta) \widehat{a}^{\dag}_{\varphi} \\ 
		&= v_{-} (\eta) \widehat{P}_{\varphi}+ v_{+} (\eta) \widehat{Q}_{\varphi}\, ,
		\label{eq:vtilde:P:Q}
	\end{align}
	where
	\begin{align}
		\widehat{P}_{\varphi} &= \frac{1}{\sqrt{2}}\left[\ee^{i\frac{\pi}{2}\left(\nu_{\varphi}+\frac{1}{2}\right)}\widehat{a}_{\varphi} + \ee^{-i\frac{\pi}{2}\left(\nu_{\varphi}+\frac{1}{2}\right)}\widehat{a}^{\dag}_{\varphi} \right] ,\\
		\widehat{Q}_{\varphi} &= \frac{i}{\sqrt{2}}\left[\ee^{i\frac{\pi}{2}\left(\nu_{\varphi}+\frac{1}{2}\right)}\widehat{a}_{\varphi} - \ee^{-i\frac{\pi}{2}\left(\nu_{\varphi}+\frac{1}{2}\right)}\widehat{a}^{\dag}_{\varphi} \right] .
	\end{align}
One can check that they constitute a set of canonical variables since $\left[\widehat{Q}_{\varphi}, \widehat{P}_{\varphi}\right] = i $. 

The idea proposed in \Refs{Boyanovsky:2015tba, Brahma:2021mng} is to insert the decomposition~\eqref{eq:vtilde:P:Q} into the TCL${}_2$ master equation~\eqref{eq:TCL2CL} in order to identify the leading late-time contribution. One finds
\bea\label{eq:MEv+v-}
		\frac{\dd \widetilde{\rho}_{\mathrm{red}}^{\mathrm{IR}}}{\dd \eta} = - \lambda^4 a^2(\eta)\Bigg\{& v_{-} (\eta) X^{*}_{-}(\eta)  v_{\chi} (\eta) \left[ \widehat{P}^{2}_{\varphi}\widetilde{\rho}_{\mathrm{red}}(\eta) - \widehat{P}_{\varphi}\widetilde{\rho}_{\mathrm{red}}(\eta)\widehat{P}_{\varphi} \right]  \\
		&+ v_{-} (\eta) X_{-}(\eta)  v^{*}_{\chi} (\eta) \left[ \widetilde{\rho}_{\mathrm{red}}(\eta)\widehat{P}^{2}_{\varphi} - \widehat{P}_{\varphi}\widetilde{\rho}_{\mathrm{red}}(\eta)\widehat{P}_{\varphi} \right]  \\
		&+ v_{+} (\eta) X^{*}_{-}(\eta)  v_{\chi} (\eta) \left[ \widehat{Q}_{\varphi}\widehat{P}_{\varphi}\widetilde{\rho}_{\mathrm{red}}(\eta) - \widehat{P}_{\varphi}\widetilde{\rho}_{\mathrm{red}}(\eta)\widehat{Q}_{\varphi} \right]   \\
		&+ v_{+} (\eta) X_{-}(\eta)  v^{*}_{\chi} (\eta) \left[ \widetilde{\rho}_{\mathrm{red}}(\eta)\widehat{P}_{\varphi}\widehat{Q}_{\varphi} - \widehat{Q}_{\varphi}\widetilde{\rho}_{\mathrm{red}}(\eta)\widehat{P}_{\varphi} \right]  \\
		&+ v_{-} (\eta) X^{*}_{+}(\eta)  v_{\chi} (\eta) \left[ \widehat{P}_{\varphi}\widehat{Q}_{\varphi}\widetilde{\rho}_{\mathrm{red}}(\eta) - \widehat{Q}_{\varphi}\widetilde{\rho}_{\mathrm{red}}(\eta)\widehat{P}_{\varphi} \right]   \\
		&+ v_{-} (\eta) X_{+}(\eta)  v^{*}_{\chi} (\eta) \left[ \widetilde{\rho}_{\mathrm{red}}(\eta)\widehat{Q}_{\varphi}\widehat{P}_{\varphi} - \widehat{P}_{\varphi}\widetilde{\rho}_{\mathrm{red}}(\eta)\widehat{Q}_{\varphi} \right]   \\
		&+ v_{+} (\eta) X^{*}_{+}(\eta)  v_{\chi} (\eta)  \left[ \widehat{Q}^{2}_{\varphi}\widetilde{\rho}_{\mathrm{red}}(\eta) - \widehat{Q}_{\varphi}\widetilde{\rho}_{\mathrm{red}}(\eta)\widehat{Q}_{\varphi} \right]   \\
		&+ v_{+} (\eta) X_{+}(\eta)  v^{*}_{\chi} (\eta) \left[ \widetilde{\rho}_{\mathrm{red}}(\eta)\widehat{Q}^{2}_{\varphi} - \widehat{Q}_{\varphi}\widetilde{\rho}_{\mathrm{red}}(\eta)\widehat{Q}_{\varphi} \right]\Bigg\},
\eea
	where
	\begin{align}
		X_{+}(\eta) &\equiv \int_{\eta_0}^{\eta} \dd \eta' a^2(\eta')  v_{+} (\eta') v_{\chi} (\eta'), \\
		X_{-}(\eta) &\equiv \int_{\eta_0}^{\eta} \dd \eta' a^2(\eta')  v_{-} (\eta') v_{\chi} (\eta').
	\end{align}
The authors of \Refs{Boyanovsky:2015tba, Brahma:2021mng} argue that dropping the decaying-mode contributions constitutes a valid approximation in the infrared (IR) limit, and for this reason hereafter we label the quantities computed in this scheme with the superscript ``IR''.
	
In the interaction picture, the configuration-configuration power spectrum reads
\bea
		\left< \widetilde{v}_{\varphi} (\eta) \widetilde{v}_{\varphi} (\eta)\right> =& v_{-} (\eta) v_{-} (\eta) \left<  \widehat{P}^{2}_{\varphi} \right> + v_{-} (\eta) v_{+} (\eta) \left< \widehat{Q}_{\varphi}  \widehat{P}_{\varphi} + \widehat{P}_{\varphi} \widehat{Q}_{\varphi}  \right> + v_{+} (\eta) v_{+} (\eta) \left<  \widehat{Q}^{2}_{\varphi} \right> \\
		\simeq& v_{+} (\eta) v_{+} (\eta) \left<  \widehat{Q}^{2}_{\varphi} \right> ,
\eea
where in the second line we have neglected the decaying mode contribution. 
The next step is to compute $\left<  \widehat{Q}^{2}_{\varphi} \right>(\eta) = \mathrm{Tr}\left[ \widehat{Q}^{2}_{\varphi}\widetilde{\rho}_{\mathrm{red}}(\eta)\right]$ with the IR master equation~\eqref{eq:MEv+v-}. Upon differentiating this expression with respect to time, one obtains
	\begin{align}
		\frac{\dd \left<  \widehat{Q}^{2}_{\varphi} \right>}{\dd \eta} = \Gamma(\eta) \left<  \widehat{Q}^{2}_{\varphi} \right>
	\end{align}
	where
	\begin{align}
		\Gamma(\eta) = 4 \lambda^4 a^2(\eta) v_{-} (\eta)  \Imag{ v_{\chi} (\eta) X^{*}_{+}(\eta) },
	\end{align}
which gives rise to
	\begin{align}
		\left<  \widehat{Q}^{2}_{\varphi} \right>(\eta) = e^{\int_{\eta_*}^{\eta}\dd \eta' \Gamma(\eta')} \left<  \widehat{Q}^{2}_{\varphi} \right>(\eta_*) . 
	\end{align}
Since we are interested in the late-time behaviour of the power spectra, we can assume $-k \eta \ll 1$ and let $\eta_*$ denote the Hubble crossing time, $\eta_* \equiv -1 / k$, if the above integral is dominated by its upper bound (hence does not depend much on the choice of the lower bound). If the effect of the interaction with the environment is small in sub-Hubble scales, as argued in \Refa{Boyanovsky:2015tba} one can evaluate $\langle  \widehat{Q}^{2}_{\varphi} \rangle(\eta_*)$ in the free theory, which simply yields
	\begin{align}
		\left<  \widehat{Q}^{2}_{\varphi} \right>(\eta_*)  \simeq 1\, .
	\end{align}
In the super-Hubble limit, using the results derived in \App{sec:TCLcoeff:superH}, one can also approximate	
	\begin{align}
		X^{*}_{+}(z) \simeq \frac{\pi }{H^2 }\frac{(-1)^{3/4}}{\nu_\varphi^2 + \mu_\chi^2} \frac{z^{-\nu_\varphi}}{\sin (\pi \nu_\varphi) \Gamma(1-\nu_\varphi)}\left[\nu_\varphi \left(\gamma^{*}_{\mu_{\chi}}+\delta^{*}_{\mu_{\chi}}\right)-i \mu_\chi \left(\gamma^{*}_{\mu_{\chi}}-\delta_{\mu^{*}_{\chi}}\right)\right] \ee^{-\pi \mu_\chi},
	\end{align}
where $\gamma_\mu$ and $\delta_\mu$ were defined in \Eq{eq:app:SHcoef2}.	This leads to
	\begin{align}
	    \Gamma(z) = \frac{\pi}{2 \nu_\varphi}\frac{1}{\nu_\varphi^2 + \mu_\chi^2}\frac{\lambda^4}{H^4} \frac{k}{z} \mu_\chi \left(\left|\gamma_{\mu_{\chi}}\right|^2 - \left| \delta_{\mu_{\chi}}\right|^2\right) \ee^{-\pi \mu_\chi},
	\end{align}
where $\vert\gamma_{\mu_\chi}\vert^2-\vert\delta_{\mu_\chi}\vert^2=2\ee^{\pi\mu_\chi}/(\pi\mu_\chi)$. One thus has
	\begin{align}
		\int_{\eta_*}^{\eta}\dd \eta' \Gamma(\eta') \simeq - \frac{1}{ \nu_\varphi}\frac{1}{\nu_\varphi^2 + \mu_\chi^2}\frac{\lambda^4}{H^4} \ln(- k \eta) ,
	\end{align}
which one can check does not depend on the detailed choice of $\eta_*$ as announced above. Combining the above results, one obtains	
	\begin{align}
		\bs{\Sigma}_{\mathrm{IR},11}(\eta) = e^{ - \frac{1}{ \nu_\varphi}\frac{1}{\nu_\varphi^2 + \mu_\chi^2}\frac{\lambda^4}{H^4} \ln(- k \eta)} |v_{\varphi} (\eta)|^2.
	\end{align}
	
The configuration-momentum and momentum-momentum power spectra can be computed along similar lines. Starting from
	\begin{align}
		\widetilde{p}_{\varphi} (\eta) = p_{+} (\eta) \widehat{Q}_{\varphi} + p_{-} (\eta) \widehat{P}_{\varphi}
	\end{align}
	and using the fact that $p_{+} =v_+'-(a'/a)v_+ $ and $p_{-}=v_-'-(a'/a)v_- $ are still growing and decaying respectively, one has 
	\begin{align}
		\left< \widetilde{v}_{\varphi} (\eta) \widetilde{p}_{\varphi} (\eta)\right> \simeq v_{+} (\eta) p_{+} (\eta) \left<  \widehat{Q}^{2}_{\varphi} \right> \, , \\ 
		\left< \widetilde{p}_{\varphi} (\eta) \widetilde{p}_{\varphi} (\eta)\right> \simeq p_{+} (\eta) p_{+} (\eta) \left<  \widehat{Q}^{2}_{\varphi} \right>\, .
	\end{align}
This implies that the same correction is obtained for all power spectra, \ie 
	\begin{align}
	\label{eq:SigmaIR:11}
		\bs{\Sigma}_{\mathrm{IR},11}(\eta) &= e^{ -\frac{1}{ \nu_\varphi}\frac{1}{\nu_\varphi^2 + \mu_\chi^2}\frac{\lambda^4}{H^4}\ln(- k \eta)} |v_{\varphi} (\eta)|^2 ,\\ 
		\bs{\Sigma}_{\mathrm{IR},12}(\eta) &= e^{ - \frac{1}{ \nu_\varphi}\frac{1}{\nu_\varphi^2 + \mu_\chi^2}\frac{\lambda^4}{H^4}\ln(- k \eta)} \Real{v_{\varphi} (\eta) p^{*}_{\varphi} (\eta)} ,\\ 
		\bs{\Sigma}_{\mathrm{IR},22}(\eta) &= e^{ -\frac{1}{ \nu_\varphi}\frac{1}{\nu_\varphi^2 + \mu_\chi^2}\frac{\lambda^4}{H^4} \ln(- k \eta)} |p_{\varphi} (\eta)|^2.
\label{eq:SigmaIR:22}
	\end{align}
	\begin{figure}[h!]
		\centering 
		\includegraphics[width=.8\textwidth]{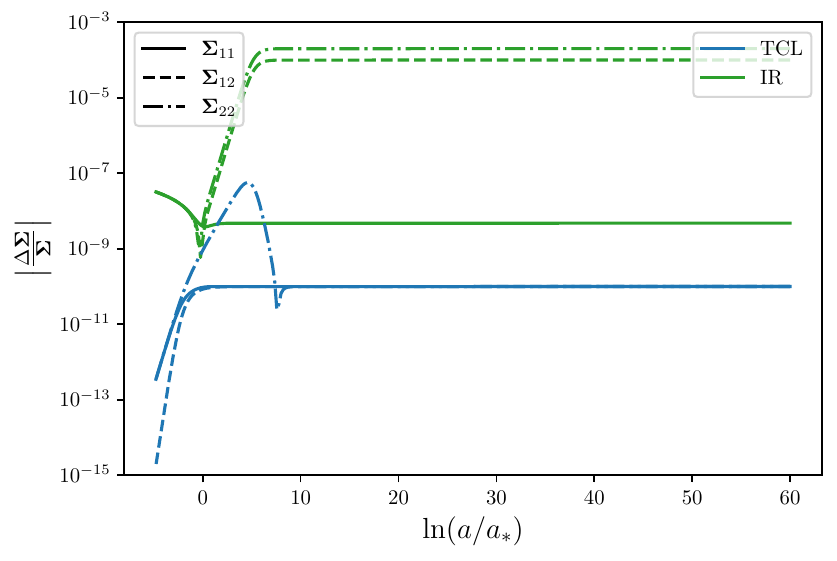}
		\caption{\label{fig:IRstrong} Relative error in the three power spectra for TCL${}_2$ (blue curves) and the IR resummation method presented in \App{app:growingdecaying} (green curves). The parameters are taken as $m^2 = 10^{-4} H^2$, $M^2 = 10^{3} H^2$ and $\lambda^2 = 10^{-3} H^2$. 
}
	\end{figure}

These expressions feature manifest resummations over powers of $\ln(a)$, which we now compare with the resummation performed by the TCL${}_2$ master equation. The relative difference between the three power spectra and their exact counterpart is displayed in \Fig{fig:IRstrong}, both for TCL${}_2$ (blue curves)\footnote{Let us note that at late time, the relative error in TCL asymptotes a constant in \Fig{fig:IRstrong}, hence it is not described by \Eq{eq:TCL:relat:error:superH}. The reason is that \Eq{eq:TCL:relat:error:superH} captures the error in the growth rate, while for the parameters displayed in \Fig{fig:IRstrong} the error in the overall amplitude provides the dominant contribution.} and IR (green curves). 

Let us first note that the growth rate of the power spectra is correctly captured in the IR approach, even at strong coupling where the perturbative result usually breaks down. This can be further understood by noting that \Eqs{eq:SigmaIR:11}-\eqref{eq:SigmaIR:22} take the same form as \Eq{eq:TCL:superH:exp:Cov} with
\bea
\label{eq:nu:IR}
\nu_{\mathrm{IR}}=\frac{1}{2\nu_\varphi} \frac{1}{\nu_\varphi^2+\mu_\chi^2}\left(\frac{\lambda}{H}\right)^4+\nu_\varphi\, ,
\eea
while according to footnote~\ref{footnote:nuLS}, in TCL${}_2$ one has
\bea
\label{eq:nu:LS:app}
\nu_{\mathrm{LS}} = \frac{3}{2}\sqrt{1-\left(\frac{2 m_{\mathrm{LS}}}{3H}\right)^2}\quad\text{where}\quad m_{\mathrm{LS}}^2=m^2-\frac{\lambda^4}{M^2-m^2}
\eea
and we recall that in the exact theory
\bea
\label{eq:nu:exact:app}
\nu_\uell=\frac{3}{2}\sqrt{1-\left(\frac{2 m_{\uell}}{3H}\right)^2}\quad\text{where}\quad m_{\uell}^2=\frac{1}{2} \left[m^2+M^2-\left(M^2-m^2\right)\sqrt{1+\left(\frac{2\lambda^2}{M^2-m^2}\right)^2}\right]\, .
\eea
Since \Eqs{eq:nu:IR},~\eqref{eq:nu:LS:app} and~\eqref{eq:nu:exact:app} coincide when expanded at first order in $\lambda^4$, one concludes that, at the level of the growth rate, the Lamb-shift renormalisation of the mass is correctly accounted for in the IR approach~\cite{Boyanovsky:2015tba} as for TCL, at least at leading order in the coupling constant. This is similar to the dynamical renormalisation group (DRG) treatment of late-time secular divergences in de Sitter performed in \Refs{Boyanovsky:1998aa, Burgess:2009bs, Green:2020txs}, as pointed out in \Refs{Boyanovsky:2015tba, Burgess:2015ajz, Brahma:2021mng}.

The IR approach however fails to reproduce the overall amplitude of the power spectra beyond the perturbative level, which explains why it does not perform as well as TCL. Let us also note that another disadvantage of the IR method is that it does not allow one to track decoherence, which as explained in \Sec{sec:decoherence} is not driven by the growing modes.

\bibliographystyle{JHEP}
\bibliography{biblio}

\end{document}

%% file: newcommands.tex
\newcommand{\ie}{\textsl{i.e.}~}

\newcommand{\eg}{\textsl{e.g.}\xspace}

\newcommand{\order}[1]{\mathcal{O}\!\left(#1\right)}

\newcommand{\Tr}{\mathrm{Tr}}

\newcommand{\dd}{\mathrm{d}}
\newcommand{\ee}{e}

\newcommand{\boldmathsymbol}[1]{{\ensuremath{\boldsymbol{#1}}}}

\newcommand{\calH}{\mathcal{H}}

\newcommand{\Rea}{\Re \mathrm{e}\,}
\newcommand{\Ima}{\Im \mathrm{m}\,}

\newcommand{\Real}[1]{\Rea \left[ #1 \right] }
\newcommand{\Imag}[1]{\Ima \left[ #1 \right] }

\newcommand{\cs}{c_{_\mathrm{S}}}

\newcommand{\beq}{\begin{equation}}
\newcommand{\eeq}{\end{equation}}
\newcommand{\bea}{\begin{equation}\begin{aligned}}
\newcommand{\eea}{\end{aligned}\end{equation}}

\newlength{\wsingfig}
\newlength{\wdblefig}
\newlength{\wquadfig}
\newlength{\wtriplefig}
\newcommand{\Eq}[1]{Eq.~(\ref{#1})}
\newcommand{\Eqs}[1]{Eqs.~(\ref{#1})}
\newcommand{\Fig}[1]{Fig.~{\ref{#1}}}

\newcommand{\Refa}[1]{Ref.~{\cite{#1}}}
\newcommand{\Refs}[1]{Refs.~{\cite{#1}}}
\newcommand{\Sec}[1]{Sec.~\ref{#1}}

\newcommand{\App}[1]{Appendix~\ref{#1}}

\newcommand{\bs}[1]{\boldsymbol{#1}}

%% file: ME.bbl
\providecommand{\href}[2]{#2}\begingroup\raggedright\begin{thebibliography}{100}

\bibitem{Planck:2018jri}
{\scshape Planck} collaboration, \emph{{Planck 2018 results. X. Constraints on
  inflation}}, \href{https://doi.org/10.1051/0004-6361/201833887}{\emph{Astron.
  Astrophys.} {\bfseries 641} (2020) A10}
  [\href{https://arxiv.org/abs/1807.06211}{{\ttfamily 1807.06211}}].

\bibitem{Martin:2013tda}
J.~Martin, C.~Ringeval and V.~Vennin, \emph{{Encyclopædia Inflationaris}},
  \href{https://doi.org/10.1016/j.dark.2014.01.003}{\emph{Phys. Dark Univ.}
  {\bfseries 5-6} (2014) 75} [\href{https://arxiv.org/abs/1303.3787}{{\ttfamily
  1303.3787}}].

\bibitem{Amendola:2016saw}
L.~Amendola et~al., \emph{{Cosmology and fundamental physics with the Euclid
  satellite}}, \href{https://doi.org/10.1007/s41114-017-0010-3}{\emph{Living
  Rev. Rel.} {\bfseries 21} (2018) 2}
  [\href{https://arxiv.org/abs/1606.00180}{{\ttfamily 1606.00180}}].

\bibitem{Maartens:2015mra}
{\scshape SKA Cosmology SWG} collaboration, \emph{{Overview of Cosmology with
  the SKA}}, \href{https://doi.org/10.22323/1.215.0016}{\emph{PoS} {\bfseries
  AASKA14} (2015) 016} [\href{https://arxiv.org/abs/1501.04076}{{\ttfamily
  1501.04076}}].

\bibitem{Lyth:2001nq}
D.H.~Lyth and D.~Wands, \emph{{Generating the curvature perturbation without an
  inflaton}}, \href{https://doi.org/10.1016/S0370-2693(01)01366-1}{\emph{Phys.
  Lett.} {\bfseries B524} (2002) 5}
  [\href{https://arxiv.org/abs/hep-ph/0110002}{{\ttfamily hep-ph/0110002}}].

\bibitem{Enqvist:2017kzh}
K.~Enqvist, R.J.~Hardwick, T.~Tenkanen, V.~Vennin and D.~Wands, \emph{{A novel
  way to determine the scale of inflation}},
  \href{https://doi.org/10.1088/1475-7516/2018/02/006}{\emph{JCAP} {\bfseries
  02} (2018) 006} [\href{https://arxiv.org/abs/1711.07344}{{\ttfamily
  1711.07344}}].

\bibitem{Ringeval:2010hf}
C.~Ringeval, T.~Suyama, T.~Takahashi, M.~Yamaguchi and S.~Yokoyama, \emph{{Dark
  energy from primordial inflationary quantum fluctuations}},
  \href{https://doi.org/10.1103/PhysRevLett.105.121301}{\emph{Phys. Rev. Lett.}
  {\bfseries 105} (2010) 121301}
  [\href{https://arxiv.org/abs/1006.0368}{{\ttfamily 1006.0368}}].

\bibitem{Kiefer:2010pb}
C.~Kiefer, F.~Queisser and A.A.~Starobinsky, \emph{{Cosmological Constant from
  Decoherence}},
  \href{https://doi.org/10.1088/0264-9381/28/12/125022}{\emph{Class. Quant.
  Grav.} {\bfseries 28} (2011) 125022}
  [\href{https://arxiv.org/abs/1010.5331}{{\ttfamily 1010.5331}}].

\bibitem{Brandenberger:1992sr}
R.H.~Brandenberger, V.F.~Mukhanov and T.~Prokopec, \emph{{Entropy of a
  classical stochastic field and cosmological perturbations}},
  \href{https://doi.org/10.1103/PhysRevLett.69.3606}{\emph{Phys. Rev. Lett.}
  {\bfseries 69} (1992) 3606}
  [\href{https://arxiv.org/abs/astro-ph/9206005}{{\ttfamily
  astro-ph/9206005}}].

\bibitem{Barvinsky:1998cq}
A.O.~Barvinsky, A.Y.~Kamenshchik, C.~Kiefer and I.V.~Mishakov,
  \emph{{Decoherence in quantum cosmology at the onset of inflation}},
  \href{https://doi.org/10.1016/S0550-3213(99)00208-4}{\emph{Nucl. Phys. B}
  {\bfseries 551} (1999) 374}
  [\href{https://arxiv.org/abs/gr-qc/9812043}{{\ttfamily gr-qc/9812043}}].

\bibitem{Lombardo:2005iz}
F.C.~Lombardo and D.~Lopez~Nacir, \emph{{Decoherence during inflation: The
  Generation of classical inhomogeneities}},
  \href{https://doi.org/10.1103/PhysRevD.72.063506}{\emph{Phys. Rev. D}
  {\bfseries 72} (2005) 063506}
  [\href{https://arxiv.org/abs/gr-qc/0506051}{{\ttfamily gr-qc/0506051}}].

\bibitem{Kiefer:2006je}
C.~Kiefer, I.~Lohmar, D.~Polarski and A.A.~Starobinsky, \emph{{Pointer states
  for primordial fluctuations in inflationary cosmology}},
  \href{https://doi.org/10.1088/0264-9381/24/7/002}{\emph{Class. Quant. Grav.}
  {\bfseries 24} (2007) 1699}
  [\href{https://arxiv.org/abs/astro-ph/0610700}{{\ttfamily
  astro-ph/0610700}}].

\bibitem{Martineau:2006ki}
P.~Martineau, \emph{{On the decoherence of primordial fluctuations during
  inflation}}, \href{https://doi.org/10.1088/0264-9381/24/23/006}{\emph{Class.
  Quant. Grav.} {\bfseries 24} (2007) 5817}
  [\href{https://arxiv.org/abs/astro-ph/0601134}{{\ttfamily
  astro-ph/0601134}}].

\bibitem{Burgess:2006jn}
C.P.~Burgess, R.~Holman and D.~Hoover, \emph{{Decoherence of inflationary
  primordial fluctuations}},
  \href{https://doi.org/10.1103/PhysRevD.77.063534}{\emph{Phys. Rev. D}
  {\bfseries 77} (2008) 063534}
  [\href{https://arxiv.org/abs/astro-ph/0601646}{{\ttfamily
  astro-ph/0601646}}].

\bibitem{Prokopec:2006fc}
T.~Prokopec and G.I.~Rigopoulos, \emph{{Decoherence from Isocurvature
  perturbations in Inflation}},
  \href{https://doi.org/10.1088/1475-7516/2007/11/029}{\emph{JCAP} {\bfseries
  11} (2007) 029} [\href{https://arxiv.org/abs/astro-ph/0612067}{{\ttfamily
  astro-ph/0612067}}].

\bibitem{Nelson:2016kjm}
E.~Nelson, \emph{{Quantum Decoherence During Inflation from Gravitational
  Nonlinearities}},
  \href{https://doi.org/10.1088/1475-7516/2016/03/022}{\emph{JCAP} {\bfseries
  03} (2016) 022} [\href{https://arxiv.org/abs/1601.03734}{{\ttfamily
  1601.03734}}].

\bibitem{Martin:2018zbe}
J.~Martin and V.~Vennin, \emph{{Observational constraints on quantum
  decoherence during inflation}},
  \href{https://doi.org/10.1088/1475-7516/2018/05/063}{\emph{JCAP} {\bfseries
  05} (2018) 063} [\href{https://arxiv.org/abs/1801.09949}{{\ttfamily
  1801.09949}}].

\bibitem{Martin:2018lin}
J.~Martin and V.~Vennin, \emph{{Non Gaussianities from Quantum Decoherence
  during Inflation}},
  \href{https://doi.org/10.1088/1475-7516/2018/06/037}{\emph{JCAP} {\bfseries
  06} (2018) 037} [\href{https://arxiv.org/abs/1805.05609}{{\ttfamily
  1805.05609}}].

\bibitem{Zurek:1981xq}
W.H.~Zurek, \emph{{Pointer Basis of Quantum Apparatus: Into What Mixture Does
  the Wave Packet Collapse?}},
  \href{https://doi.org/10.1103/PhysRevD.24.1516}{\emph{Phys. Rev. D}
  {\bfseries 24} (1981) 1516}.

\bibitem{Zurek:1982ii}
W.H.~Zurek, \emph{{Environment induced superselection rules}},
  \href{https://doi.org/10.1103/PhysRevD.26.1862}{\emph{Phys. Rev. D}
  {\bfseries 26} (1982) 1862}.

\bibitem{Joos:1984uk}
E.~Joos and H.~Zeh, \emph{{The Emergence of classical properties through
  interaction with the environment}},
  \href{https://doi.org/10.1007/BF01725541}{\emph{Z. Phys. B} {\bfseries 59}
  (1985) 223}.

\bibitem{Hollowood:2017bil}
T.J.~Hollowood and J.I.~McDonald, \emph{{Decoherence, discord and the quantum
  master equation for cosmological perturbations}},
  \href{https://doi.org/10.1103/PhysRevD.95.103521}{\emph{Phys. Rev. D}
  {\bfseries 95} (2017) 103521}
  [\href{https://arxiv.org/abs/1701.02235}{{\ttfamily 1701.02235}}].

\bibitem{Martin:2021znx}
J.~Martin, A.~Micheli and V.~Vennin, \emph{{Discord and decoherence}},
  \href{https://doi.org/10.1088/1475-7516/2022/04/051}{\emph{JCAP} {\bfseries
  04} (2022) 051} [\href{https://arxiv.org/abs/2112.05037}{{\ttfamily
  2112.05037}}].

\bibitem{Koks:1996ga}
D.~Koks, A.~Matacz and B.~Hu, \emph{{Entropy and uncertainty of squeezed
  quantum open systems}},
  \href{https://doi.org/10.1103/PhysRevD.55.5917}{\emph{Phys. Rev. D}
  {\bfseries 55} (1997) 5917}
  [\href{https://arxiv.org/abs/quant-ph/9612016}{{\ttfamily
  quant-ph/9612016}}].

\bibitem{Anastopoulos:2013zya}
C.~Anastopoulos and B.L.~Hu, \emph{{A Master Equation for Gravitational
  Decoherence: Probing the Textures of Spacetime}},
  \href{https://doi.org/10.1088/0264-9381/30/16/165007}{\emph{Class. Quant.
  Grav.} {\bfseries 30} (2013) 165007}
  [\href{https://arxiv.org/abs/1305.5231}{{\ttfamily 1305.5231}}].

\bibitem{Fukuma:2013uxa}
M.~Fukuma, Y.~Sakatani and S.~Sugishita, \emph{{Master equation for the
  Unruh-DeWitt detector and the universal relaxation time in de Sitter space}},
  \href{https://doi.org/10.1103/PhysRevD.89.064024}{\emph{Phys. Rev. D}
  {\bfseries 89} (2014) 064024}
  [\href{https://arxiv.org/abs/1305.0256}{{\ttfamily 1305.0256}}].

\bibitem{Cheung:2007st}
C.~Cheung, P.~Creminelli, A.L.~Fitzpatrick, J.~Kaplan and L.~Senatore,
  \emph{{The Effective Field Theory of Inflation}},
  \href{https://doi.org/10.1088/1126-6708/2008/03/014}{\emph{JHEP} {\bfseries
  03} (2008) 014} [\href{https://arxiv.org/abs/0709.0293}{{\ttfamily
  0709.0293}}].

\bibitem{Chen:2009zp}
X.~Chen and Y.~Wang, \emph{{Quasi-Single Field Inflation and
  Non-Gaussianities}},
  \href{https://doi.org/10.1088/1475-7516/2010/04/027}{\emph{JCAP} {\bfseries
  04} (2010) 027} [\href{https://arxiv.org/abs/0911.3380}{{\ttfamily
  0911.3380}}].

\bibitem{Senatore:2010wk}
L.~Senatore and M.~Zaldarriaga, \emph{{The Effective Field Theory of Multifield
  Inflation}}, \href{https://doi.org/10.1007/JHEP04(2012)024}{\emph{JHEP}
  {\bfseries 04} (2012) 024} [\href{https://arxiv.org/abs/1009.2093}{{\ttfamily
  1009.2093}}].

\bibitem{Assassi:2013gxa}
V.~Assassi, D.~Baumann, D.~Green and L.~McAllister, \emph{{Planck-Suppressed
  Operators}}, \href{https://doi.org/10.1088/1475-7516/2014/01/033}{\emph{JCAP}
  {\bfseries 01} (2014) 033} [\href{https://arxiv.org/abs/1304.5226}{{\ttfamily
  1304.5226}}].

\bibitem{Arkani-Hamed:2015bza}
N.~Arkani-Hamed and J.~Maldacena, \emph{{Cosmological Collider Physics}},
  \href{https://arxiv.org/abs/1503.08043}{{\ttfamily 1503.08043}}.

\bibitem{Shandera:2017qkg}
S.~Shandera, N.~Agarwal and A.~Kamal, \emph{{Open quantum cosmological
  system}}, \href{https://doi.org/10.1103/PhysRevD.98.083535}{\emph{Phys. Rev.
  D} {\bfseries 98} (2018) 083535}
  [\href{https://arxiv.org/abs/1708.00493}{{\ttfamily 1708.00493}}].

\bibitem{Akhtar:2019qdn}
S.~Akhtar, S.~Choudhury, S.~Chowdhury, D.~Goswami, S.~Panda and A.~Swain,
  \emph{{Open Quantum Entanglement: A study of two atomic system in static
  patch of de Sitter space}},
  \href{https://doi.org/10.1140/epjc/s10052-020-8302-2}{\emph{Eur. Phys. J. C}
  {\bfseries 80} (2020) 748}
  [\href{https://arxiv.org/abs/1908.09929}{{\ttfamily 1908.09929}}].

\bibitem{Burgess:2020tbq}
C.P.~Burgess, \emph{{Introduction to Effective Field Theory}}, Cambridge
  University Press (12, 2020),
  \href{https://doi.org/10.1017/9781139048040}{10.1017/9781139048040}.

\bibitem{Banerjee:2020ljo}
S.~Banerjee, S.~Choudhury, S.~Chowdhury, R.N.~Das, N.~Gupta, S.~Panda et~al.,
  \emph{{Indirect detection of Cosmological Constant from interacting open
  quantum system}},
  \href{https://doi.org/10.1016/j.aop.2022.168941}{\emph{Annals Phys.}
  {\bfseries 443} (2022) 168941}
  [\href{https://arxiv.org/abs/2004.13058}{{\ttfamily 2004.13058}}].

\bibitem{Pinol:2021aun}
L.~Pinol, S.~Aoki, S.~Renaux-Petel and M.~Yamaguchi, \emph{{Inflationary flavor
  oscillations and the cosmic spectroscopy}},
  \href{https://arxiv.org/abs/2112.05710}{{\ttfamily 2112.05710}}.

\bibitem{Oppenheim:2022xjr}
J.~Oppenheim, C.~Sparaciari, B.~\v{S}oda and Z.~Weller-Davies,
  \emph{{Gravitationally induced decoherence vs space-time diffusion: testing
  the quantum nature of gravity}},
  \href{https://arxiv.org/abs/2203.01982}{{\ttfamily 2203.01982}}.

\bibitem{Pimentel:2022fsc}
G.L.~Pimentel and D.-G.~Wang, \emph{{Boostless Cosmological Collider
  Bootstrap}},  \href{https://arxiv.org/abs/2205.00013}{{\ttfamily
  2205.00013}}.

\bibitem{Jazayeri:2022kjy}
S.~Jazayeri and S.~Renaux-Petel, \emph{{Cosmological Bootstrap in Slow
  Motion}},  \href{https://arxiv.org/abs/2205.10340}{{\ttfamily 2205.10340}}.

\bibitem{Brahma:2022yxu}
S.~Brahma, A.~Berera and J.~Calder\'on-Figueroa, \emph{{Quantum corrections to
  the primordial tensor spectrum: Open EFTs \& Markovian decoupling of UV
  modes}},  \href{https://arxiv.org/abs/2206.05797}{{\ttfamily 2206.05797}}.

\bibitem{Breuer:2002pc}
H.P.~Breuer and F.~Petruccione, \emph{{The theory of open quantum systems}},
  Oxford University Press (2002),
  \href{https://doi.org/10.1093/acprof:oso/9780199213900.001.0001}{10.1093/acprof:oso/9780199213900.001.0001}.

\bibitem{Calzetta:2008iqa}
E.A.~Calzetta and B.-L.B.~Hu, \emph{{Nonequilibrium Quantum Field Theory}},
  Cambridge Monographs on Mathematical Physics, Cambridge University Press (9,
  2008),
  \href{https://doi.org/10.1017/CBO9780511535123}{10.1017/CBO9780511535123}.

\bibitem{Burgess:2014eoa}
C.P.~Burgess, R.~Holman, G.~Tasinato and M.~Williams, \emph{{EFT Beyond the
  Horizon: Stochastic Inflation and How Primordial Quantum Fluctuations Go
  Classical}}, \href{https://doi.org/10.1007/JHEP03(2015)090}{\emph{JHEP}
  {\bfseries 03} (2015) 090} [\href{https://arxiv.org/abs/1408.5002}{{\ttfamily
  1408.5002}}].

\bibitem{Boyanovsky:2015tba}
D.~Boyanovsky, \emph{{Effective field theory during inflation: Reduced density
  matrix and its quantum master equation}},
  \href{https://doi.org/10.1103/PhysRevD.92.023527}{\emph{Phys. Rev.}
  {\bfseries D92} (2015) 023527}
  [\href{https://arxiv.org/abs/1506.07395}{{\ttfamily 1506.07395}}].

\bibitem{Boyanovsky:2015jen}
D.~Boyanovsky, \emph{{Effective field theory during inflation. II. Stochastic
  dynamics and power spectrum suppression}},
  \href{https://doi.org/10.1103/PhysRevD.93.043501}{\emph{Phys. Rev. D}
  {\bfseries 93} (2016) 043501}
  [\href{https://arxiv.org/abs/1511.06649}{{\ttfamily 1511.06649}}].

\bibitem{Burgess:2015ajz}
C.~Burgess, R.~Holman and G.~Tasinato, \emph{{Open EFTs, IR effects \&
  late-time resummations: systematic corrections in stochastic inflation}},
  \href{https://doi.org/10.1007/JHEP01(2016)153}{\emph{JHEP} {\bfseries 01}
  (2016) 153} [\href{https://arxiv.org/abs/1512.00169}{{\ttfamily
  1512.00169}}].

\bibitem{Kaplanek:2021fnl}
G.~Kaplanek, C.P.~Burgess and R.~Holman, \emph{{Qubit heating near a hotspot}},
  \href{https://doi.org/10.1007/JHEP08(2021)132}{\emph{JHEP} {\bfseries 08}
  (2021) 132} [\href{https://arxiv.org/abs/2106.10803}{{\ttfamily
  2106.10803}}].

\bibitem{Chaykov:2022zro}
S.~Chaykov, N.~Agarwal, S.~Bahrami and R.~Holman, \emph{{Loop corrections in
  Minkowski spacetime away from equilibrium 1: Late-time resummations}},
  \href{https://arxiv.org/abs/2206.11288}{{\ttfamily 2206.11288}}.

\bibitem{Chaykov:2022pwd}
S.~Chaykov, N.~Agarwal, S.~Bahrami and R.~Holman, \emph{{Loop corrections in
  Minkowski spacetime away from equilibrium 2: Finite-time results}},
  \href{https://arxiv.org/abs/2206.11289}{{\ttfamily 2206.11289}}.

\bibitem{Martin:2007bw}
J.~Martin, \emph{{Inflationary perturbations: The Cosmological Schwinger
  effect}}, \href{https://doi.org/10.1007/978-3-540-74353-8_6}{\emph{Lect.
  Notes Phys.} {\bfseries 738} (2008) 193}
  [\href{https://arxiv.org/abs/0704.3540}{{\ttfamily 0704.3540}}].

\bibitem{Hsiang:2021vgx}
J.-T.~Hsiang and B.-L.~Hu, \emph{{Fluctuation-Dissipation Relation for a
  Quantum Brownian Oscillator in a Parametrically Squeezed Thermal Field}},
  \href{https://arxiv.org/abs/2107.13343}{{\ttfamily 2107.13343}}.

\bibitem{Colas:2021llj}
T.~Colas, J.~Grain and V.~Vennin, \emph{{Four-mode squeezed states: two-field
  quantum systems and the symplectic group $\mathrm {Sp}(4,{\mathbb {R}})$}},
  \href{https://doi.org/10.1140/epjc/s10052-021-09922-y}{\emph{Eur. Phys. J. C}
  {\bfseries 82} (2022) 6} [\href{https://arxiv.org/abs/2104.14942}{{\ttfamily
  2104.14942}}].

\bibitem{Banerjee:2021lqu}
S.~Banerjee, S.~Choudhury, S.~Chowdhury, J.~Knaute, S.~Panda and K.~Shirish,
  \emph{{Thermalization Phenomena in Quenched Quantum Brownian Motion in De
  Sitter Space}},  \href{https://arxiv.org/abs/2104.10692}{{\ttfamily
  2104.10692}}.

\bibitem{Lindblad:1975ef}
G.~Lindblad, \emph{{On the Generators of Quantum Dynamical Semigroups}},
  \href{https://doi.org/10.1007/BF01608499}{\emph{Commun. Math. Phys.}
  {\bfseries 48} (1976) 119}.

\bibitem{Kaplanek:2022xrr}
G.~Kaplanek and E.~Tjoa, \emph{{Mapping Markov: On effective master equations
  for two accelerated qubits}},
  \href{https://arxiv.org/abs/2207.13750}{{\ttfamily 2207.13750}}.

\bibitem{Boyanovsky:1998aa}
D.~Boyanovsky, H.J.~de~Vega, R.~Holman and M.~Simionato, \emph{{Dynamical
  renormalization group resummation of finite temperature infrared
  divergences}}, \href{https://doi.org/10.1103/PhysRevD.60.065003}{\emph{Phys.
  Rev. D} {\bfseries 60} (1999) 065003}
  [\href{https://arxiv.org/abs/hep-ph/9809346}{{\ttfamily hep-ph/9809346}}].

\bibitem{Burgess:2009bs}
C.P.~Burgess, L.~Leblond, R.~Holman and S.~Shandera, \emph{{Super-Hubble de
  Sitter Fluctuations and the Dynamical RG}},
  \href{https://doi.org/10.1088/1475-7516/2010/03/033}{\emph{JCAP} {\bfseries
  03} (2010) 033} [\href{https://arxiv.org/abs/0912.1608}{{\ttfamily
  0912.1608}}].

\bibitem{Green:2020txs}
D.~Green and A.~Premkumar, \emph{{Dynamical RG and Critical Phenomena in de
  Sitter Space}}, \href{https://doi.org/10.1007/JHEP04(2020)064}{\emph{JHEP}
  {\bfseries 04} (2020) 064}
  [\href{https://arxiv.org/abs/2001.05974}{{\ttfamily 2001.05974}}].

\bibitem{Brahma:2021mng}
S.~Brahma, A.~Berera and J.~Calder\'on-Figueroa, \emph{{Universal signature of
  quantum entanglement across cosmological distances}},
  \href{https://arxiv.org/abs/2107.06910}{{\ttfamily 2107.06910}}.

\bibitem{Caldeira:1981rx}
A.O.~Caldeira and A.J.~Leggett, \emph{{Influence of dissipation on quantum
  tunneling in macroscopic systems}},
  \href{https://doi.org/10.1103/PhysRevLett.46.211}{\emph{Phys. Rev. Lett.}
  {\bfseries 46} (1981) 211}.

\bibitem{Caldeira:1982uj}
A.O.~Caldeira and A.J.~Leggett, \emph{{Quantum tunneling in a dissipative
  system}}, \href{https://doi.org/10.1016/0003-4916(83)90202-6}{\emph{Annals
  Phys.} {\bfseries 149} (1983) 374}.

\bibitem{Caldeira:1982iu}
A.O.~Caldeira and A.J.~Leggett, \emph{{Path integral approach to quantum
  Brownian motion}},
  \href{https://doi.org/10.1016/0378-4371(83)90013-4}{\emph{Physica A}
  {\bfseries 121} (1983) 587}.

\bibitem{Choudhury:2022btc}
S.~Choudhury, S.~Panda, N.~Pandey and A.~Roy, \emph{{Four-mode squeezed states
  in de Sitter space: A study with two field interacting quantum system}},
  \href{https://arxiv.org/abs/2203.15815}{{\ttfamily 2203.15815}}.

\bibitem{Bunch:1978yq}
T.S.~Bunch and P.C.W.~Davies, \emph{{Quantum Field Theory in de Sitter Space:
  Renormalization by Point Splitting}},
  \href{https://doi.org/10.1098/rspa.1978.0060}{\emph{Proc. Roy. Soc. Lond.}
  {\bfseries A360} (1978) 117}.

\bibitem{Grain:2019vnq}
J.~Grain and V.~Vennin, \emph{{Squeezing formalism and canonical
  transformations in cosmology}},
  \href{https://doi.org/10.1088/1475-7516/2020/02/022}{\emph{JCAP} {\bfseries
  2002} (2020) 022} [\href{https://arxiv.org/abs/1910.01916}{{\ttfamily
  1910.01916}}].

\bibitem{Hu:1993qa}
B.L.~Hu and A.~Matacz, \emph{{Quantum Brownian motion in a bath of parametric
  oscillators: A Model for system - field interactions}},
  \href{https://doi.org/10.1103/PhysRevD.49.6612}{\emph{Phys. Rev. D}
  {\bfseries 49} (1994) 6612}
  [\href{https://arxiv.org/abs/gr-qc/9312035}{{\ttfamily gr-qc/9312035}}].

\bibitem{PhysRevD.45.2843}
B.L.~Hu, J.P.~Paz and Y.~Zhang, \emph{Quantum brownian motion in a general
  environment: Exact master equation with nonlocal dissipation and colored
  noise}, \href{https://doi.org/10.1103/PhysRevD.45.2843}{\emph{Phys. Rev. D}
  {\bfseries 45} (1992) 2843}.

\bibitem{PhysRevD.53.2012}
J.J.~Halliwell and T.~Yu, \emph{Alternative derivation of the hu-paz-zhang
  master equation of quantum brownian motion},
  \href{https://doi.org/10.1103/PhysRevD.53.2012}{\emph{Phys. Rev. D}
  {\bfseries 53} (1996) 2012}.

\bibitem{Huang:2022hru}
Y.-W.~Huang and W.-M.~Zhang, \emph{{Exact Master Equation for Quantum Brownian
  Motion with Generalization to Momentum-Dependent System-Environment
  Couplings}},  \href{https://arxiv.org/abs/2204.09965}{{\ttfamily
  2204.09965}}.

\bibitem{Ferialdi_2016}
L.~Ferialdi, \emph{Exact closed master equation for gaussian non-markovian
  dynamics},
  \href{https://doi.org/10.1103/physrevlett.116.120402}{\emph{Physical Review
  Letters} {\bfseries 116} (2016) }.

\bibitem{Di_si_2014}
L.~Diósi and L.~Ferialdi, \emph{General non-markovian structure of gaussian
  master and stochastic schrödinger equations},
  \href{https://doi.org/10.1103/physrevlett.113.200403}{\emph{Physical Review
  Letters} {\bfseries 113} (2014) }.

\bibitem{2008JPhA...41q5304W}
R.S.~{Whitney}, \emph{{Staying positive: going beyond Lindblad with
  perturbative master equations}},
  \href{https://doi.org/10.1088/1751-8113/41/17/175304}{\emph{Journal of
  Physics A Mathematical General} {\bfseries 41} (2008) 175304}
  [\href{https://arxiv.org/abs/0711.0074}{{\ttfamily 0711.0074}}].

\bibitem{RevModPhys.88.021002}
H.-P.~Breuer, E.-M.~Laine, J.~Piilo and B.~Vacchini, \emph{Colloquium:
  Non-markovian dynamics in open quantum systems},
  \href{https://doi.org/10.1103/RevModPhys.88.021002}{\emph{Rev. Mod. Phys.}
  {\bfseries 88} (2016) 021002}.

\bibitem{Moustos:2016lol}
D.~Moustos and C.~Anastopoulos, \emph{{Non-Markovian time evolution of an
  accelerated qubit}},
  \href{https://doi.org/10.1103/PhysRevD.95.025020}{\emph{Phys. Rev. D}
  {\bfseries 95} (2017) 025020}
  [\href{https://arxiv.org/abs/1611.02477}{{\ttfamily 1611.02477}}].

\bibitem{Nicacio:2022djs}
F.~Nicacio and R.N.P.~Maia, \emph{{Gauge Quantum Thermodynamics of Time-local
  non-Markovian Evolutions}},
  \href{https://arxiv.org/abs/2204.02966}{{\ttfamily 2204.02966}}.

\bibitem{Prudhoe:2022pte}
S.~Prudhoe and S.~Shandera, \emph{{Classifying the non-Markovian,
  non-time-local, and entangling dynamics of an open quantum system}},
  \href{https://arxiv.org/abs/2201.07080}{{\ttfamily 2201.07080}}.

\bibitem{Spaventa:2022hks}
G.~Spaventa and P.~Verrucchi, \emph{{Nature and origin of the operators
  entering the master equation of an open quantum system}},
  \href{https://arxiv.org/abs/2209.14209}{{\ttfamily 2209.14209}}.

\bibitem{Chruscinski:2022hvy}
D.~Chru\'sci\'nski, \emph{{Dynamical maps beyond Markovian regime}},
  \href{https://arxiv.org/abs/2209.14902}{{\ttfamily 2209.14902}}.

\bibitem{Brasil_2013}
C.A.~Brasil, F.F.~Fanchini and R.d.J.~Napolitano, \emph{A simple derivation of
  the lindblad equation},
  \href{https://doi.org/10.1590/s1806-11172013000100003}{\emph{Revista
  Brasileira de Ensino de Física} {\bfseries 35} (2013) 01–09}.

\bibitem{Manzano_2020}
D.~Manzano, \emph{A short introduction to the lindblad master equation},
  \href{https://doi.org/10.1063/1.5115323}{\emph{AIP Advances} {\bfseries 10}
  (2020) 025106}.

\bibitem{Baumann:2011su}
D.~Baumann and D.~Green, \emph{{Equilateral Non-Gaussianity and New Physics on
  the Horizon}},
  \href{https://doi.org/10.1088/1475-7516/2011/09/014}{\emph{JCAP} {\bfseries
  09} (2011) 014} [\href{https://arxiv.org/abs/1102.5343}{{\ttfamily
  1102.5343}}].

\bibitem{Garcia-Saenz:2018vqf}
S.~Garcia-Saenz and S.~Renaux-Petel, \emph{{Flattened non-Gaussianities from
  the effective field theory of inflation with imaginary speed of sound}},
  \href{https://doi.org/10.1088/1475-7516/2018/11/005}{\emph{JCAP} {\bfseries
  11} (2018) 005} [\href{https://arxiv.org/abs/1805.12563}{{\ttfamily
  1805.12563}}].

\bibitem{Lombardo:2004fr}
F.C.~Lombardo, \emph{{Influence functional approach to decoherence during
  inflation}},
  \href{https://doi.org/10.1590/S0103-97332005000300005}{\emph{Braz. J. Phys.}
  {\bfseries 35} (2005) 391}
  [\href{https://arxiv.org/abs/gr-qc/0412069}{{\ttfamily gr-qc/0412069}}].

\bibitem{Jackson:2010cw}
M.G.~Jackson and K.~Schalm, \emph{{Model Independent Signatures of New Physics
  in the Inflationary Power Spectrum}},
  \href{https://doi.org/10.1103/PhysRevLett.108.111301}{\emph{Phys. Rev. Lett.}
  {\bfseries 108} (2012) 111301}
  [\href{https://arxiv.org/abs/1007.0185}{{\ttfamily 1007.0185}}].

\bibitem{Jackson:2012qp}
M.G.~Jackson, \emph{{Integrating out Heavy Fields in Inflation}},
  \href{https://arxiv.org/abs/1203.3895}{{\ttfamily 1203.3895}}.

\bibitem{Boyanovsky:2018fxl}
D.~Boyanovsky, \emph{{Information loss in effective field theory: entanglement
  and thermal entropies}},
  \href{https://doi.org/10.1103/PhysRevD.97.065008}{\emph{Phys. Rev. D}
  {\bfseries 97} (2018) 065008}
  [\href{https://arxiv.org/abs/1801.06840}{{\ttfamily 1801.06840}}].

\bibitem{Boyanovsky:2018soy}
D.~Boyanovsky, \emph{{Imprint of entanglement entropy in the power spectrum of
  inflationary fluctuations}},
  \href{https://doi.org/10.1103/PhysRevD.98.023515}{\emph{Phys. Rev. D}
  {\bfseries 98} (2018) 023515}
  [\href{https://arxiv.org/abs/1804.07967}{{\ttfamily 1804.07967}}].

\bibitem{Burrage:2018pyg}
C.~Burrage, C.~K\"ading, P.~Millington and J.~Min\'a\v{r}, \emph{{Open quantum
  dynamics induced by light scalar fields}},
  \href{https://doi.org/10.1103/PhysRevD.100.076003}{\emph{Phys. Rev. D}
  {\bfseries 100} (2019) 076003}
  [\href{https://arxiv.org/abs/1812.08760}{{\ttfamily 1812.08760}}].

\bibitem{Burrage:2019szw}
C.~Burrage, C.~K\"ading, P.~Millington and J.~Min\'a\v{r}, \emph{{Influence
  functionals, decoherence and conformally coupled scalars}},
  \href{https://doi.org/10.1088/1742-6596/1275/1/012041}{\emph{J. Phys. Conf.
  Ser.} {\bfseries 1275} (2019) 012041}
  [\href{https://arxiv.org/abs/1902.09607}{{\ttfamily 1902.09607}}].

\bibitem{Pinol:2020cdp}
L.~Pinol, S.~Renaux-Petel and Y.~Tada, \emph{{A manifestly covariant theory of
  multifield stochastic inflation in phase space: solving the discretisation
  ambiguity in stochastic inflation}},
  \href{https://doi.org/10.1088/1475-7516/2021/04/048}{\emph{JCAP} {\bfseries
  04} (2021) 048} [\href{https://arxiv.org/abs/2008.07497}{{\ttfamily
  2008.07497}}].

\bibitem{Choudhury:2022ati}
S.~Choudhury, S.~Dey, R.M.~Gharat, S.~Mandal and N.~Pandey,
  \emph{{Schwinger-Keldysh path integral formalism for a Quenched Quantum
  Inverted Oscillator}},  \href{https://arxiv.org/abs/2210.01134}{{\ttfamily
  2210.01134}}.

\bibitem{Kading:2022jjl}
C.~K\"ading and M.~Pitschmann, \emph{{A new method for directly computing
  reduced density matrices}},
  \href{https://arxiv.org/abs/2204.08829}{{\ttfamily 2204.08829}}.

\bibitem{2002quant.ph..9153B}
H.-P.~{Breuer}, A.~{Ma} and F.~{Petruccione}, \emph{{Time-local master
  equations: influence functional and cumulant expansion}}, {\emph{arXiv
  e-prints} (2002) quant}
  [\href{https://arxiv.org/abs/quant-ph/0209153}{{\ttfamily
  quant-ph/0209153}}].

\bibitem{Boyanovsky:2015xoa}
D.~Boyanovsky, \emph{{Effective Field Theory out of Equilibrium: Brownian
  quantum fields}},
  \href{https://doi.org/10.1088/1367-2630/17/6/063017}{\emph{New J. Phys.}
  {\bfseries 17} (2015) 063017}
  [\href{https://arxiv.org/abs/1503.00156}{{\ttfamily 1503.00156}}].

\bibitem{Kamenev_2009}
A.~Kamenev and A.~Levchenko, \emph{Keldysh technique and non-linear sigma
  model: basic principles and applications},
  \href{https://doi.org/10.1080/00018730902850504}{\emph{Advances in Physics}
  {\bfseries 58} (2009) 197–319}.

\bibitem{doi:10.1119/1.2957889}
W.B.~Case, \emph{Wigner functions and weyl transforms for pedestrians},
  \href{https://doi.org/10.1119/1.2957889}{\emph{American Journal of Physics}
  {\bfseries 76} (2008) 937}
  [\href{https://arxiv.org/abs/https://doi.org/10.1119/1.2957889}{{\ttfamily
  https://doi.org/10.1119/1.2957889}}].

\bibitem{Burgess:2007pt}
C.P.~Burgess, \emph{{Introduction to Effective Field Theory}},
  \href{https://doi.org/10.1146/annurev.nucl.56.080805.140508}{\emph{Ann. Rev.
  Nucl. Part. Sci.} {\bfseries 57} (2007) 329}
  [\href{https://arxiv.org/abs/hep-th/0701053}{{\ttfamily hep-th/0701053}}].

\bibitem{Burgess:2021luo}
C.P.~Burgess, R.~Holman and G.~Kaplanek, \emph{{Quantum Hotspots: Mean Fields,
  Open EFTs, Nonlocality and Decoherence Near Black Holes}},
  \href{https://doi.org/10.1002/prop.202200019}{\emph{Fortsch. Phys.}
  {\bfseries 2022} (2021) 2200019}
  [\href{https://arxiv.org/abs/2106.10804}{{\ttfamily 2106.10804}}].

\bibitem{Pi:2012gf}
S.~Pi and M.~Sasaki, \emph{{Curvature Perturbation Spectrum in Two-field
  Inflation with a Turning Trajectory}},
  \href{https://doi.org/10.1088/1475-7516/2012/10/051}{\emph{JCAP} {\bfseries
  10} (2012) 051} [\href{https://arxiv.org/abs/1205.0161}{{\ttfamily
  1205.0161}}].

\bibitem{PhysRevA.36.3868}
R.~Simon, E.C.G.~Sudarshan and N.~Mukunda, \emph{Gaussian-wigner distributions
  in quantum mechanics and optics},
  \href{https://doi.org/10.1103/PhysRevA.36.3868}{\emph{Phys. Rev. A}
  {\bfseries 36} (1987) 3868}.

\bibitem{NIST:DLMF}
``{\it NIST Digital Library of Mathematical Functions}.''
  http://dlmf.nist.gov/, Release 1.1.6 of 2022-06-30.

\end{thebibliography}\endgroup
